\documentclass[5p]{elsarticle}
\usepackage{graphicx}
\usepackage{amsmath,amssymb} 
\usepackage{wrapfig}
\usepackage{color}
\usepackage{soul} 
\usepackage{fixltx2e,hyperref}

\makeatletter \define@key{Gin}{dpi}{\pdfpxdimen=\dimexpr 1in/(#1)\relax} \makeatother

\usepackage{algorithm}
\usepackage{algpseudocode} 

\usepackage{tabu}
\usepackage{booktabs}

\usepackage{array}
\usepackage{dcolumn}
\newcolumntype{.}{D{.}{.}{-1}}
\newcolumntype{d}[1]{D{.}{.}{#1}}

\usepackage{dsfont} 

\title{\bf Practical Shape Analysis and Segmentation Methods for Point Cloud Models}

\begin{document}
\begin{frontmatter}



\author[uconn]{Reed M. Williams}
\author[uconn]{Horea T. Ilie\c{s}\corref{cor1}}
\ead{ilies@engr.uconn.edu}
\cortext[cor1]{Corresponding author}
\address[uconn]{Department of Mechanical Engineering\\ University of Connecticut\\ Storrs, CT 06269}


\begin{abstract}
Current point cloud processing algorithms do not have the capability to automatically extract semantic information from the observed scenes, except in very specialized cases. Furthermore, existing mesh analysis paradigms cannot be directly employed to \textit{automatically} perform typical shape analysis tasks directly on point cloud models.

We present a potent framework for shape \textit{analysis}, \textit{similarity} and \textit{segmentation} of noisy point cloud models for real objects of engineering interest, models that may be incomplete. The proposed framework relies on spectral methods and the heat diffusion kernel to construct compact shape signatures, and we show that the framework supports a variety of clustering techniques that have traditionally been applied only on mesh models. We developed and implemented one practical and convergent estimate of the Laplace-Beltrami operator for point clouds as well as a number of clustering techniques adapted to work directly on point clouds to produce geometric features of engineering interest. The key advantage of this framework is that it supports practical shape analysis capabilities that operate directly on point cloud models of objects \textit{without} requiring surface reconstruction or global meshing. We show that the proposed technique is robust against typical noise present in possibly incomplete point clouds, and segment point clouds scanned by depth cameras (e.g. Kinect) into semantically-meaningful sub-shapes.  

\end{abstract}

\end{frontmatter}


\section{Introduction}

3D cameras are now being produced commercially in larger numbers and at lower cost than ever before. Such sensors provide a low entry barrier to the field of computer vision, and are allowing practically everyone to capture and integrate digital models of reality directly into their applications or engineering design processes. A depth camera generates a point cloud model, a structure which, though less `complete' than a mesh model, provides a useful representation of real objects of engineering interest \cite{pauly2006point}. 

Traditionally, practical shape analysis for graphics and engineering applications has largely relied on geometric representations endowed with some kind of topological structure, especially polygonal surface meshes \cite{iyer2005shape}. Although the explicit topological information borne by mesh representations lends itself to simple discrete formulations, creating a mesh from a point cloud is an \textit{ill-posed} problem without unique solutions, and the process relies in practice on costly approximation algorithms \cite{amenta1999surface}. In fact, the general surface reconstruction problem is hard with many remaining challenges \cite{berger2013benchmark, berger2014state, dannenhoffer2017creation}. At the same time, automatically constructing a  \textit{valid} surface mesh (with guarantees on the geometric approximation and topological validity) of a point cloud is far from being a solved problem, particularly in the presence of noise, sharp features, sampling anisotropy, and incomplete point clouds \cite{berger2014state, alliez2008recent}. Consequently, the initial meshes can contain crude approximations, element degeneracies, overlaps and self-intersections, surface holes, as well as other `mesh flaws.' A good review of the typical flaws in the resulting meshes is provided in \cite{attene2013polygon}.  

\begin{figure}[ht]
\begin{center}
    \begin{tabular}{cc}
\vspace{5pt}
\hspace{0pt}   \includegraphics[totalheight=180pt]{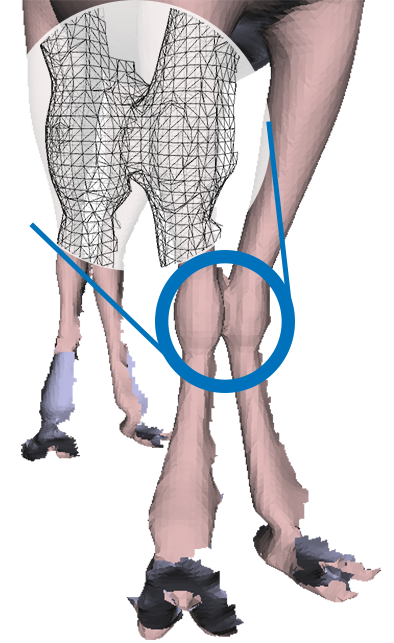} 
& \hspace{0pt} \includegraphics[totalheight=180pt]{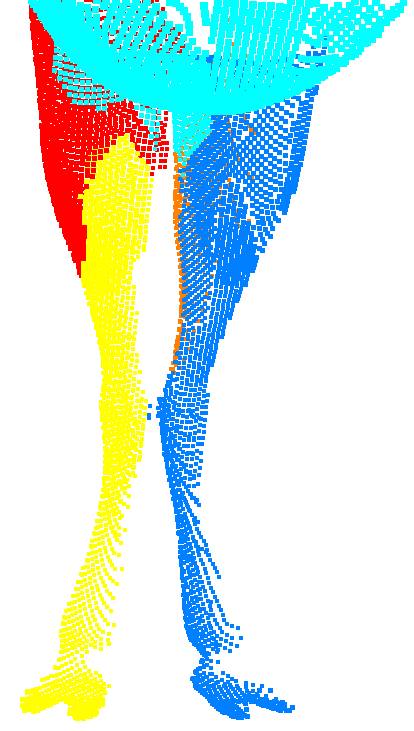} 
\\
    {\small (a) } & {\small (b) }
    \end{tabular}
\end{center}
\caption{(a) The \textit{incomplete} camel model of \cite{rosa_sig09} -- shown also in Figure \ref{camel2} --  meshed by RIMLS Marching Cubes implemented in Meshlab conjoins the legs at the knee where points get close; (b) The legs of the original point cloud remain decoupled when employing the techniques presented in this paper.} \label{camel}
\end{figure}

For example, meshing the point cloud camel model of \cite{rosa_sig09} with the popular RIMLS Marching Cubes \cite{oztireli2009feature} implemented in Meshlab conjoins the legs at the knee, as shown in Figure \ref{camel}(a), where the points in the cloud corresponding to two different ``legs'' of the camel model get relatively close to each other. This, in turn, leads to drastic topological changes in the meshed model that are not found in the original physical model. Clearly, such topological errors propagate through any further geometry processing, geodesic algorithm, or analysis, such as mesh segmentation, subsequently developed on this mesh. At the same time, the framework proposed in this paper directly and robustly develops algorithms on and segments the point cloud model itself without producing such topological changes, as illustrated in Figure \ref{camel}(b). This behavior is not specific to any one meshing algorithm. In fact, different meshing algorithms may produce meshes of the \textit{same} point cloud having \textit{different} geometric and topological properties. Operating on and analyzing the point cloud directly, rather than a mesh, avoids introducing topological errors to the analysis pipeline at an early stage where such errors will cascade into the rest of the analysis.

Because of these difficulties with mesh generation, see also \cite{pfister2004point}, the initial meshing step is almost always followed by an application-dependent mesh repair process, which introduces additional approximations that may or may not conform to the original physical model \cite{attene2013polygon}. In fact, arguments supporting direct processing of point clouds in the field of CAD/CAM have appeared for some time \cite{Cripps2003Algorithms, Azariadis2005Drawing, rosen2016using}. Furthermore, many important engineering applications require geometric information in higher dimensional spaces, e.g., in space-time (4D) \cite{foteinos4d} or configuration spaces (6D) \cite{nelaturi2013solving}, but meshing higher dimensional point clouds gets even harder as the dimension goes up. 

It is apparent that understanding the semantics of the output of a range camera \textit{without} requiring meshing\footnote{It is worth noting here that robust surface reconstruction algorithms do require {\it a priori} estimates of \textit{differential operators} on the point cloud data \cite{digne2014diffop} in order to construct the geometry and topology of the meshed model.} or global surface reconstruction would avoid a critical bottleneck in a host of key application contexts relying on the fundamental concept of shape similarity, including engineering design, scene recognition, digital shape reconstruction, functional co-robots, autonomous navigation, and part sorting, as well as virtual and augmented reality. At the core of shape analysis for point clouds are \textit{compact shape descriptors} tailored to this discrete representation, as well as shape similarity, comparison, and segmentation capabilities, which are omnipresent in applications as varied as industrial product design, assistive technologies, medical diagnosis, and quality control \cite{iyer2005shape}.

In this paper, we formulate a potent framework for performing shape analysis and segmentation directly on point clouds, including those of engineering significance obtained from engineering components and systems, that may be noisy and/or incomplete. To this end, we have developed \cite{williams2012towards,williams2014shape} and implemented the Symmetric Point Cloud Laplacian (SPCL), a symmetric version of the PCD Laplacian (PCDL) \cite{belkin2009constructing}, which we show retains the convergence guarantees of the PCDL, and allows us to confidently apply physics-based signature methods to point cloud models. Furthermore, we investigate several segmentation schemes including Heat Walks and a novel point cloud clustering method based on the Vietoris-Rips filtration. We apply these methods on signature values over point cloud models to produce segmentations of shapes into geometric features of engineering interest.

\subsection{Prior Work}

\subsubsection{Description and Similarity}

Recently, high-quality ``physics-based'' methods aimed at providing compact shape description (e.g., shape signatures) and similarity have been developed for triangular surface mesh models. We focus on physics-based because they are most related to the work presented in this manuscript. These methods are founded in the mathematical firmament of geometry-dependent physical processes (e.g., thermal conduction). Diffusion-type processes, behavior corresponding to second-order partial differential equations in space and time, are intrinsically dependent upon the local and global geometry and topology of the continuous shapes over which they act, linking known mathematical descriptions of diffusion processes to the geodesic distances on that shape. The physics-based methods apply discrete versions of the mathematics of these processes to mesh models in order to construct shape descriptors that are used in downstream applications. Observe that reliance on pseudo-geodesic distances on the meshes results in robust signatures for noisy or incomplete models \cite{dey2010persistent}. 

The physics-based methods showing such promise in mesh application (such as the Wave Kernel Signature \cite{aubry2011WKS}, ShapeDNA \cite{reuter2006laplace}, and Heat Kernel Signature \cite{sun2009concise}) rely on the Laplace-Beltrami operator (LBO), which is a continuous differential operator defined on Riemannian manifolds. 
This past decade has seen much development in the area of discrete representations of the LBO, which arises from and with application to the study of diffusion \cite{knyazev2017signed}. 
In mesh-based discretizations, the ``cotan method'' \cite{hildebrandt2006convergence} or newer, more convergent Mesh Laplacian operator \cite{belkin2008discrete} may be used. On the one hand, shape descriptors that leverage the locally-descriptive power of the Laplacian via its eigensystem to compute a comparable description of shape are known as ``spectral'' methods, and are currently being explored on \textit{mesh models} by a number of groups \cite{sun2009concise, bronstein2011shape, aubry2011WKS}. At the same time, \textit{direct} point cloud model-based shape analysis has not received as much attention.

Part of the reason for this discrepancy is that surface reconstruction from point cloud models (especially to mesh models) is a thriving research field \cite{berger2014state}. However, meshing remains a challenging and ill-posed problem and cannot even be solved without a number of assumptions (what Berger et al. call ``priors'' in their analysis \cite{berger2013benchmark}). In fact, different meshing algorithms will produce different meshes, if at all, for the same point cloud, which can impact the output of any similarity or segmentation algorithm that would process the mesh. By avoiding global meshing, we effectively side-step this bottleneck.

Some research groups have developed point cloud shape analysis strategies for specific applications, but the existing research has not focused on the kinds of shapes or shape analysis tasks that an engineering analyst or designer might find useful. For example, Pokrass et al. \cite{pokrass2013partial} developed a ``bag of features''-type partial similarity method for deformable shapes based on diffusion physics, which was developed for partial similarity-based shape matching without considering semantically meaningful segmentation or internal matching. Similarly, Bronstein et al. mention in \cite{bronstein2011shape} that it may be possible to make use of a LBO for point clouds in order to perform HKS and some related analyses on them, but they do not develop the idea further beyond providing a symmetric Laplacian estimate.

Certain industrial researchers have also begun to develop point cloud-based methods for specific applications, such as GE's work \cite{ramamurthy2015geometric} to fit linear and arc segments to the point cloud output of a scanner in order to measure manufactured parts against tolerances. This work, however, remains limited in scope and deals only with 2D cross-sections of 3D point clouds and a small number of primitives. Other approaches, such as deep learning-based methods, are also beginning to show promise for classifying and segmenting point clouds, though of course learning methods have limitations such as the availability of quality training data \cite{PointNet, PointNet++}. SyncSpecCNN \cite{SyncSpecCNN} proposes to combine Laplacian eigenfunctions with a convolutional neural network, operating on a 3D graph representation of scenes.

The Point Cloud Data Laplacian (PCDL) \cite{belkin2009constructing} provides one possible estimate of the Laplacian operator on point clouds with desirable theoretical guarantees. It has been shown that the PCDL converges to the true LBO of any given shape under mild conditions on sampling. However, unlike the continuous Laplacian, the PCDL operator is \textit{not} naturally symmetric, as required by spectral methods. 

Consequently, despite its theoretical guarantees, the PCDL estimate cannot be used in any application in which the innate symmetry of the operator is essential. The spectral shape signature methods require a symmetric LBO estimate (a real matrix will have real eigenvalues and orthogonal eigenvectors if it is Hermitian, i.e., symmetric) in order to faithfully approximate the eigensystem of the Laplacian \cite{CalcBook}.

\subsubsection{Clustering and Segmentation}

Human observers easily produce, in general, high quality shape segmentations. Indeed, the current state-of-the-art for ground truth for a given (mesh) segmentation is consensus of human observers \cite{chen2009benchmark}. In order to develop good automatic segmentation methods that approach human accuracy without active human participation, the recent literature has begun exploring concepts from algebraic topology. A variety of segmentation methods have been proopsed recently, such as \cite{wang2013projective,yuan2016space}, and a recent survey of mesh-based segmentations can be found in \cite{rodriguespart}. Here we review those that are most closely related to our work. 

Rustamov, et al. \cite{GPS} showed one example of using simple k-means clustering on their Global Point Signature to produce segmentations of shapes. Skraba, et al. \cite{skraba2010persistence} have proposed using persistence-based clustering of the heat kernel signature for shape segmentation. Similarly, Dey, et al. \cite{dey2010persistent} examine persistence of local maxima of the HKS at several of the HKS's multiple  ``intrinsic scales'' in order to characterize shapes for matching, which  produces a segmentation of the shape as a byproduct. These methods all rely on surface mesh models and their innate data structure for their formulations.  By contrast, the state-of-the-art for segmenting point cloud models remains so limited that the large-scale open-source point cloud processing project, i.e., Point Cloud Library, does not include any methods for segmenting point cloud models into semantically-meaningful sub-shapes. One brief article was published recently on the topic of feature identification from laser scan data, but that article focuses only on approximate normal-based segmentation and doesn't provide any theoretical backing for their proposed techniques \cite{yoshioka2016automatic}.

\subsection{Framework Overview}

\begin{figure*}[h]
\begin{center}
    \begin{tabular}{cccc}
\hspace{0pt}   \includegraphics[trim=50px 0px 20px 20px,totalheight=80pt]{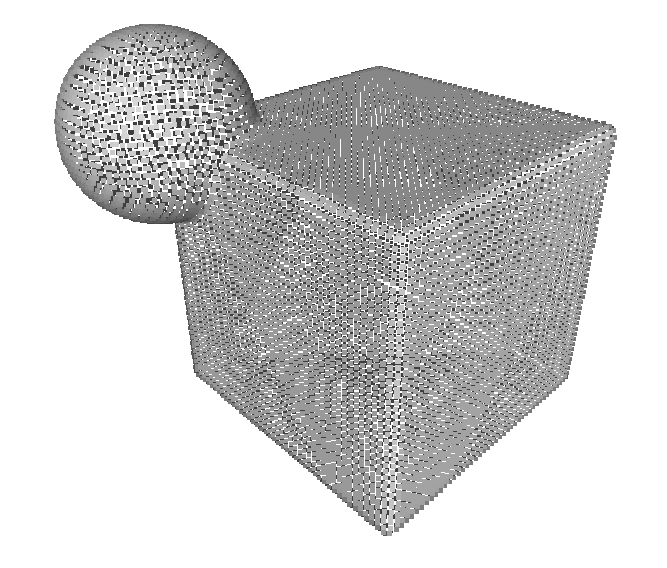} 
& \hspace{0pt} \includegraphics[trim=20px 0px 20px 20px,totalheight=80pt]{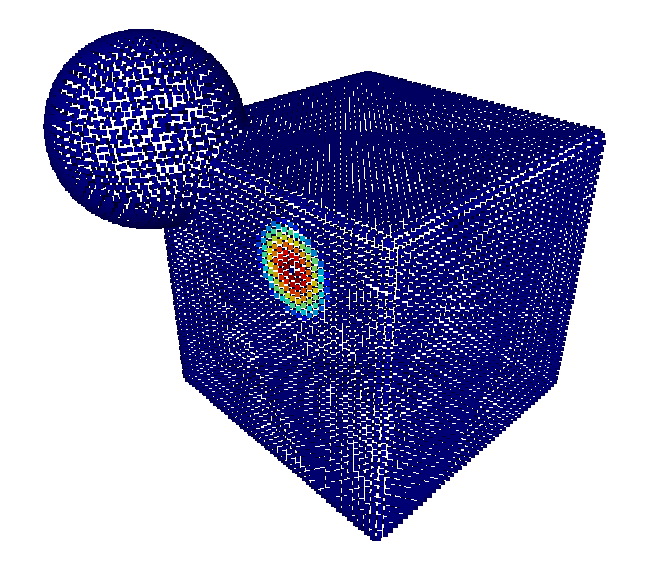} 
& \hspace{0pt} \includegraphics[trim=20px 0px 20px 20px,totalheight=80pt]{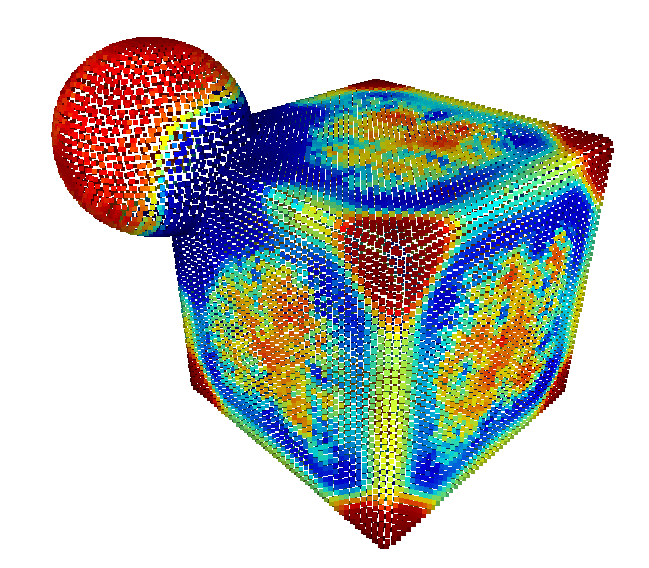}
& \hspace{0pt} \includegraphics[trim=20px 0px 20px 20px,totalheight=80pt]{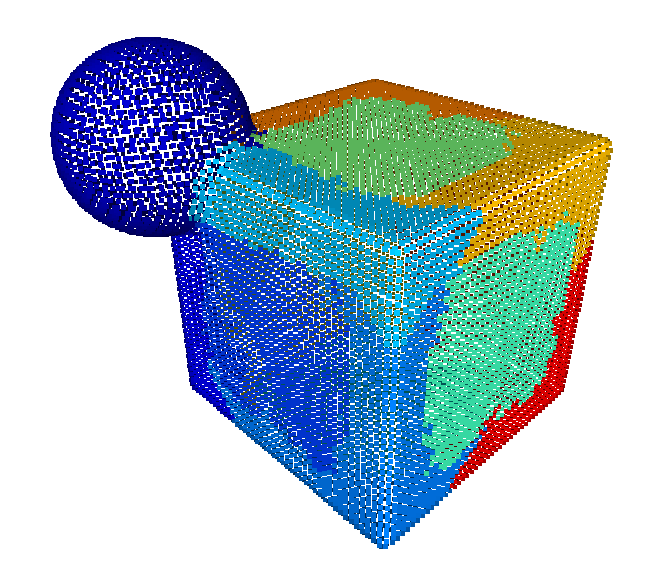} 
        \\
        {\small (a) Point cloud model} & {\small(b) Local description} & {\small c) Similarity measure} & {\small d) Automatic segmentation}
    \end{tabular}
\end{center}
\caption{The stages of our point cloud model analysis procedure, from model to SPCL to an HKS vector to one of many possible segmentations.} \label{Planfig}
\end{figure*}

Our shape analysis procedure for point cloud models, whose main steps are illustrated in Figure \ref{Planfig}, may be understood as a feed-forward network of three analyses: 
\begin{enumerate}
\item \textbf{Local shape description:} Describe the local neighborhoods on the shape via the SPCL (Section \ref{Description}). 
\item \textbf{Shape similarity measure:} Compute spectral shape signatures to enable shape matching and discrimination (Section \ref{Similarity}).
\item \textbf{Segmentation:} Segmente the shape by clustering signature values and other information from the preceding steps over the model's point cloud geometry (Section \ref{Segmentation}).
\end{enumerate}

The procedure starts with the computation of a discrete estimate of the Laplace-Beltrami operator from the input point cloud produced by the output of a depth camera. This provides a local description of our input surface at each point in the cloud. A spectral shape signature is then computed for the point cloud model using the guidelines we provide in Section \ref{HKS}, obtaining a measure of shape similarity which can be used for matching and discriminating between arbitrary shapes. 
Finally, the similarity measure and other analysis information is clustered over model points. This ensures that shape segments retain context as part of the overall shape in the form of their signature values.

\subsection{Scope and Contributions}

The main contributions of this paper are as follows.
\begin{itemize}
\item We show that spectral methods applied directly on point clouds can be effectively used construct compact shape signatures of noisy and incomplete point clouds for point cloud analysis and segmentation. We have developed and implemented one such convergent estimate with known error bounds, namely the Symmetric Point Cloud Laplacian (SPCL) \cite{williams2012towards,williams2014shape}, but any other convergent estimate of this differential operator could be used in principle.
\item We develop point cloud clustering tools for shape segmentation, including equivalences between intrinsic neighborhood sizes on meshes and point clouds and a novel point cloud formulation of a persistent homological segmentation.
\item We formulate and implement the first published unified analysis framework to perform shape description, similarity, and segmentation directly on point cloud data that does not rely on surface reconstruction or meshing. 
\item We show that the proposed techniques are robust against typical noise present in possibly incomplete point clouds, and segment them into semantically meaningful sub-shapes for point clouds scanned by depth cameras (e.g. Kinect).
\item We introduce a new clustering method based on the Vietoris-Rips filtration for grouping segmented point cloud model sub-shapes (i.e., features) into similarity classes. This technique could be used to explore geometric factorizations of and solid model reconstruction from point cloud models.
\end{itemize}

Together, the work we present provides a highly-\\automatable integrated analysis procedure for performing direct shape comparison and segmentation of point cloud models. 


\section{Local Shape Description}
\label{Description}

\subsection{The Laplace-Beltrami Operator}
\label{Laplacian}

The Laplace operator is a second-order differential operator $\Delta{}f$ which describes the variation of a differentiable function $f$ within a space. It is defined as the divergence of the gradient of the function 
\begin{equation*}
\Delta f = \nabla \cdot \nabla f
\end{equation*}
which is equivalent to the sum of the unmixed second-order partial derivatives \cite{mathbook3}. Intuitively, this operator describes the flux of the gradient field of a function in that space. The equivalent form on a Riemannian (i.e., real, smooth, equipped with an inner product) manifold is called the Laplace-Beltrami operator (LBO):
\begin{equation}
\Delta_M f = tr(H(f))
\label{trace}
\end{equation}
The Hessian $H(f)$ of the function is a square matrix of second-order partial derivatives that describes the local curvature of the function $f$ over the manifold. Taking the trace of the Hessian in Equation (\ref{trace}) keeps only the unmixed second derivatives, as in the definition of the standard Laplacian.

\begin{figure*}[!ht]
\begin{center}
 \includegraphics[dpi=96,trim=10px 0px 0px 0px,totalheight=200pt]{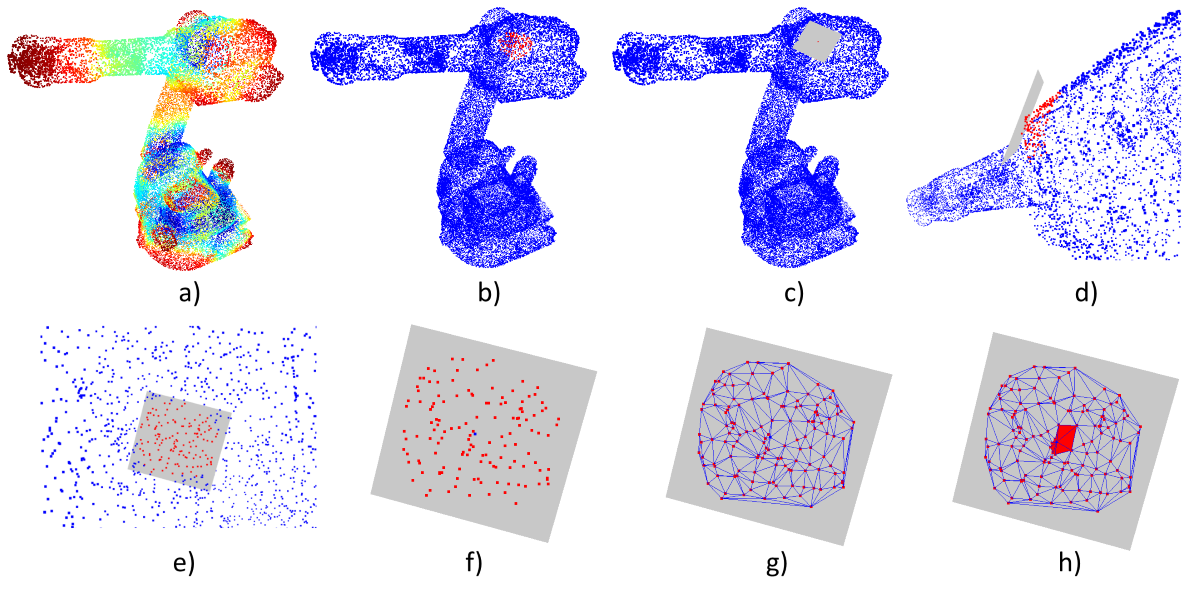} 
\end{center}
\caption{An important part of the SPCL's construction: Projecting a query point's three-dimensional noisy point cloud neighborhood into a locally-computed two-dimensional tangent space and triangulating the points in that lower-dimensional space: (a) A noised robot point cloud model; (b) The neighborhood of a point in a corner of the third linkage; (c) The local approximate tangent plane computed for the neighborhood; (d) Another angle showing the local approximate tangent plane and the neighborhood (camera is behind robot and looking toward the end effector along the left side); (e) A view of the tangent plane from inside the robot model (viewed from the -z direction of the tangent plane); (f) The neighborhood points projected into the tangent plane; (g) A 2D local Delaunay triangulation of the projected neighborhood; (h) The triangles adjacent to the query point (the triangles whose areas sum to $A_{p_{i}}$ in Equations \ref{Weq} and \ref{W-hat}) are here highlighted for clarity.} \label{SPCLneigh}
\end{figure*}

This property of describing local curvature makes the LBO of a surface a valuable tool in shape analysis. Discretizations of the LBO for various discrete representations of a surface have been the subject of intense academic interest \cite{wardetzky2007discrete}. A recently developed discretization (and the first for point clouds) is the Point Cloud Data Laplacian (PCDL) operator \cite{belkin2009constructing}. This operator is of particular interest because of its stronger than usual convergence bounds.

A native definition of the LBO on a Riemannian manifold $M$ with metric $g$ has the form
\begin{equation}
\Delta_{M} = \frac{1}{\sqrt{|\det g|}}\sum_{ij = 1}^n\frac{\partial}{\partial x_i}\left(g^{i,j}\sqrt{|\det g|}\frac{\partial }{\partial x_j}\right)
\end{equation}
where $\det$ is the determinant. The PCDL on point cloud $P$ sampled from manifold $M$ at scale $t$ takes function $f$ as an input and has a similar form:
\begin{align}
\label{Ltp}
L^t_{P}f(p) = & \frac{1}{4\pi t^2}\sum_{\sigma \in K_{d}}\frac{A_{\sigma}}{3}\sum_{q \in V(\sigma)} e^{-\delta} \left(f(p)-f(\Phi(q))\right)\\
\nonumber {\rm where} \hspace{5pt} \delta= & {||p-\Phi(q)||}^2 (4t)^{-1}
\end{align}
Here the intrinsic dimension of the manifold is 2, $\Phi$ is the projection onto an approximate local tangent plane, and $A_{\sigma}$ and $V(\sigma)$ are the area associated to and the vertices of a given simplex $\sigma$ in the local triangulation $K_d$ on that tangent plane.

\subsubsection{What Makes the Laplacian Special?}

The LBO has been used for more than estimating curvatures. First, we observe that the Laplacian on $\mathds{R}^n$ commutes with isometries on general Riemannian manifolds, which is exactly what is needed in processes whose underlying physics are independent of position and direction, such as heat diffusion and wave propagation in $\mathds{R}^n$. Hence, the eigensystem of the LBO arises naturally in spectral solutions to various physical problems on these manifolds \cite{wetzler2013laplace}.

\subsubsection{Construction of the PCDL/Matrix}

The PCDL construction is point-wise agglomerative, echoing the summations in the manifold-native form. The operator is built row-by-row from local neighborhood estimates in reduced-dimension tangent spaces approximated from the point cloud. Figure \ref{SPCLneigh} shows the creation of one such reduced-dimension tangent space and subsequent local triangulation from PCA (principal component analysis) of neighbors about the centroid of the local neighborhood. It has been shown \cite{belkin2009constructing} that, given a sampling fine enough to capture the highest-curvature features of the manifold, these local neighborhood estimates approximate the actual surface to a third order term. 

Consider a sampling of points from a Riemannian manifold such that no point on the manifold is farther than $\varepsilon$ from a point in the sampling. Let the reach $\rho$ of the surface be defined as the the radius of the largest ball which can roll to touch every point on the surface \cite{Federer1969Geometric}. This factor may be considered the ``size'' of the highest-curvature features of the surface. The angle between the actual tangent space to the surface and the approximate tangent space into which points are projected by the projection $\Phi$ in equation (\ref{Ltp}) is bounded to the order of ${\rm O}(f(\varepsilon)/\rho)$ and for points which are near one another (within ${\rm O}(\rho/2)$), the projected approximate tangent plane distance approximates the geodesic distance up to a third order term. As sampling becomes finer ($\lim_{\varepsilon \to 0}L^t_Pf(p)$), the value of the PCDL approaches that of the Laplace-Beltrami operator. This is the essence of the convergence proofs for the PCDL \cite{belkin2009constructing}.

\subsection{Symmetrizing the Point Cloud Data Laplacian}

In order for the eigensystem of an LBO estimate to be real, the estimate itself must be a real Hermitian (therefore symmetric) matrix \cite{CalcBook}. The PCDL estimate of the LBO proposed in \cite{belkin2009constructing} and summarized above is not symmetric for a general point cloud. Its asymmetry arises chiefly due to the row-wise (equivalent to point-wise) computation of representative areas and point-to-point distances in its computation.  The error between distance estimates in the PCDL is small, bounded by a third order term in the distance, but neighborness of a pair of points is discrete, and any discrepancy between distance measures can propagate inconsistency of neighborness (e.g., point \textit{i}'s row implies that point \textit{j} and it are neighbors, but point \textit{j}'s row implies they are not).

To restore the LBO's natural symmetry to the PCDL estimate, we include \cite{williams2012towards,williams2014shape} the row-point's area along with the neighbor point's area (as developed independently in \cite{liu2012point}), then average the operator across its matrix diagonal and compute the eigensystem by the generalized eigenvalue problem. Thus, we compute the SPCL matrix and eigensystem $\Phi$ and $\Lambda$ for a point cloud representing an $\varepsilon$-sampled 2D surface embedded in 3D space as 

\begin{align}
W_{i,j} = & \begin{cases}
 -S(\varepsilon) \cdot \frac{A_{p_{i}}A_{p_{j}}}{9} \cdot e^{-\delta(\varepsilon,i,j)}, & j \ne i \\ 
\\
-\sum_j{W_{i,j\neq i}}, & j = i 
\end{cases} \label{Weq}\\ 
\nonumber {\rm where \hspace{30pt}} &\\
\nonumber S(\varepsilon) = & \hspace{5pt} 4\left(\pi (2\cdot \varepsilon)^4 \right)^{-1}\\
\nonumber \delta(\varepsilon,i,j) = & \hspace{5pt} {{||p_j-p_i||}^2 4^{-1}\varepsilon^{-2}}\\
\nonumber {\rm Then \hspace{35pt}} &\\
\hat{W} = & \hspace{5pt} 2^{-1}\left(W+W^{T}\right) \\ 
\hat{W} \Phi = & diag(A_{p_i}/3) \Lambda \Phi \label{W-hat}
\end{align}

Here, $A_{p_{j}}$ is the total area of the simplices adjacent to point $p_j$ in the local triangulation near that point. The diagonal element of each row of the SPCL is the negative sum of the other elements of that row, since the Laplace operator is by definition an averaging operator \cite{mathbook3}. The local tangent spaces are approximated by PCA of each local ball of points about the neighborhood's centroid.

\subsubsection{Error Bounds and Guarantees Retained by the SPCL}

In order to improve robustness with respect to noise, the distances used in each row of the PCDL between the corresponding points were specified as post-projection Euclidean distances in the tangent plane of the point associated with that row $d_{T_i}(\Phi(p_i),\Phi(p_j))$. It has been shown that the error between $d_{T_i}(\Phi(p_i),\Phi(p_j))$ and the geodesic distance in the manifold $d_{M}(p_i,p_j)$ is bounded by a third order term for points closer than $\rho/2$ \cite{belkin2009constructing}:
\begin{equation}
d_{T_i}(u,v) \leq d_M(u,v) \leq{} d_{T_i}(u,v) + O(d^3)
\end{equation}
The size of neighborhood specified for construction of the local triangulations is $\lambda < \rho/4$. This means that the difference in tangent-space Euclidean distances between two points, each of which appearing in the tangent space belonging to the other (i.e., appearing in each others' rows in the operator), will be bounded by a third order term. 
\begin{align}
0 \leq d_M(u,v) - d_{T_i}(u,v)\leq{}& O(d^3) \\
0 \leq d_{T_2}(u,v) - d_{T_1}(u,v)\leq{}& O(d^3)
\end{align}
The error of the average of those two distance estimates will therefore also be bounded by a third order term, making the SPCL no worse an estimate than the PCDL in terms of geodesic distances approximation.
\begin{equation}
avg(d_{T_1},d_{T_2}) \leq d_M(u,v) \leq avg(d_{T_1},d_{T_2}) + O(d^3)
\end{equation}

Similarly, the representative areas associated to points appearing in one another's tangent spaces is computed from the local triangulations developed on the projected points. $A_{p_j}$ is the sum of triangle areas for simplices containing $p_j$. Each of those triangle area terms is a multiplication of two point-to-point distances in the tangent space, a term which must therefore be bound by $O(d^3)$ error from the same area in the manifold. Averaging associated areas then, just as averaging distances above, does not increase the order of the error term. Thus, the SPCL retains the error guarantees and convergence properties of the PCDL.

\subsubsection{Normals at no additional computational cost}

\begin{figure}[thb]
\begin{center}
        \includegraphics[totalheight=150pt]{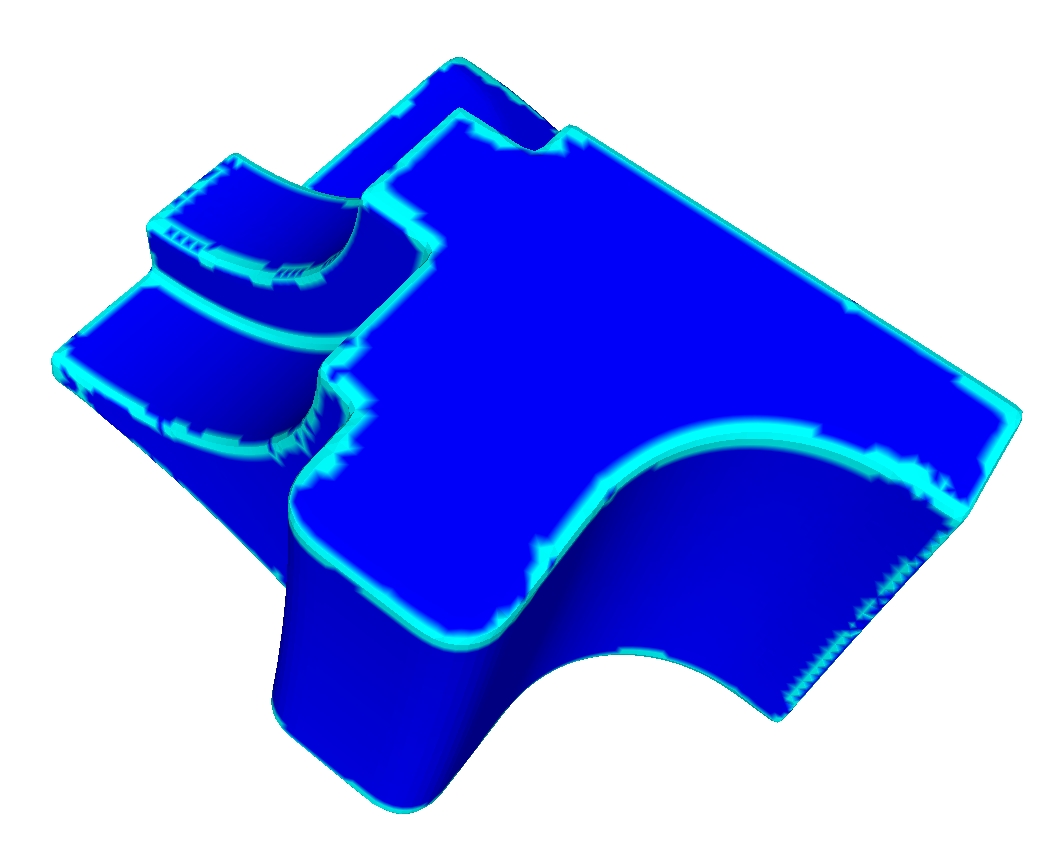} 
\end{center}
\caption{A 26525 point sampling of the ``fandisk'' model showing edges computed from point normal vector estimates obtained as a byproduct of constructing the SPCL.} \label{fig_edgefigure}
\end{figure}

The construction of the SPCL requires the estimation of approximate tangent spaces (i.e., of tangent planes for a 3D model) at each point in the sampled surface. This tangent plane is computed with Principal Components Analysis (PCA) and the eigenvector associated with the least eigenvalue corresponds to the estimated normal vector of the surface at that point. This vector value computed at each point as the SPCL is built provides an approximate normal at no additional computational cost. Figure \ref{fig_edgefigure} shows the good results of finding edges in a point cloud model by examining the maximum angle between normal vectors for the points in the local neighborhood of each point.

\section{Shape Similarity Measure}
\label{Similarity}

\subsection{Spectral Shape Signatures}
\label{HKS}

A \textbf{shape signature} is a compact representation that retains relevant information about a shape.
A useful signature should retain enough information to discriminate completely between any two general shapes or classes of shapes, while allowing straightforward computations of degree of similarity and remaining a manageable size. Many shape signatures have been proposed in the literature, but have almost exclusively operated on meshes or parametric surfaces, not point clouds \cite{GPS,sun2009concise,spinimages}.

The particular class of shape signatures known as ``spectral'' shape signatures consists of signatures whose values are computed by reference to the spectrum of the Laplacian of a shape. Many of these signatures derive meaning by analogy to physical processes that are governed by the intrinsic geometry of the space in which they act. The spectral shape signature we use in our investigations is the Heat Kernel Signature, although we observe that other shape signatures could be used within the proposed framework instead, such as the Wave Kernel Signature, Global Point Signature, ShapeDNA, or the Giaquinta--Hildebrandt operator \cite{GiaHilde} (though reformulating this last option would be more involved than for those dependent on the Laplace-Beltrami operator only). In each case, the use of the operator needs to be freed from any dependence on mesh structure, as we discuss and demonstrate in the following Sections.

\subsubsection{The Heat Kernel Signature}
\label{subsec:HKS}
The Heat Kernel Signature (HKS) is a spectral shape signature founded in the physical process of heat diffusion \cite{sun2009concise}. It has a number of desirable properties: it is invariant up to model isometry, intrinsically multi-scale, and stable under perturbations on the scale of typical depth camera noise.

In order to get a physical sense for the meaning of the HKS of a shape, consider a point source of heat applied to a point on a surface.  As time passes, the heat will diffuse on the surface away from that point. The heat kernel signature's value $k_t$ at that point is the sum total of all of the heat which has diffused away by time $t$. The heat equation on a manifold is defined as
\begin{equation}
\frac{\partial{u}}{\partial{t}} - \alpha \Delta_M u = 0 
\end{equation}
where $\alpha$ is a positive constant, $u$ is the thermal energy as a function of time and location on the surface \cite{mathbook2}, and $\Delta_M$ is the Laplace-Beltrami operator.

The heat kernel is a fundamental solution to the general heat equation \cite{hsu2002stochastic}. 
Consider an operator $H_t$ that maps any initial heat distribution $u_0(x)$ on a surface onto the distribution of heat on that surface at any time $t$
\begin{equation}
H_t(u_0(x)) = u(x,t) 
\end{equation}
A unique solution to the heat equation above may be written
\begin{equation}
H_t(u_0(x)) = \int_M k_t(x,y)u_0(y)dy 
\end{equation}
where $k_t(x,y)$ is called the \textbf{heat kernel}.
Since $H_t$ is compact, positive semi-definite, and self-adjoint, the spectral theorem from linear algebra, allows us to recast it in terms of its eigensystem \cite{elghaouilivebookopt2015}:
\begin{equation}
k_t(x,y)_H = \sum \lambda^H_i\phi^H_i(x)\phi^H_i(y).
\end{equation}
$H_t$ being a solution to the heat equation, it also has the form $H_t = e^{-t\Delta_M}$. The heat operator $H_t$ therefore shares the same eigenvectors as $\Delta_M$, and their eigenvalues are related by $\lambda_{H} = e^{-t\lambda_M}$. This relationship allows us to write the heat kernel in terms of the eigensystem of the Laplace-Beltrami operator as 
\begin{equation}
\label{HK}
k_t(x,y) = \sum e^{-\lambda_it}\phi_i(x)\phi_i(y)
\end{equation}
The quantity $k_t(x,y)$  may be considered equivalent to a measurement, for time $t$, of the amount of heat transferred from point $x$ to point $y$, for some initial distribution of heat energy on the surface $u_0(x)$. Using this quantity as a measure for similarity would require mappings between each of the neighborhoods, which would be difficult or time-consuming to define between models.

\begin{figure*}[!htb]
\begin{center}
    \begin{tabular}{ccc}
        \includegraphics[totalheight=100pt]{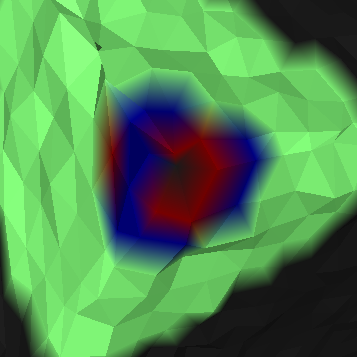} 
& \hspace{10pt} \includegraphics[totalheight=100pt]{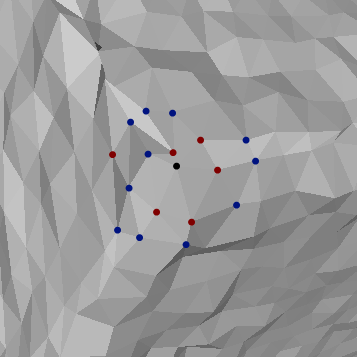} 
& \hspace{10pt} \includegraphics[totalheight=100pt]{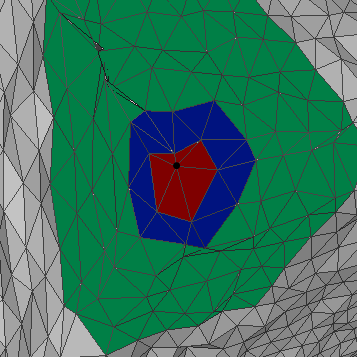}
        \\
        (a) &(b) & c)        \\
    \end{tabular}
\end{center}
\caption{(a) The non-zero points of a row of the SPCL centered about the corner of a cube, colored according to segment after being clustered into three segments by SPCL value, and shown on the noisy mesh from which this point cloud is sampled for clarity; (b) The points contained in the first two segments of the clustering shown on the mesh for clarity; (c) The actual 1-, 2-, and 6-ring mesh neighborhoods around the point of interest, colored to show correspondence with the point cloud estimate of 1-, 2-, and 6-ring neighborhoods.} \label{fig_neighs}
\end{figure*}

The heat kernel signature (HKS) of a shape is a more compact description of a shape than the heat kernel itself; it is a restriction of the heat kernel to $k_t(x,x)$, i.e., the diagonal of the heat kernel. This restriction captures the ``amount of heat'' that has diffused away from point $x$ by ``time'' $t$, and is sufficient to describe the local area of point $x$ for the purposes of similarity \cite{sun2009concise}. Restricting the heat kernel to the ``time'' domain over the model reduces the computational complexity of the signature and obviates the need to develop these local mappings for similarity. This form of the heat kernel on $M$ has the eigendecomposition
\begin{equation}
k_t(x,x)_M = \sum e^{-\lambda_it}\phi_i(x)\phi_i(x)
\label{hks:def}
\end{equation}
where $\lambda$ and $\phi$ are the eigenvalues and eigenvectors of the LBO of $M$. 

Because of its dependence on the factor $t$, the HKS is innately multi-scale and may be calculated over a range of scales, which can be considered analogous to neighborhood sizes. This allows the capture of information about a surface over different degrees of locality, e.g., examining local curvature information vs. examining the global shape of the surface (extremity, global convexity/concavity, etc.).

\subsection{Feature Points and Feature Vectors}
For matching shapes and for efficient signature storage, signatures are often queried for ``feature points.'' A typical feature point from a signature is an extreme point in some way, either locally or globally, representing a location on the surface possessing some particular interesting properties. 
In the case of a signature whose values correspond roughly to local curvature (such as HKS at low $t$-scale), a feature point in the signature may correspond to a projection from or indentation into the surface, being a point of extreme local curvature. Generally, the extreme values of a signature on a shape have shown themselves useful for discriminating between shapes and for matching \cite{bronstein2010scale}. Using lists of feature points, either as a group of features themselves or in combination as a ``feature vector,'' reduces the overhead for matching by reducing the number of values that must be compared or computed.

Numerous features and feature vectors have been defined from shape signatures, and they are typically  chosen experimentally to produce good results on some example set of models. For example, one can select all of the local maxima \cite{sun2009concise}, a fixed number of extrema \cite{aubry2011WKS}, or multiple values per feature point for a set of local maxima \cite{dey2010persistent}.

The original work proposing HKS for meshes recommended finding vertices that are locally maximal in the HKS space by examining 2-ring neighborhoods in the mesh considering those maximal points to be ``feature points.'' Since we are dealing with point clouds rather than meshes, we have at least two options for converting this 2-ring mesh neighborhood. Specifically, we can:
\begin{enumerate}
\item use those points in the point cloud model that are within the $n$ nearest neighbors of each of the $n$ nearest neighbors of a point, which can exploit, for example, the ``nearest neighborness'' property of a local Delaunay triangulation;
\item exploit the neighborness information offered by SPCL, as discussed further in Section \ref{ClustSeg}. Specifically, because of the way we define the SPCL with respect to the $\varepsilon$ ball, the points corresponding to non-zero values in a row of the SPCL are those points which, if the point cloud were well-meshed, would be vertices in the 6-ring neighborhood of the vertex which corresponds to that row. Those values are computed from local areas and distances from the point corresponding to the row, so for a relatively uniformly-sampled point cloud, simple k-means clustering the points by value into three clusters produces a reasonable approximation to the 1-ring neighborhood, the 2-ring neighborhood, and the 6-ring neighborhood as illustrated in Figure \ref{fig_neighs}.
\end{enumerate}

Additionally, it has been recently proposed \cite{dey2010persistent,skraba2010persistence} that examining the values of a signature in terms of topological persistence may yield a high quality feature vector. Such methods are based on ideas from algebraic topology, such as the Vietoris-Rips filtration, which computes linkages between elements as a network is grown between them \cite{hausmann1995vietoris}. Dey et al. \cite{dey2010persistent} proposed a feature-point selection method based on homological persistence. The method seeks to grow segments around mesh elements by a procedure that begins with the segment (initially, a single triangle) of least persistence and joins it to the region adjacent to a particular edge of that triangle. This growing and merging of regions to seek persistent feature points coincidentally produces a segmentation of the shape under examination, resulting in segments labeled under the name of the ``central triangle'' of highest value in each region. They observe experimentally that 15 feature points defined by their region-merging method are ``usually sufficient'' to differentiate between models, and then calculate the HKS values at the triangles so selected for each model at fifteen different $t$ scales. This 15-dimensional feature vector for each feature point in the model exploits the multi-scale nature of the HKS to aid in model discrimination. 

The algorithm described relies heavily on mesh structure and properties. To adapt this $n \times n$ feature vector for matching on point clouds, we reformulate the segmentation algorithm proposed in \cite{skraba2010persistence} to produce feature points. We find a point of maximum HKS value within each region of the segmentation (at a particular $t$ scale). This point provides a natural analogue to the ``central triangle'' of that algorithm.


\section{Segmentation}
\label{Segmentation}


Humans are typically very good at segmenting shapes into semantically meaningful sub-shapes \cite{benhabiles2009framework}. Therefore, a high quality automatic segmentation of an engineering model should consist of a very similar set of segments to a human segmentation of the same model \cite{chen2009benchmark}. Unfortunately, the decomposition of a shape into semantically meaningful ``features'' is a widely open problem not only because the concept of a feature is almost always context/application dependent \cite{han1998feature}, but also because the so-called intersecting features still pose a significant challenge to the state of the art feature recognition methods. 

\subsection{Clustering for Segmentation}
\label{ClustSeg}

We develop a new segmentation which may be conceptually considered to be a 0-dimensional Vietoris-Rips filtration and additionally explore the implementation within our framework of two established methods: Heat Walks \cite{heatwalks} and a particular instance of curvature-aware segmentation \cite{lavoue2005new}. These and other segmentation methods can be implemented within this framework and used to segment point clouds corresponding to organic and engineering shapes characterized by intricate feature patterns including through holes, countersinks, chamfers, etc.


Defining a general unsupervised clustering method that produces only engineering-relevant design features as clusters without matching to a finite set of primitives is a very challenging problem. Our solution is to apply tools from algebraic topology and homology to develop a segmentation method that, with some parameter adjustments, produces segmentations separating large-scale engineering features, such as fins or holes, regardless of the shape of the model. 

Guibas et al. \cite{skraba2010persistence} developed a shape segmentation method that uses the heat kernel signature of a shape calculated on a surface mesh model. Their method was intended for use on ``deformable shapes,'' which could be a very broad class of models; their demonstrations, however, were on only ``organic'' models regularly observed under various deformations. Specifically, the method described in \cite{skraba2010persistence} is specified as an examination of persistence of regions grown around HKS maxima defined over 1-ring mesh neighborhoods. In a point cloud model, there is no mesh and therefore no $n$-ring neighborhood structure. However, we propose to use the SPCL itself to find equivalent neighborhoods.

A recent clustering method that can be applied to both meshes as well as point clouds is described by van Kaick et al \cite{vankaick2014shape} and uses the approximate convexity-based method. As discussed above in Section \ref{HKS}, the HKS is proportional to curvature at low $t$-values and encodes more global information for higher $t$-values. This allows segmentation methods based on the HKS to segment shapes based on local curvature, just as in \cite{vankaick2014shape}, but also by more heuristic measures that allow segmentations to develop more naturally as in so-called ``part-based'' (or ``higher-level'') segmentations \cite{agathos20073D}. It is also important to note that the segmentation described in \cite{vankaick2014shape} produces segments, but does not produce the semantics of these segments. This is the convexity-based segmentation method developed in that work does not rely on a similarity measure. 

On the other hand, our segmentation relies on a specific similarity metric so that we not only produce segmented point clouds, but retain information about what shape those point clouds are. This information could be used in downstream processing, for example for reconstructing parametrized solid models of the point cloud.

\subsubsection{From Mesh Clustering to Point Cloud Clustering}

The SPCL matrix described in Section \ref{Laplacian} may be understood as a weighted adjacency graph for the point cloud it describes. Local balls of points in sufficiently dense point clouds may be considered in some important ways analogous to {\em n}-ring neighborhoods in a mesh representation of an object. It has been shown that the Euclidean distance between the tangent space neighbors as computed in the construction of the SPCL approximates the geodesic distance between those same points on the surface to a third order error term. This means that the projection of the local neighborhood of points into the approximate tangent space is a good local approximation of the surface under consideration. We can exploit this approximation to estimate geodesic distances locally from Euclidean distances, to estimate neighboring regions and cliques, and to find the associated representative surface areas of the points in the cloud.

Since neighbors in the local tangent spaces are neighbors in the surface, the local neighborhoods used to create the SPCL are neighborhoods in the surface represented by the point cloud model, as well.  Thus, we can treat the LBO as a weighted neighborhood graph of the point cloud. If we ignore the weights in this graph, and consider only the connectivity information,  we have a matrix of neighborhood information in which the non-zero points in each LBO row are analogous to the points in a particular size of neighborhood in an equivalent mesh. Constructing the SPCL using the same neighborhood size suggested for the PCDL, each non-zero point in a given row in the LBO is a member of the 6-ring mesh neighborhood of the point corresponding to that row (see Figure \ref{fig_neighs}). This correspondence holds because the neighborhood size used in computing the LBO rows is based on the intrinsic scale of the point cloud (which is equivalent to the scale of a mesh with vertices at those points).

Importantly, this implied structure allows us to compute locality and connectivity for regions on the surface by simple query, something simple in a complete surface mesh, but previously not straightforward on point clouds. We can also put this structure to work in defining clusterings that operate not just on the signature values on a point cloud, but locally over the surface because we can examine the way the signature values change across the local neighborhoods on the point cloud.

Our persistent clustering method (See Algorithm \ref{algofig_clust}) takes as inputs the point cloud model itself, the SPCL and the HKS thereof, and a user-selected scale $\tau$. The highest $\nu$ values (we note that $\nu=10$ seems to give good results) of the weight matrix of the SPCL are taken as neighbors of the point corresponding to that row. That is, $\nu$ represents the approximate number of one-ring neighbors of the point in question in an equivalent mesh surface. The values of the HKS vector selected as clustering criterion are sorted in descending order. Beginning with the point in the cloud with the highest associated HKS value, points with maximum values within their neighborhood are assigned to their own cluster. Whenever a point is found to not be a maximum within its neighborhood, that point is assigned to the cluster of its highest-valued neighbor. Whenever two clusters are adjacent to a point that is being investigated, the maximum HKS value of each cluster is compared. If the difference between maximums is less than or equal to the $\tau$ selected, the points belonging to both clusters are assigned to the cluster with the highest maximum HKS value and the smaller-valued cluster is removed. The clustering parameter $\tau$ can be tuned to control the number of desired segments for a particular model as discussed in Section \ref{param-dependence-appendix}.


\begin{algorithm}
\caption{Clustering}
\label{algofig_clust}
\begin{algorithmic}
\Function{Clustering}{PC,SPCL,$\nu$,HKS,$\tau$}
    \State Find $\nu$ largest SPCL weight values for each row OR nearest $\nu$ neighbors for each point
    \State sorthks $\gets$ Sort points by HKS descending.
    \State Pop sorthks
    \If{maximum in neighborhood}
        \State Assign Self-Cluster 
    \Else 
        \State Assign highest-value adjacent cluster 
    \EndIf
    \If{more than one adjacent cluster} 
        \If{$max(cluster1) - max(cluster2) \leq \tau$}
            \State Merge the smaller-valued cluster into the higher-valued cluster 
        \EndIf 
    \EndIf
\EndFunction
\end{algorithmic}
\end{algorithm}

Algorithm \ref{algofig_clust} exploits the nearest neighborness property of the SPCL to perform model segmentation, and follows a similar procedure to that introduced in \cite{skraba2010persistence}, but excises all dependence on mesh structure in favor of {\em ad-hoc} neighborhoods provided by the SPCL. We consider this persistence-based segmentation method as a \textit{0-dimensional persistent homological filtration over a restriction of the neighborhood graph induced by the SPCL to the top ten points in each SPCL row}. We use the HKS value difference between points as a distance measure in the implied graph. Defining the segmentation in this way avoids reliance on mesh structure, allowing us to meaningfully define this segmentation on a point cloud model without disregarding local connectivity information. In practice we find it expeditious to seek a particular number of segments \textit{a priori} to track the births and deaths of segments, sort by lifespan, then assign $\tau$ based on the number of segments desired/expected survive during the merging process.

\subsection{Heat Walks}
\label{heatwalk_section}

The Heat Walk \cite{heatwalks} is a segmentation method that uses the full-sized heat kernel in contrast with the HKS's restriction to the diagonal of that matrix. It assumes that the Heat Walk algorithm develops knowledge about the pathways of maximum heat flow capacity on the surface and leverages that knowledge to segment the surface resulting in a robust and stable segmentation method. 

The Heat Walk algorithm operates as follows: The heat kernel is computed for the shape, then a voting step is used to find ``exemplar'' vertices, while non-exemplar vertices are merged into accumulator regions represented by those exemplars. After this merging, there will be portions of the surface which are merged into accumulator regions but which may be more accurately understood as ``dissipator'' regions (rather than simply low-accumulation regions). Vertices of this kind are split off into their own dissipator region by a step which computes a difference between two quantities: (a) the difference between the value at each vertex and the exemplar for the accumulator region to which is belongs and (b) the difference between the value at that vertex and a uniform distribution over the model. Vertices which are ``closer'' to a uniform value over the model than to their accumulator exemplar are considered dissipative.

We implemented a version of this algorithm that does not rely on the mesh connectivity and hence can be applied directly to a point cloud model, as follows:

\begin{enumerate}
\item Construct the SPCL of the point cloud as described in Section \ref{Laplacian}.

\item Compute the heat kernel signature on the point cloud. The heat kernel at scale $t$ for points $x$ and $y$, $k_t(x,y)$, is computed as in Equation \ref{HK}.

\item Initialize the value of the ``heat potential'' $s^{1}(x)$ for each point $x$ to be $k_t(x,x)$.

\item Find for each point an exemplar point to represent that point. For a point $x$, for step $m+1$, its exemplar $e^{m+1}(x)$ is defined as the point $y$ which maximizes $min(k_t(x,y),s^m(y))$. 

\item Update $s^{m+1}(x)$ using the set of exemplars, that is
\begin{equation}
s^{m+1}(x) = max\left( min(k_t(x,y),s^m(y)) \right)
\end{equation}
where $y\in{e^m}$, the set of exemplars at the current step.

\item Update $e^{m+1}$ and $s^{m+1}$ until the set $e^{m+1}$ does not change from $e^m$, i.e., until the algorithm converges to one set of exemplars. This completes the merging into accumulator regions portion of the heat walk. See Figure \ref{heatwalkex}b.

\item Next, to find the dissipator region, compute the probability density function (PDF) for each point $p_x$, for the uniform heat distribution $p_U$, and, for each agglomerative region $i$, the ``mean cluster density'' $p_i$ (i.e, the pdf of the ``average'' point in a given accumulator region). These are defined as
\begin{equation}
p_x(y) = \frac{k_t(x,y)}{\sum_yk_t(x,y)}
\end{equation}
\begin{equation}
p_U(y) = \frac{1}{n}
\end{equation}
\begin{equation}
p_i(y) = \frac{\sum_{x\in{i}}k_t(x,y)}{\sum_{x\in{i}}\sum_yk_t(x,y)}
\end{equation}
for each accumulator region $i$ and where $n$ is the number of points in the point cloud. 

\item Once these are computed, for each point, compute the Kullback-Liebler divergence (KLD) between the PDFs, defined as:
\begin{equation}
KLD(j|k) = \sum_ij(i)log\frac{j(i)}{k(i)}
\end{equation}
The divergence of the PDF of each point with the uniform distribution and with the average distribution for the region that point has been assigned are compared. If $KLD(p_x|p_U) < KLD(p_x|p_{i(x)})$ where $i(x)$ is the agglomerative cluster in which point $x$ resides, then point $x$ is reassigned to a new dissipative cluster. That is, points $x$ with $k_t(x,y)$ distributions closer to the uniform distribution than to the distributions of the other points in its assigned cluster is mis-labeled and should instead be part of the dissipative cluster.

\end{enumerate}

Although in \cite{heatwalks} it is claimed that this segmentation method is ``fast'', in practice computing the entire heat kernel for even a medium-sized point cloud is time consuming and the storage requirements are daunting. Even a modest 10,000-point model has a heat kernel of size $10000\times10000 = 1E8$ floats (4E8 bytes) or doubles (8E8 bytes). In double precision, that requires storage on the order of 760MB. Decimation or other data reduction schemes may be used to reduce the impact of these drawbacks.

Segmentation results of the Heat Walks algorithm implemented to operate directly on point clouds are presented in Section \ref{Results}. Importantly, since the heat walk retains a record of the exemplar point of that cluster for all accumulator clusters, and since each of those points possesses a heat kernel signature value at $s^{1}(x)$, heat walks outputs can be clustered for segment types as described next.


\subsection{Clustering Segments by Type}
\label{reclust}

The segments produced by the heat walks and the persistence segmentation method described above are constrained to local relevance by the neighborhood graph over which we allow the segmentation to act. This constraint is typically helpful: it requires that segments be connected in the neighborhood graph, a necessary condition for a real shape segment. 
However, in cases where a complex shape is made up of a smaller number of distinct shape classes, it is of interest to determine how many ``types'' of shapes actually comprise the model. Since our persistent homological segmentation method relies on a physics-based signature and retains that distinctive physicality, we can compare the values of the maximum HKS values contained by the segments in order to group computed segments into a smaller number of ``sub-shapes''. We compute the number of ``types'' of shapes by using the Vietoris-Rips filtration, which can be followed by a similarity evaluation to semantically identify each class of sub-shapes. The 0-dimensional homology of the V-R filtration can be calculated quickly over the very small and low-dimensional space of the locally maximum HKS values of the segments returned by any other segmentation method. In fact, in this case, the V-R filtration is acting essentially as a nearest neighbor grouping algorithm acting on a scattering of points on a number line. This procedure can, however, produce useful results, as illustrated in Figure \ref{fig_mols}, where a segmentation showing that a number of segments detected within a shape can sometimes be reduced to show that there are some smaller number of distinct sub-shapes (or types of segment) present. 

In grouping clustered segments together to produce a more relevant result, our method here is similar to that presented in Chazal et al.'s paper on persistence in Riemannian manifolds \cite{chazal2013persistence}; as the kinds of surfaces engineers are typically interested in for design and analysis are Riemannian, there may be additional utility to be found by combining their research with our method as described above. The 0-dimensional homology method we describe and which we consider equivalent to the mesh method of Guibas \cite{skraba2010persistence} is certainly useful. The 1-dimensional or even 2-dimensional homology of that same filtration may also be of interest, but these investigations are outside the scope of this paper and the subject of ongoing research.




\section{Experimental Validations}
\label{Results}


In this Section we demonstrate the quality results obtained by our point cloud analysis and segmentation on a variety of models beyond those discussed in the previous Section. The proposed framework can be used not only to measure the similarity between whole point cloud models, but to segment these point cloud models into subsets that correspond to features.
The similarity of these features can be measured in turn, allowing the individual features to then be matched with known features from a database. Furthermore, our method can in principle detect the number of distinct shape classes that make up a point cloud model, as illustrated in Figure \ref{fig_mols}, which could potentially be used as a way to explore geometric factorizations as well as solid/geometric model reconstructions of point cloud models.

\begin{figure}[!htbp]
\begin{center}
    \begin{tabular}{cc}
  \hspace{0pt}   \includegraphics[totalheight=70pt]{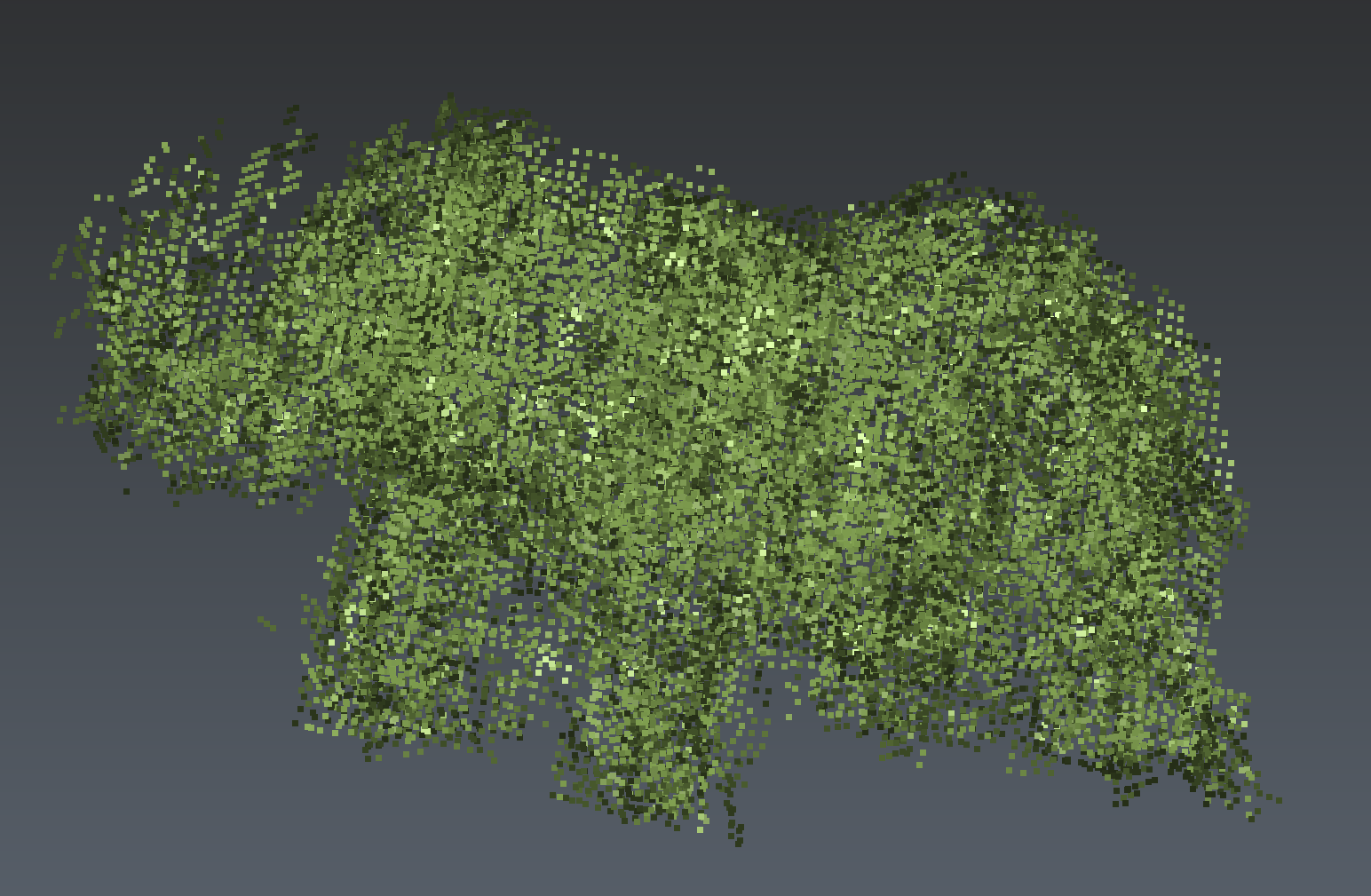} & 
  \hspace{0pt} \includegraphics[totalheight=70pt]{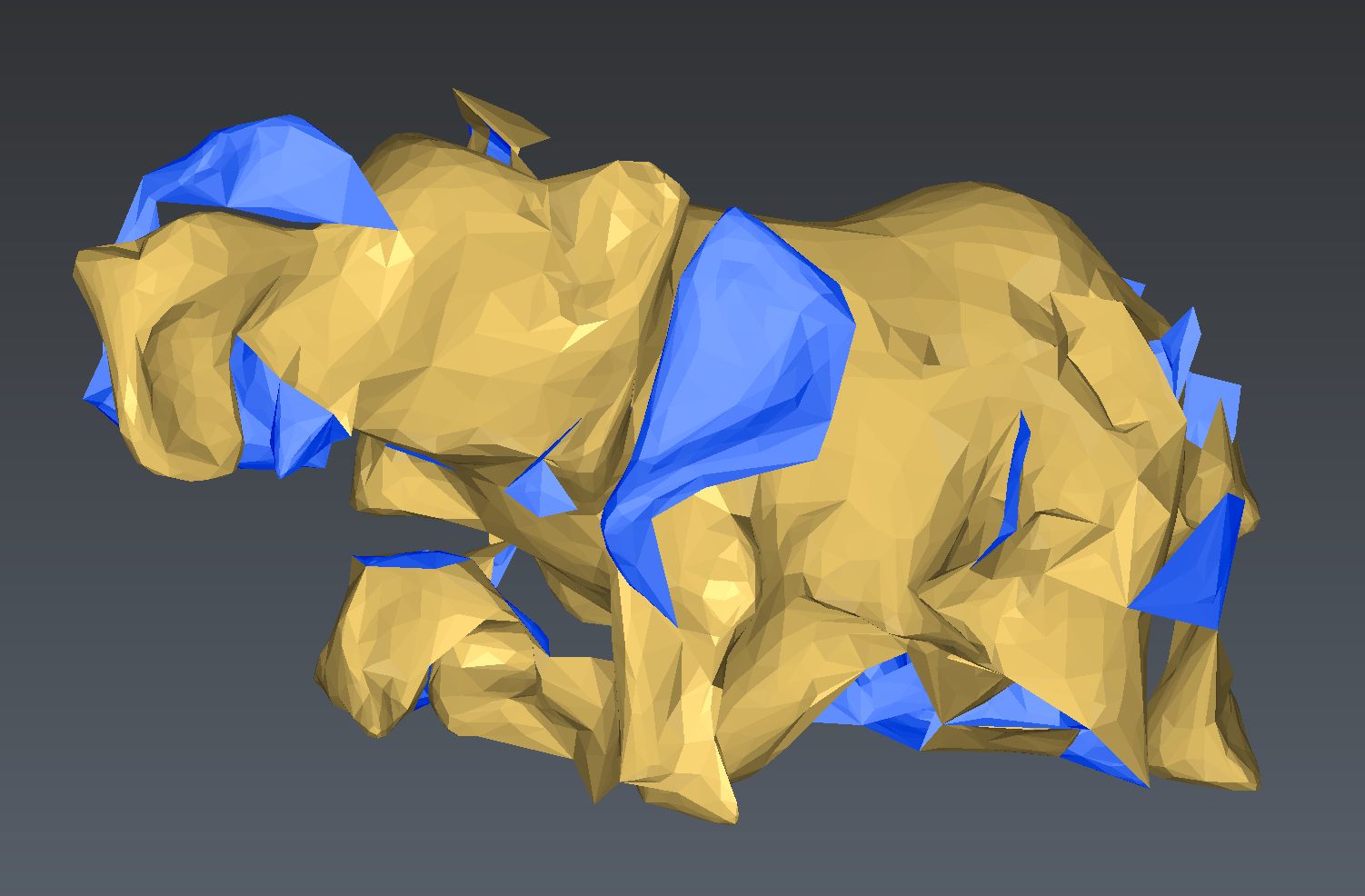} \\
  \hspace{0pt}   \includegraphics[totalheight=75pt]{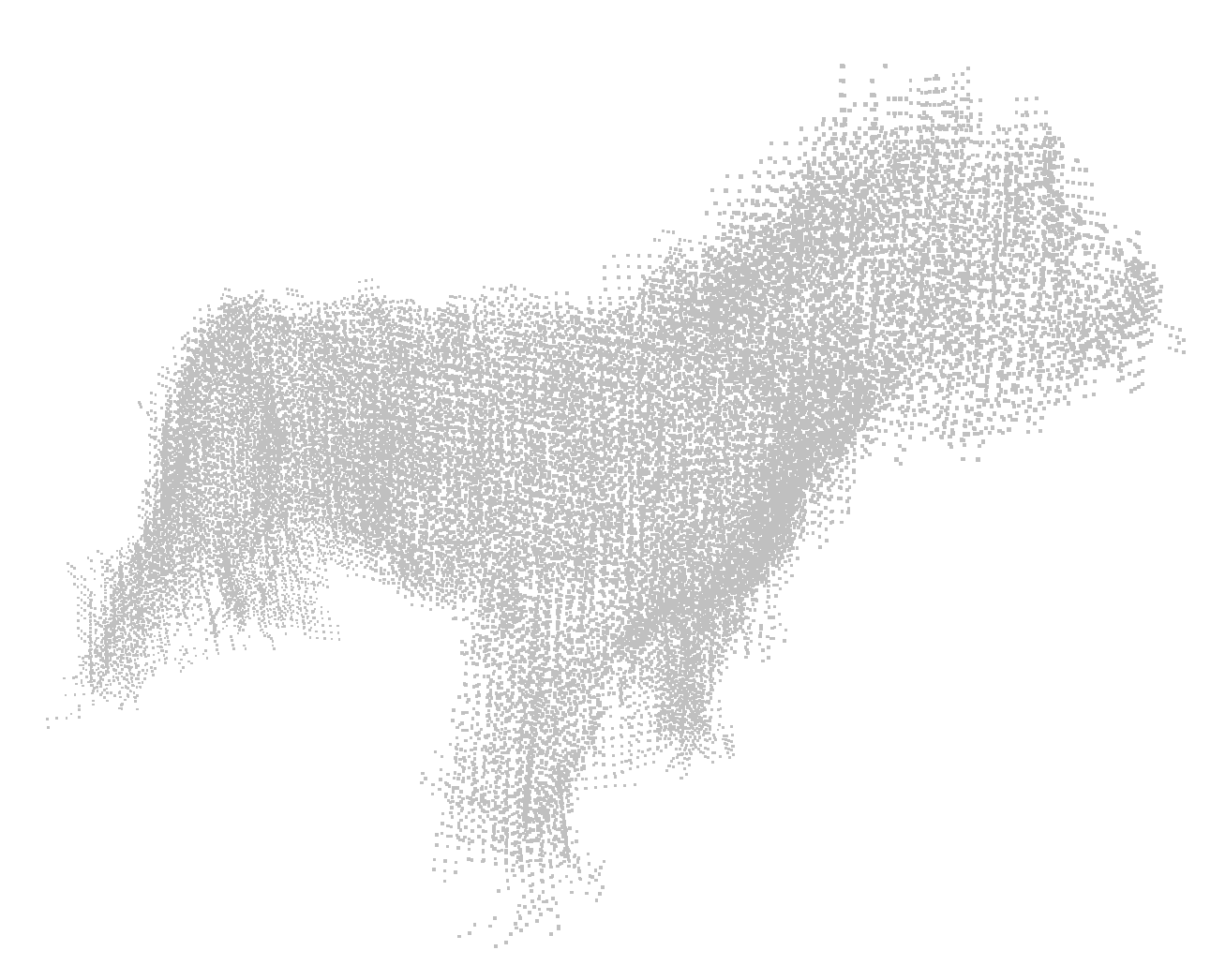} & 
  \hspace{0pt} \includegraphics[totalheight=75pt]{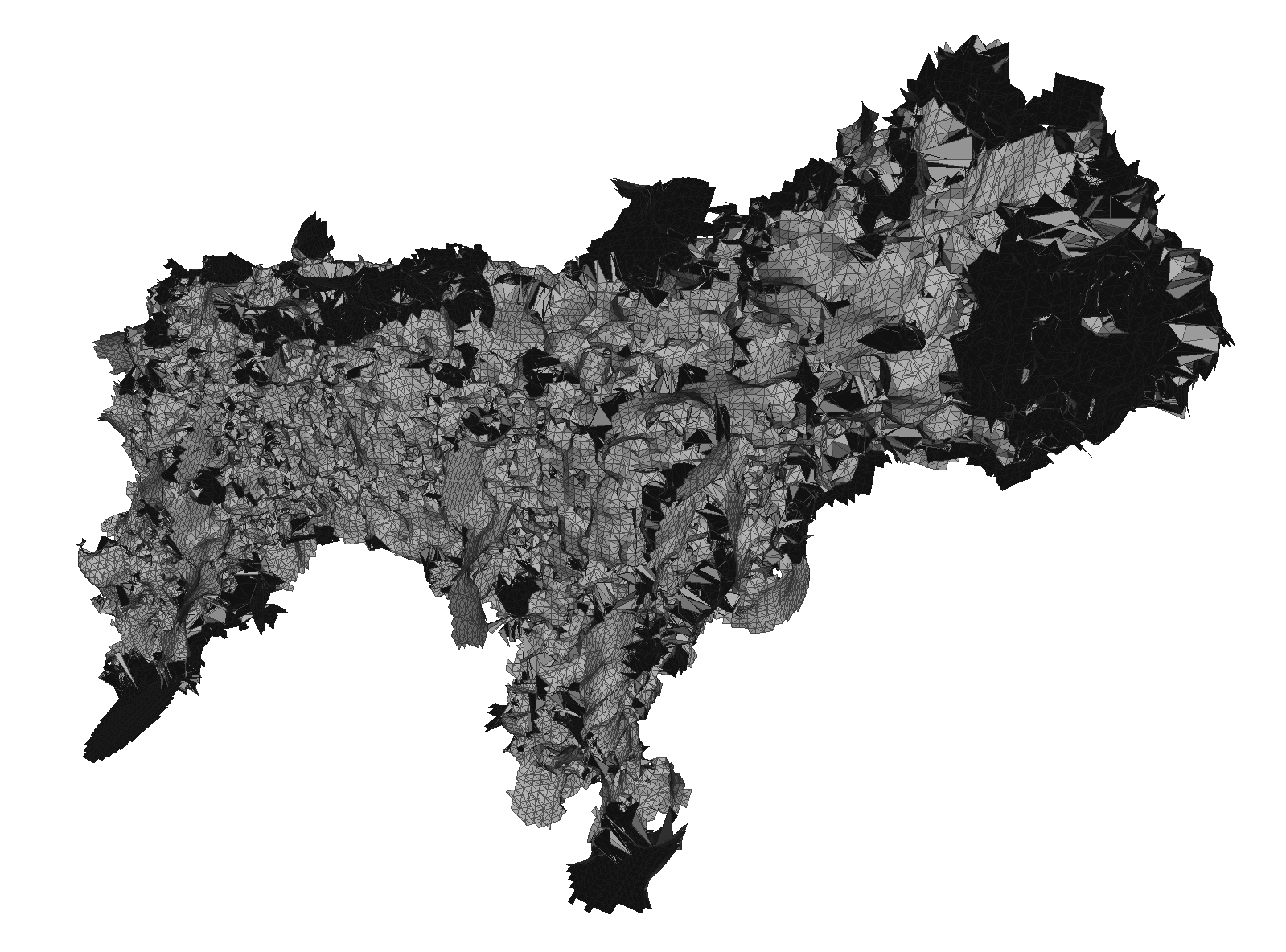} \\
    \end{tabular}
\end{center}
\caption{The noisy, poorly-aligned Kinect scans of the CERTH/ITI database are highly resistant to quality meshing, as these two models (with a naive meshing for each) demonstrate. The elephant mesh is produced by the Surface Mesh function of 3DReshaper Meteor. The dog mesh is produced by MeshLab's RIMLS method.} \label{CERTHmeshes}
\end{figure}

\begin{figure*}[!htbp]
\begin{center}
   \includegraphics[width=0.8\textwidth]{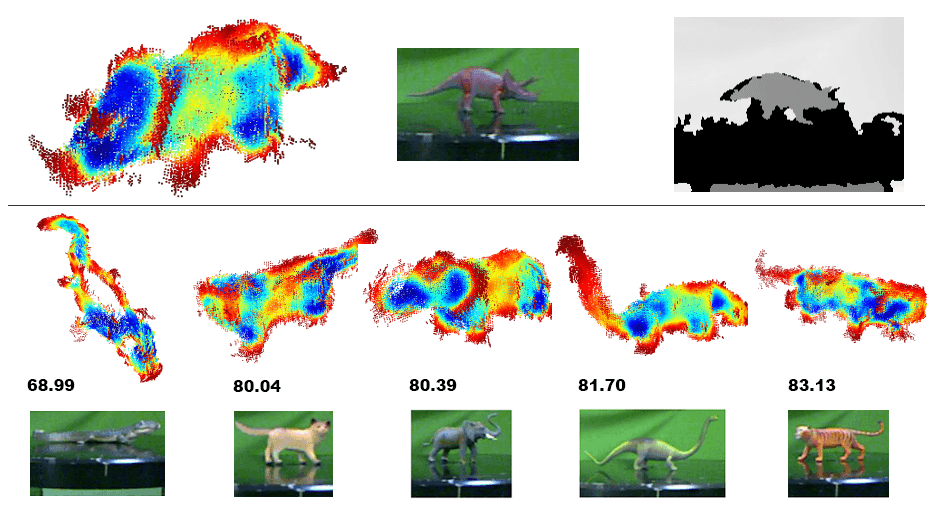} 
\end{center}
\caption{A query model from the CERTH/ITI dataset (left) and its five closest matches in the set according to matching HKS vectors from noisy point cloud models. The numbers aligned with the matching models from the database are the distance scores computed from the HKS feature vectors used to determine the best matches. Note the geometric and shape class similarities between query shape and matches.} \label{CERTHTop5}
\end{figure*}



Throughout this Section, adding model-scale or proportional Gaussian noise to a model means that each point in the cloud of that model has been displaced in the $x$, $y$, and $z$ directions by three samples drawn from a normal distribution with $\mu=0$ and $\sigma=p\epsilon$. Recall from Section \ref{HKS} that $\epsilon$ is the average distance between points in the model's point cloud. 

\subsection{Similarity and Classification for the \\CERTH/ITI Range Scan Dataset}
\label{ExR:CERTH}

We present the following example on the CERTH/ITI Range Scan Dataset, which is a freely-available database of scanned point cloud models of a variety of small objects \cite{Doumanoglou2013dataset} produced by the original Microsoft Kinect sensor with a depth resolution of about 1 cm. 


The 59 objects in the database, categorized into classes as ``17 land mammals, 6 dinosaurs, 11 sea mammals, 10 objects with humans, 2 guns, 2 bugs, 5 cars and 6 uncategorized objects'' \cite{Doumanoglou2013dataset}, were scanned in eighteen different rotations of a turntable.  The database provides an .XYZ file of the set of scans of each object rotated into a common coordinate system (so-called ``registered'' scans). Minimal cleaning has been performed to remove points outside the bounds of the turntable (i.e., background removal), but outliers remain, and the points of the aligned scans are often positioned so that a reconstructed surface through the points would result in self-intersections and other surface degeneracies. These scans, which are similar to those produced by industrial and hobby range scan systems, would be challenging to mesh without human operator intervention. 
Figure \ref{CERTHmeshes} shows the low-quality meshes which result from using established research and commercial meshing techniques on two representative point clouds from the dataset.

We demonstrate the efficacy of our point cloud method for grouping shapes of this Kinect model database of 54 shapes. To this end, we remove the outlier points that remained after background removal by discarding the points belonging to non-merged segments of a given model that contain fewer than 1\% of the total model points. This design is intended to prevent the discarding of true disjoint models if present while discarding unconnected patches of points from the background, turntable, or very severe alignment errors, although more sophisticated outlier detection and removal methods \cite{Shenspectral,Wangoutlier} could be employed as well.

We normalize the size of the models to a unit bounding box, and then compute the SPCL. The HKS at $t = 0.001$ was used to find 15 persistent clusters and subsequently a $15 \times 15$ feature vector was produced as described above. The HKS values comprising the feature vector were scaled to a max value of 1. Matching quality can be examined with the models representing the top matches for a given query. In Figure \ref{CERTHTop5}, we show an example of the top five models matched to a particular query model using our point cloud-based techniques.

The ``natural'' classification suggested by the CERTH dataset authors would label the query model a member of the class ``dinosaurs'' and only one of the top five matches is also so classified (other categories represented include ``land mammals'' and ``other''). However, all of the top five matches appear to belong to a slightly broader class of ``quadrupedal caudate land animals'', a more geometrically defined class than the geometrically variable but more deeply contextualized ``dinosaurs'' class. 

The top-5 hit rate of our current CPU-based implementation used for the CERTH database of noisy and incomplete point cloud models is $>61$\%. As a comparison, a top-5 hit rate of 88\% is reported in \cite{dey2010persistent} for a database of 50 \textit{noiseless} and \textit{meshed} models, in different poses and different levels of completeness. The difference in the two top-5 hit rates is due to the fact that the CERTH dataset is not only more general, but also has significantly more noise than the dataset used in \cite{dey2010persistent}, and that the categories given by the CERTH dataset authors are, as noted above, somewhat narrow and non-geometric. 

Perhaps a more apt comparison, running our techniques on point cloud models from the SHREC 2015 dataset and an equivalent mesh pipeline results in an E metric \cite{SHREC2015Dataset} score of 0.90 for point cloud and for mesh on a 32-model query. This is an E score of near the middle of the pack for results reported for model-retrieval routines tested on the SHREC dataset \cite{SHREC2015Results}. The models in this dataset are very incomplete, as well.

\subsection{Persistence-Based Segmentation}
\label{ExR:segs}

\begin{figure*}[!htb]
\begin{center}
    \begin{tabular}{ccccc}
        \includegraphics[width=0.1\textwidth]{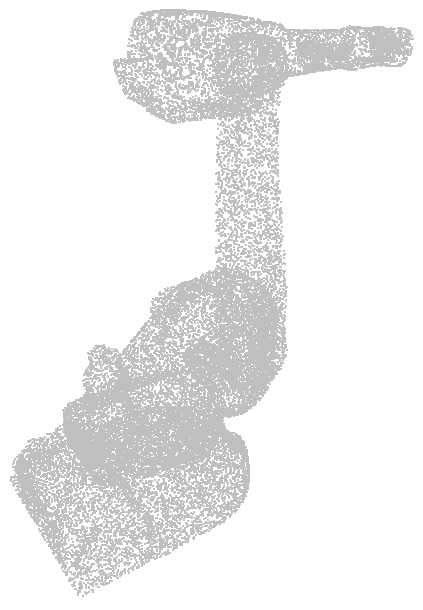} & 
        \includegraphics[width=0.1\textwidth]{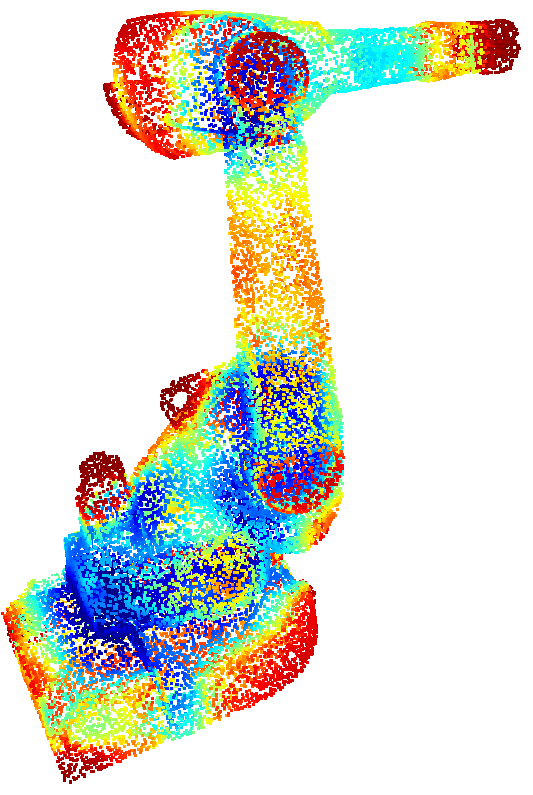} & 
        \includegraphics[width=0.15\textwidth]{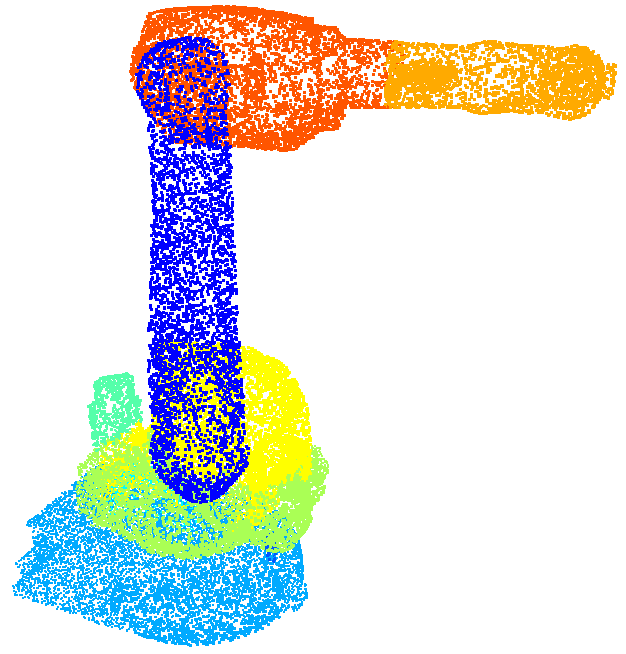} & 
        \includegraphics[width=0.15\textwidth]{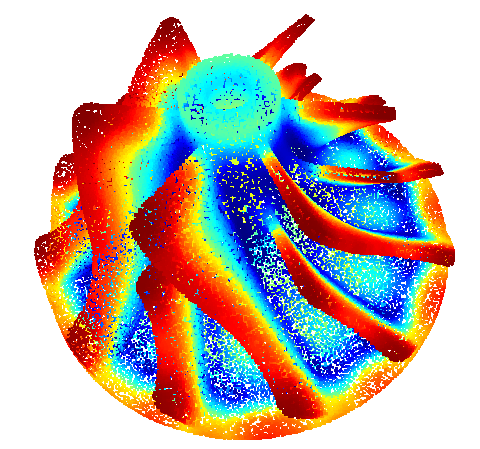} &
        \includegraphics[width=0.15\textwidth]{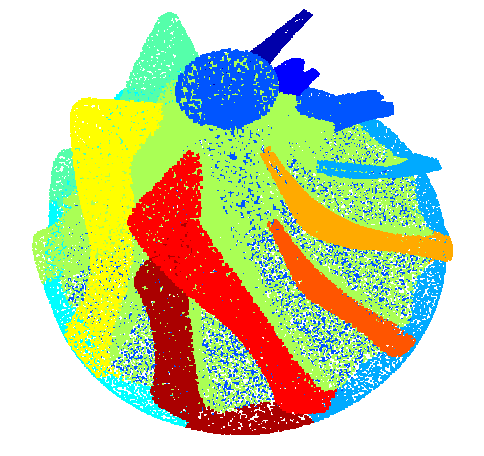} \\
        (a) &(b) & (c) & (d) & (e) \vspace{-10pt}
    \end{tabular}
\end{center}
\caption{(a) A point cloud Monte Carlo-sampled from an STL file of a robot courtesy of ABB \cite{abbwebsite} with model-scale noise ($\mu=\epsilon/2$, $p=0.125$) added to each point; (b) The HKS values for at $t = 4\lambda_{300}^{-1}ln10$; (c) One of the many possible automatic segmentations of the model by using Algorithm \ref{algofig_clust}; (d) The HKS vector for a turborotor model at $t = 0.001$; (e) An automatic persistent homological segmentation of the turborotor model output by Algorithm \ref{algofig_clust}.} \label{segex}
\end{figure*}

The following examples demonstrate the quality segmentation results which can be obtained using the techniques proposed in this paper.

Figure \ref{segex}(a)-(c) show the point cloud, HKS vector, and automatic segmentation of the ABB robot model first introduced in Figure \ref{SPCLneigh}. The proposed automatic persistent homological segmentation clearly delineates the point cloud into regions roughly corresponding to bulk design features that an engineer would find meaningful. Note especially the very well-defined middle rotational link and base link.

Figure \ref{segex}(d)-(e) show the HKS and persistent homological segmentation for an approximately 150,000 point turborotor-type industrial part. This point cloud is a high-quality sampling at good resolution, demonstrating the efficacy of the presented framework on high-quality, industrial-type point cloud models. Note the good definition of the numerous fins projecting from the central cone and bore of the model. 

Despite their different forms, resolutions, and constituent features, the ABB robot and turborotor models are both models of real engineering artifacts which can be segmented by the techniques introduced in this article directly from their point cloud information without either explicit connectivity or normal vector information.

\subsection{Heat Walks-based Segmentation}
\label{hw}
\begin{figure}[!h]
    \begin{tabular}{ccc}
        \includegraphics[width=0.15\textwidth]{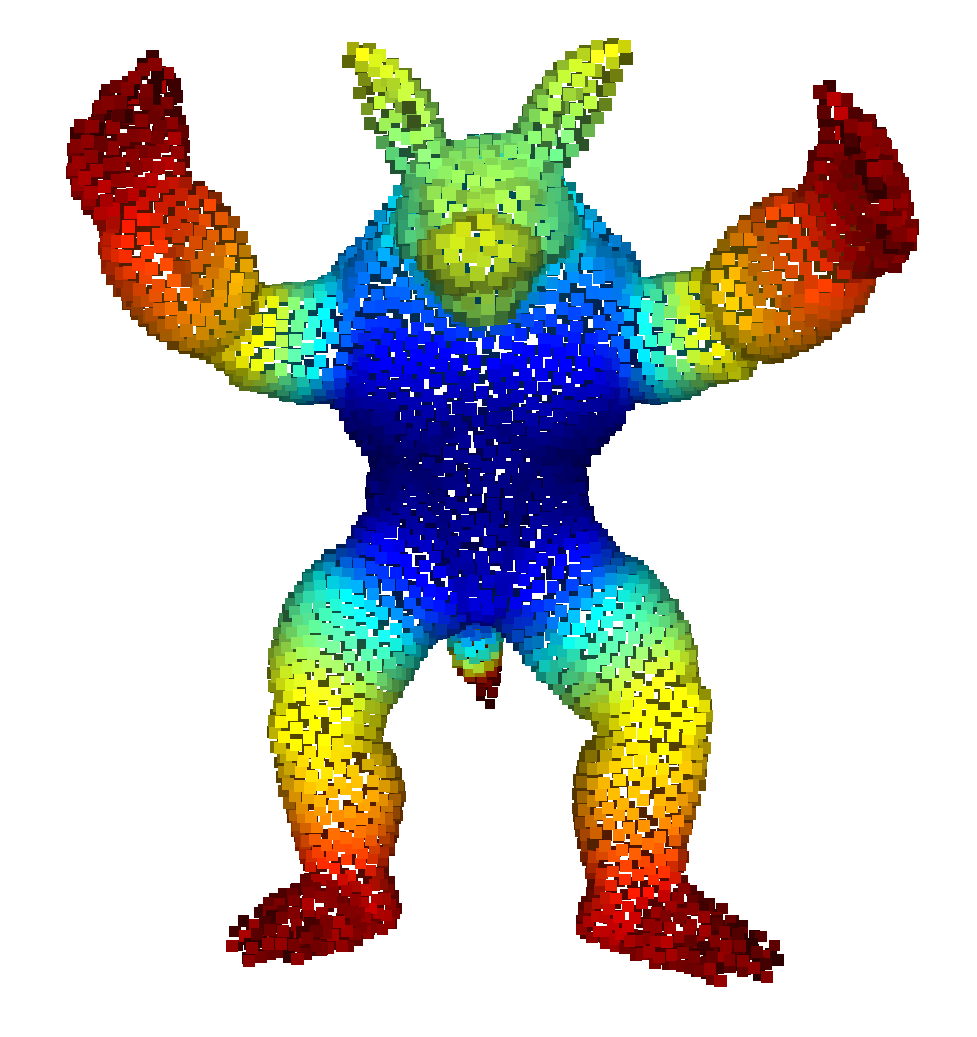} 
        &
        \includegraphics[width=0.15\textwidth]{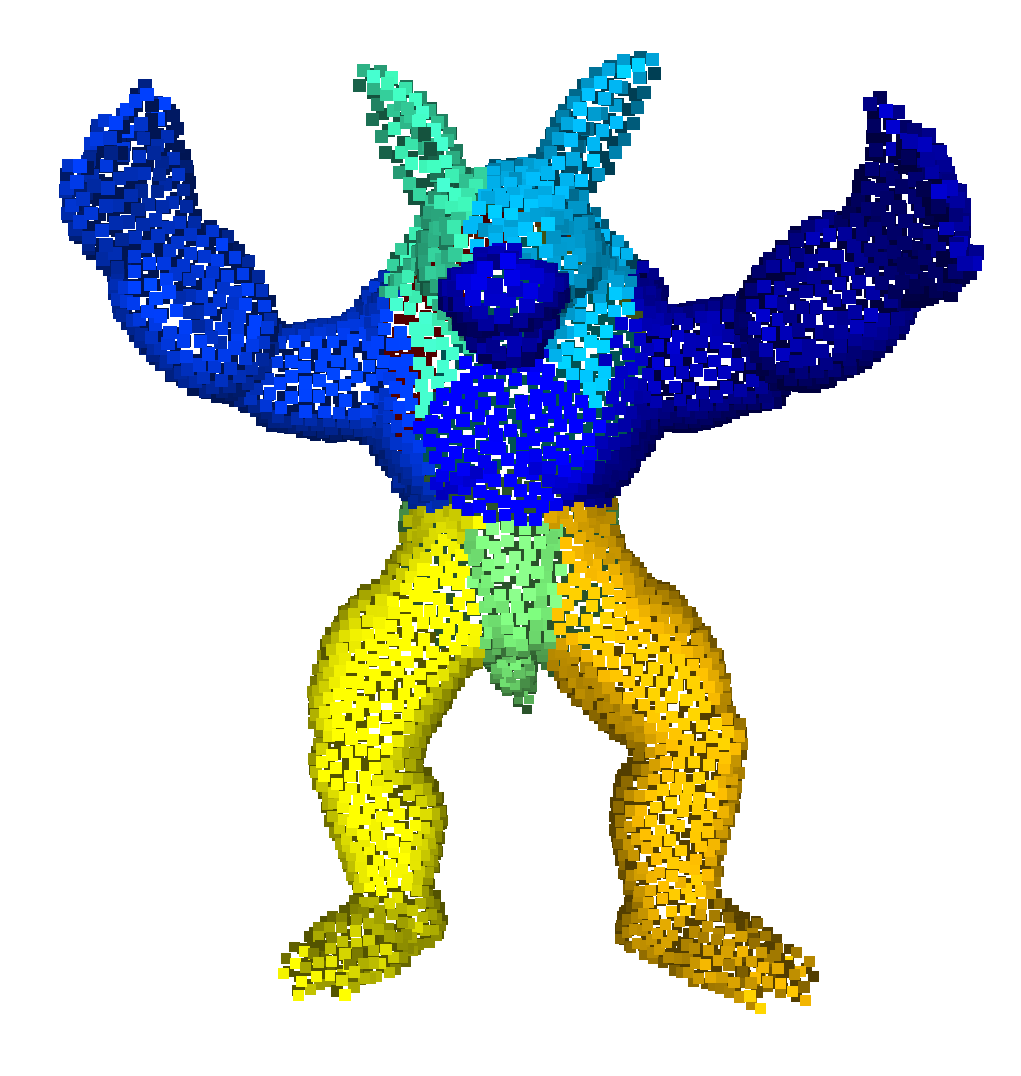}   
        &
        \includegraphics[width=0.15\textwidth]{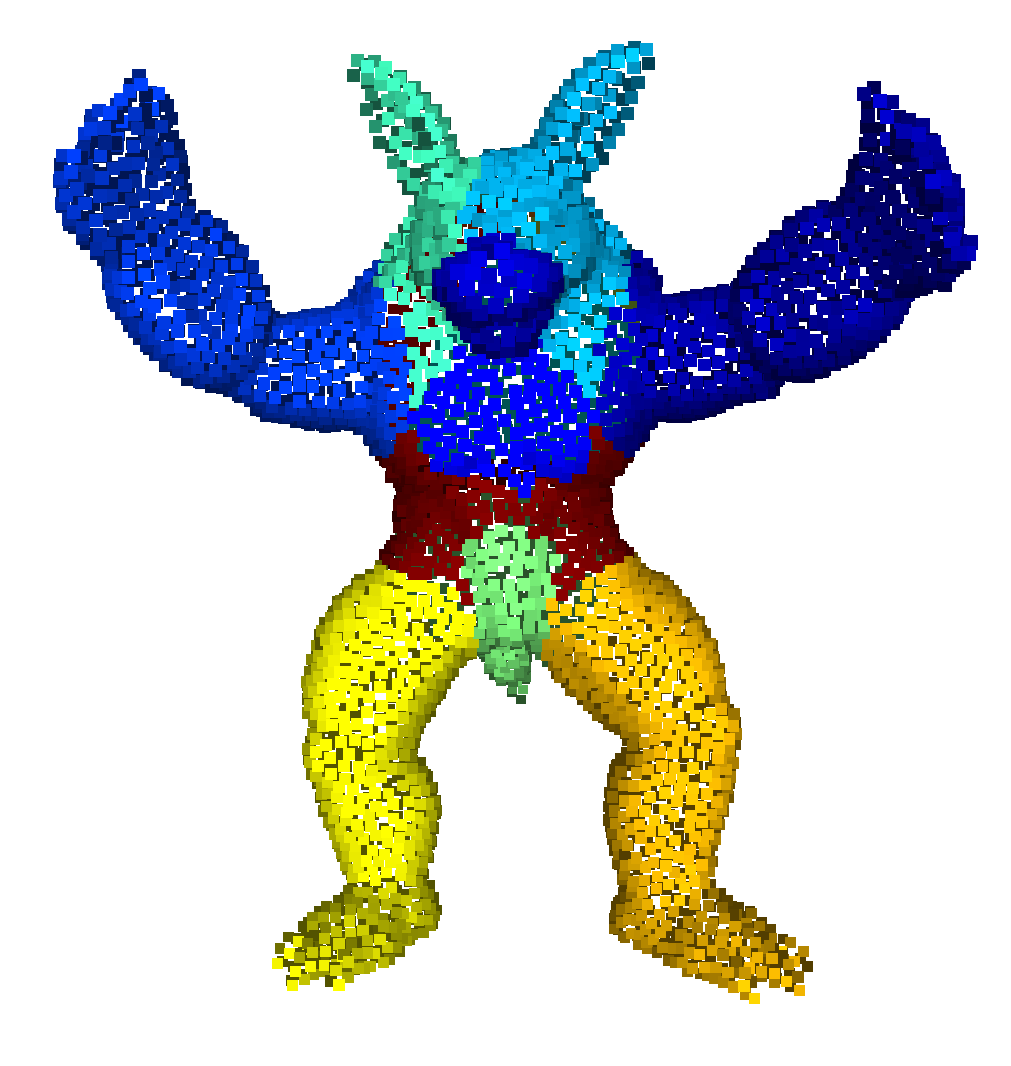} 
        \\   (a)   &  (b)   & (c)
 \end{tabular}
\vspace{-10pt} \caption{A point cloud of the classic Armadillo model, first: (a) colored with the HKS vector $k_0.1$; (b) after running the accumulator steps of the Heat Walk algorithm for $t=0.1$ (through step 6 above), and c) after the complete Heat Walk is finished, showing accumulator regions and a dissipator region, as well, around the midsection} \label{heatwalkex}
\end{figure}

The core of our techniques can be extended to related methods, such as enabling the use of the Heat Walk segmenter \cite{heatwalks} on point cloud models. 
Figure \ref{heatwalkex}(b)-(c) show examples of the result of the Heat Walk segmentation method used for a point cloud model as we discuss in Section \ref{heatwalk_section}. 
The near identical results (shown in Figure \ref{heatpcmesh}) between the point cloud model method run with the techniques we introduce in this paper and the mesh model method using traditional MeshLP, Heat Kernel, and Heat Walks serve as an example of the usability and correspondence of our methods.

\subsection{Curvature-Aware Segmentation}
\label{curveA}
Our techniques can be combined with methods which seek to improve shape segmentation accuracy for ``sharp'' shapes by incorporating explicit curvature information. 

\begin{figure}[!h]
\begin{center}
    \begin{tabular}{cc}
                \includegraphics[totalheight=115pt]{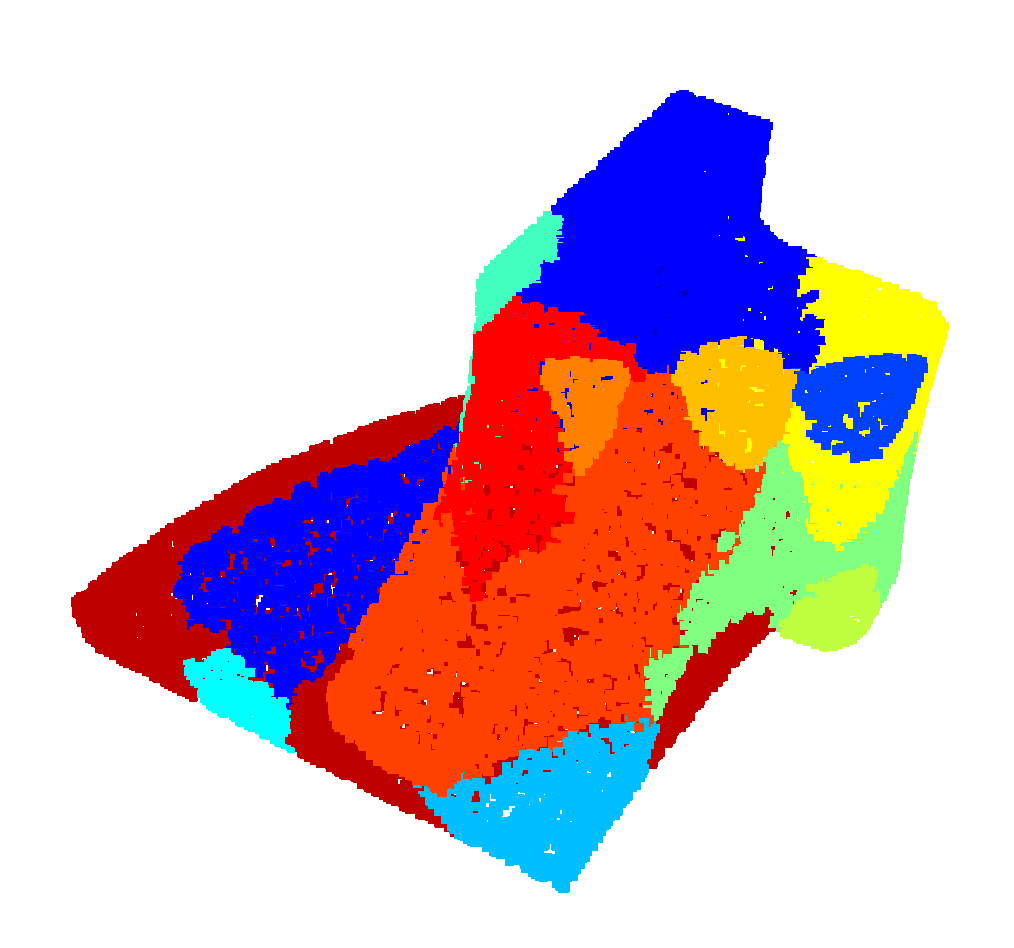} 
       \includegraphics[totalheight=115pt]{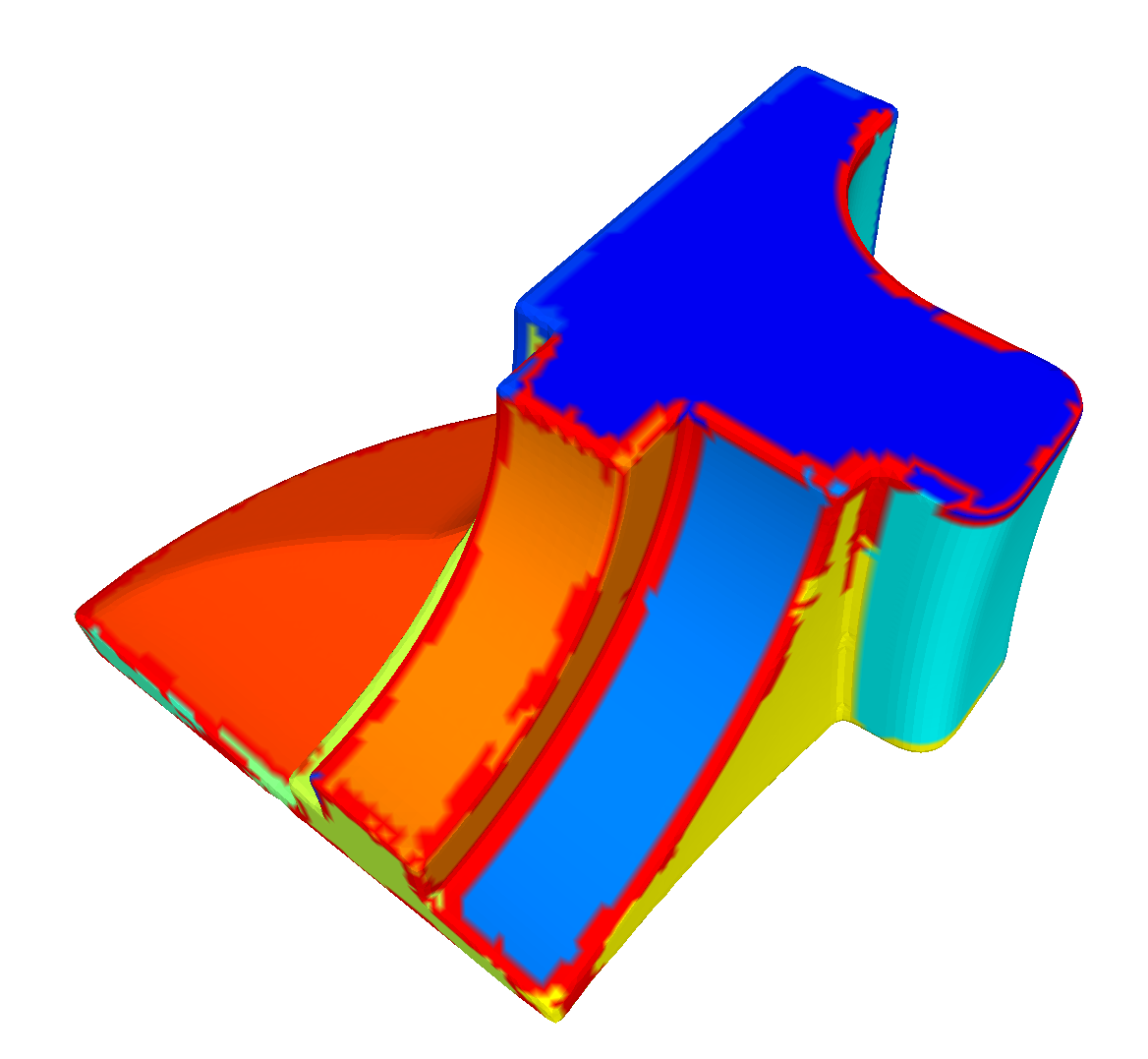} 
        \end{tabular}
\end{center}
\vspace{-10pt} \caption{a) Poor results on the fandisk model from segmenting without understanding of the various sharp edges, contrasted with (b) a segmentation that includes explicit information about the local curvature, derived from local normals as computed during the construction of the SPCL.} \label{newseg}
\end{figure}

Figure \ref{newseg} shows an example of combining spectral signature information with explicit local curvature information for segmentations. In this example, we have used the pointwise normal information which is computed as a byproduct of local tangent space approximation during SPCL construction (see Section \ref{Laplacian}) in reformulating the ``seeded'' segmentation procedure of \cite{lavoue2005new}.

Our formulation 1) determines approximate curvature values by examining the maximum difference in normal direction between points in local neighborhoods, 2) finds sharp edges by finding vertices with particularly drastic changes of normal direction (i.e., discontinuities) in their neighborhood (see Figure \ref{fig_edgefigure} for an example of what this looks like on the Fandisk model), 3) performs a k-means clustering of the curvature values and treats local neighbors with the same k-means cluster value as seeds, growing those seed clusters without letting them cross sharp edges, 4) builds a region adjacency graph for the clusters based on the mean curvature of the regions and the mean curvatures of the boundaries between regions, then 5) merges clusters across the smallest graph edges in that adjacency graph (i.e., the most similar adjacent clusters on the surface) until some desired number of clusters or the minimum graph edge value is exceeds a prescribed threshold.

This example formulation underscores the flexibility of our method for point cloud shape analysis, which allows any analysis technique relying on a shape signature computed on mesh models to be converted to operate on a point cloud models.

\subsection{Re-Clustering and Segment Type}
\label{ExR:types}
The techniques proposed in this article can be used not only to obtain segmentations, but to group the identified segments by geometric similarity. By exploiting the similarity information (computed in the HKS) that is retained to identify the point clusters which make up the model segments, we can, with little additional cost, understand easily which segments represent similar features.

\begin{figure}[!htb]
\begin{center}
    \begin{tabular}{cc}
        \includegraphics[totalheight=130pt]{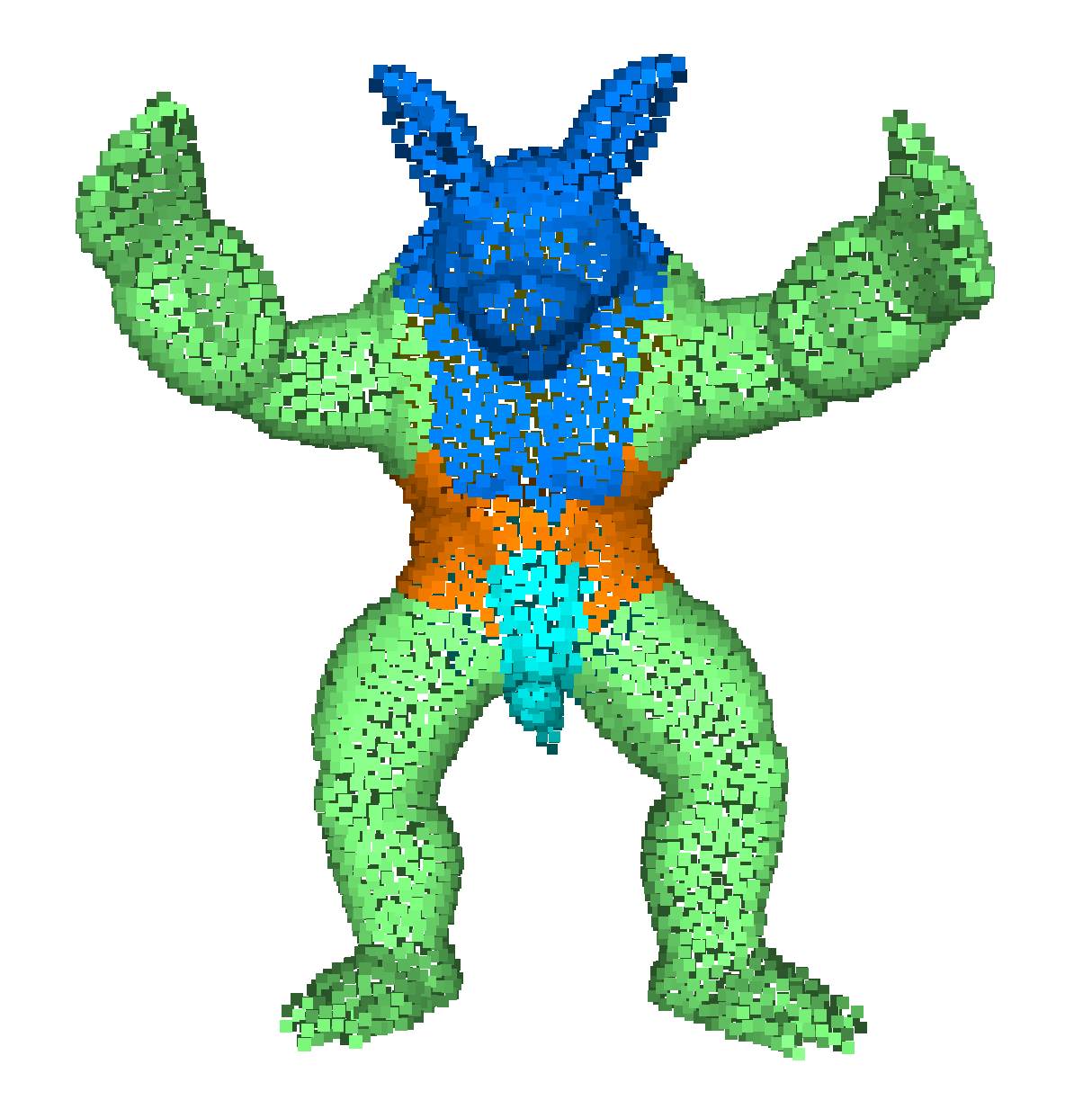} 
        \includegraphics[totalheight=130pt]{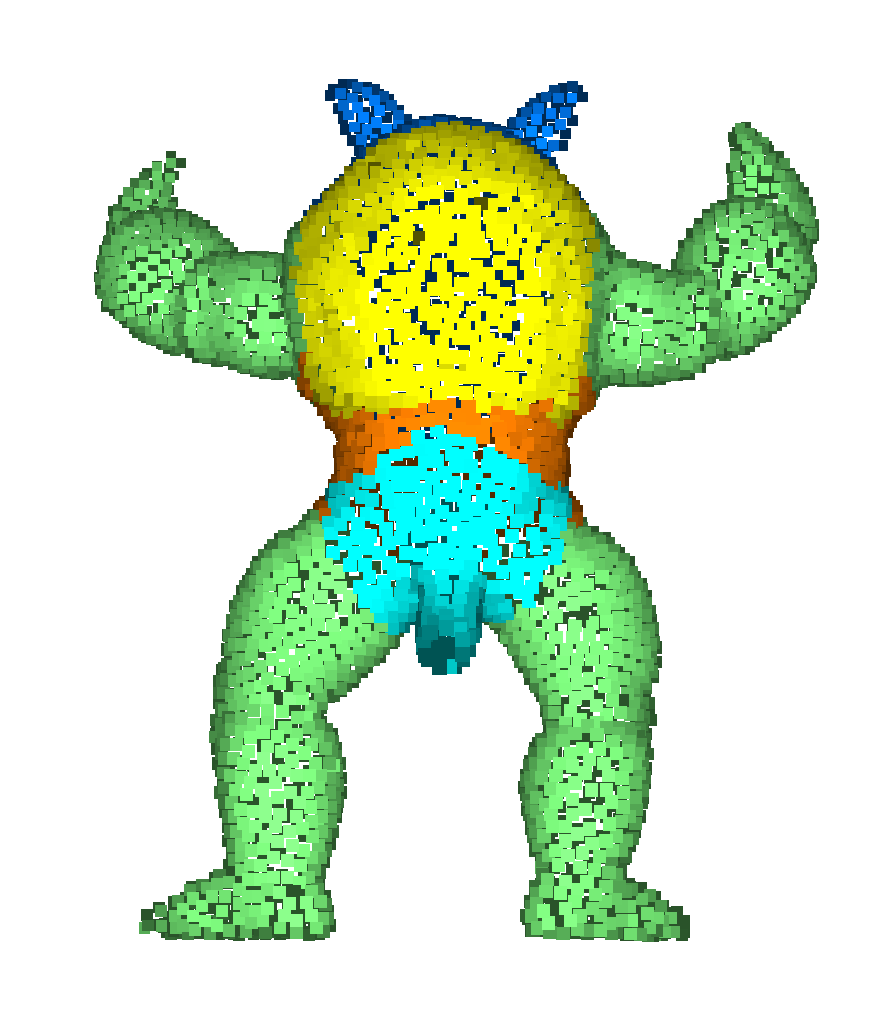} 
     \end{tabular}
\end{center}
\vspace{-15pt} \caption{The Armadillo model with the heat walks output segments clustered by our method from  Section \ref{reclust} by initial heat potential value (i.e., heat kernel signature value) at the exemplar point for that segment. For the dissipator cluster, we assign the mean of the HKS values of the rest of the points (as that cluster is closer to an ideal dissipator than to the other clusters). Note that all of the limbs are assigned to the same segment type and all three head clusters are also assigned to their own group. These groupings may both happen in one clustering because the HKS value used to regroup the segments by type includes local and global geometry information.} \label{ArmaReclust}
\end{figure}

Figure \ref{ArmaReclust} demonstrates the segment clustering technique of Section \ref{reclust} on the Heat Walk segmentation from Figure \ref{heatwalkex}. 
Recall that in this technique, the segments identified by the Heat Walk are clustered by Heat Kernel values to identify similarity between segmentation sub-shapes. 
Note that the limbs have all been classified as the same kind of cluster and that the three head sub-clusters have been identified as one cluster. 

\begin{figure}[!htb]
\begin{center}
    \begin{tabular}{ccc}
        \includegraphics[width=0.14\textwidth]{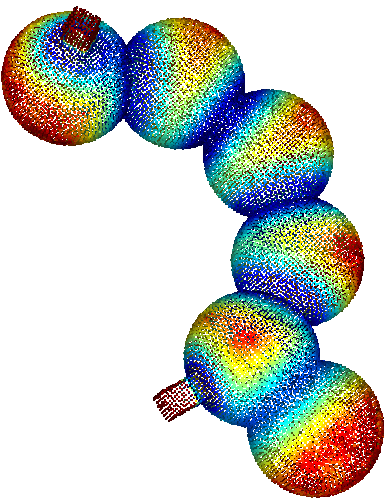} &
        \includegraphics[width=0.14\textwidth]{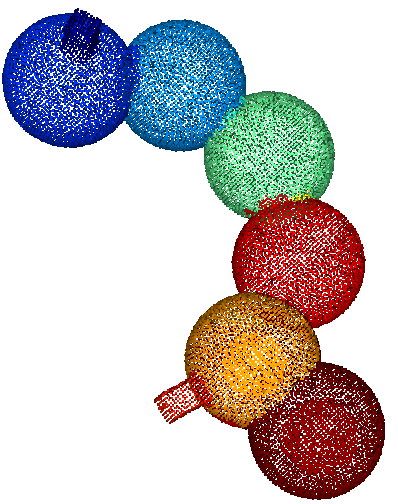} &
        \includegraphics[width=0.14\textwidth]{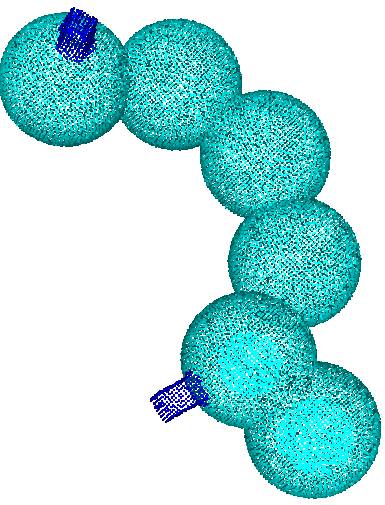}
        \\   (a)  &  (b)  &(c)
    \end{tabular}
\end{center}
\vspace{-10pt} \caption{(a) The first HKS vector for this synthetic model of fused spheres with Gaussian noise ($\mu=\epsilon/2$, $p=0.125$) is used in (b) to segment the model by the method described in Section \ref{ClustSeg} for a particular set of parameters, showing eight individual segments; (c) That same segmentation's clusters is then merged down by hierarchical clustering to only two clusters, showing the two distinct shapes present in the model.} \label{fig_mols}
\end{figure}

\begin{figure}[!htb]
\begin{center}
        \includegraphics[totalheight=170pt]{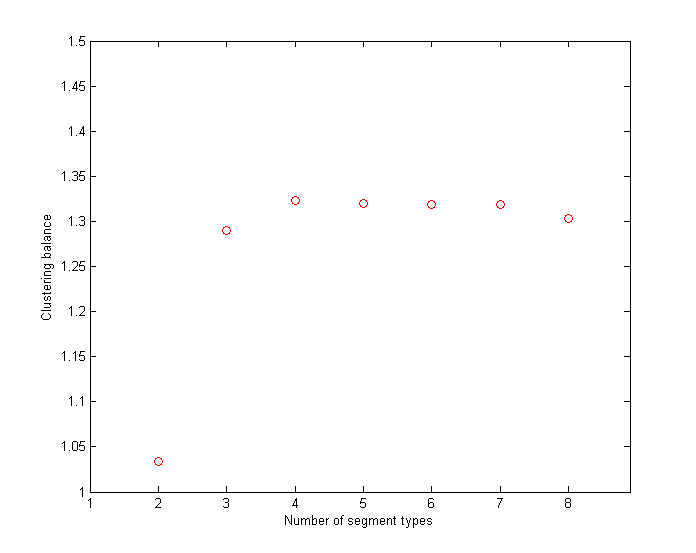}
\end{center}
\vspace{-10pt} \caption{The Clustering Balance computed for the HKS values for the fused spheres model from Figure \ref{fig_mols} points to only two distinct types of features. Note the steep drop off to two clusters.} \label{mols_gain}
\end{figure}

In Figure \ref{fig_mols}, the method identifies both shapes present and in fact identifies two clusters as the point at which the Clustering Balance \cite{Jung2003decision} is minimized, implying that this is the ``best'' clustering by that metric, as shown in Figure \ref{mols_gain}.

These examples demonstrate that this technique can be used to identify the {\em{number of feature types}} present in a model. This information can, in turn, be used in downstream applications where high-level semantic information is critical, such as in manufacturing planning (e.g., holes bored during in one operation, slots cut in a different operation) or robot task planning (are there handles and knobs in a part or just handles?).

\subsection{Resistance to Noise and Model Incompleteness}
\label{ExR:noise}

Section \ref{ExR:CERTH} showed the robustness against noise of our similarity computations. In those examples, the noise was due to the sensing limitations of specific depth-camera used for imaging. In this Section, we provide additional evidence to various levels of noise that the proposed integrated analysis and segmentation framework provides, by revisiting the incomplete camel model of Figure \ref{camel} with the proximal points in the knee areas, as well as the Armadillo model with additions of Gaussian noise. 

\begin{figure}[!ht]
    \begin{tabular}{ccc}
            \includegraphics[width=0.14\textwidth]{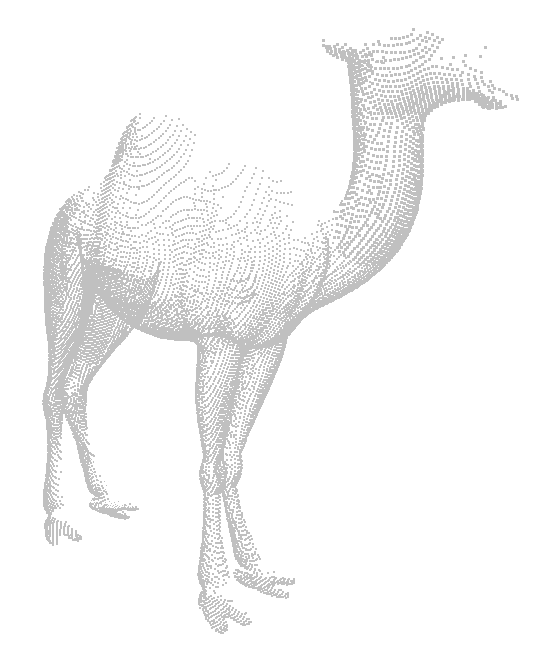} & 
            \includegraphics[width=0.14\textwidth]{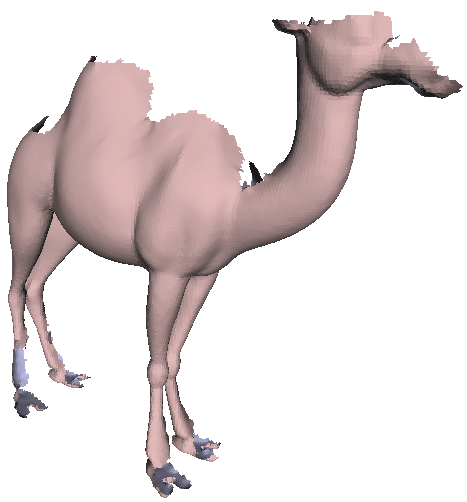} &
            \includegraphics[width=0.14\textwidth]{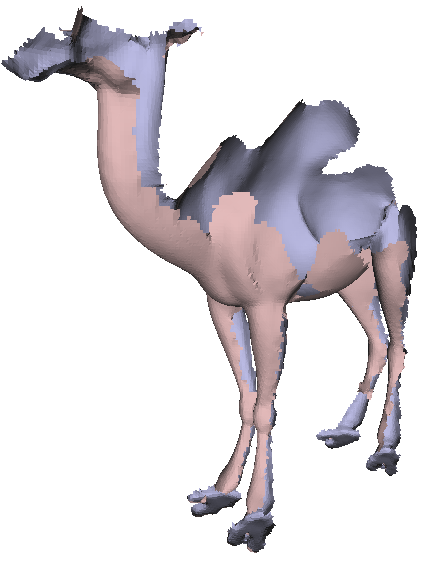} \\
 (a) &(b) & c)  \vspace{-5pt}\\
            \includegraphics[width=0.14\textwidth]{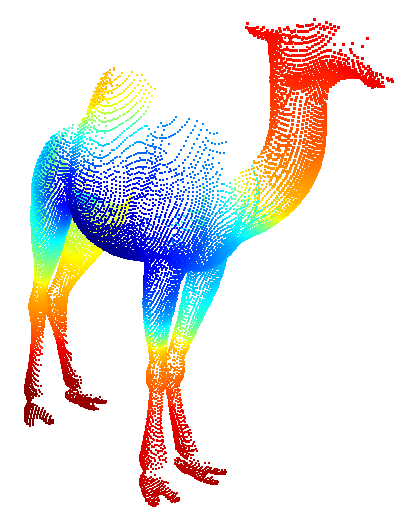} &
            \includegraphics[width=0.14\textwidth]{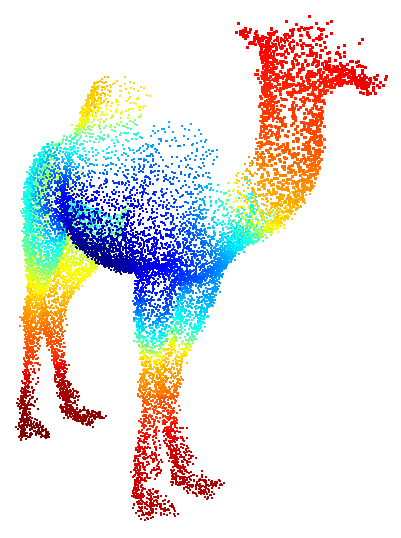} &                     
            \includegraphics[width=0.14\textwidth]{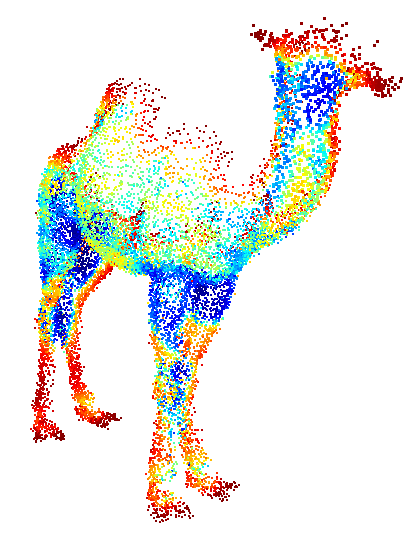} \\
 d) & e) & f)  \vspace{-5pt}\\
        \includegraphics[width=0.14\textwidth]{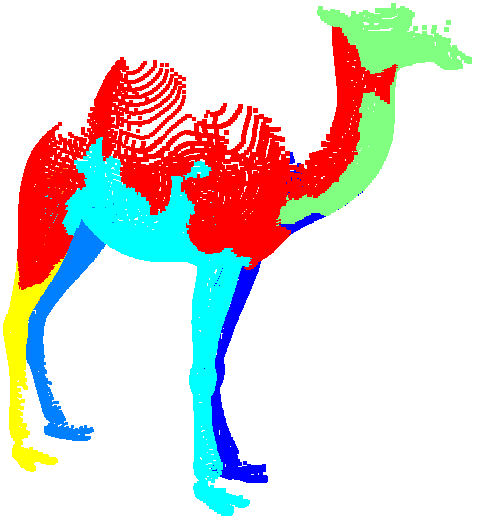} & 
            \includegraphics[width=0.14\textwidth]{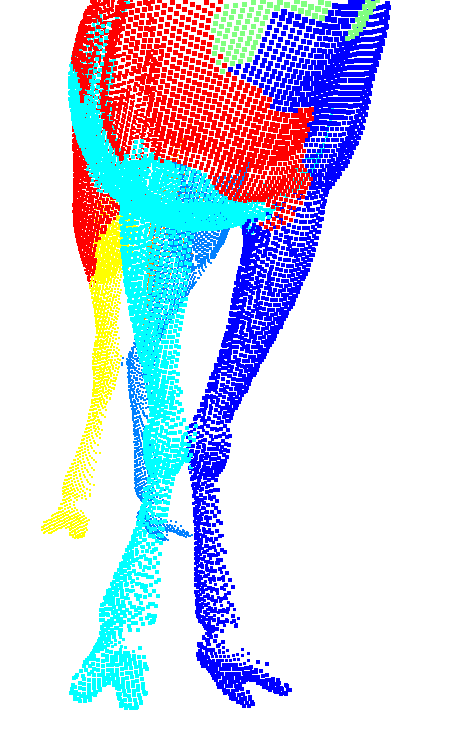} &
            \includegraphics[width=0.14\textwidth]{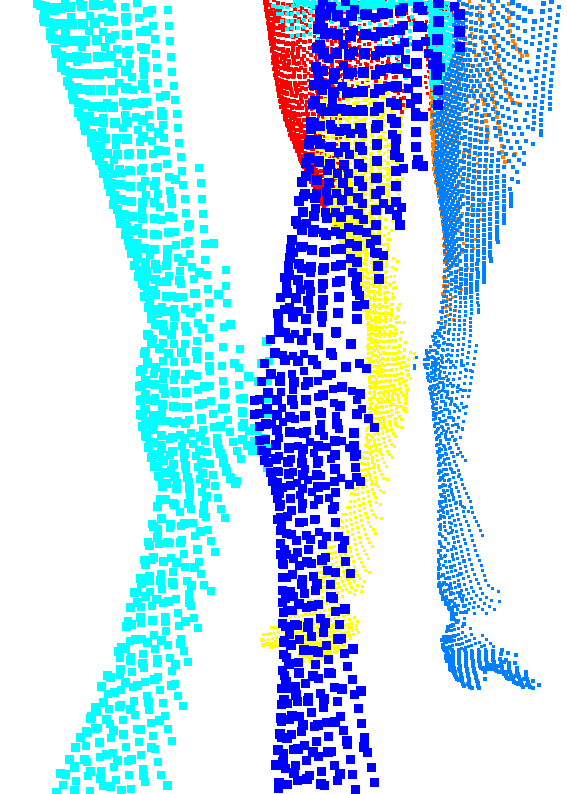} \\
 g) & h) & i) \vspace{-5pt} \\
    \end{tabular}
\caption{a) The noiseless and incomplete camel model;(b) An RIMLS meshing of the point cloud; c) Another view of the RIMLS mesh showing the model incompleteness; d) The $t = 0.1$ HKS vector for the noiseless camel model; e) The ($t = 0.1$) HKS vector for the camel model with model-scale Gaussian noise ($\mu=\epsilon/2$, $p=0.125$); f) The HKS vector for $t = 0.001$ for the noised camel model, showing additional definition at local scales but still retaining some of the global understanding evinced by the higher $t$-value HKS vectors; g-i) An automatic segmentation produced at $t = 0.0001$ into 7 segments. Note especially the separation of the legs despite topological noise at the knees.} 
\label{camel2}
\end{figure}

The original noiseless point cloud for the camel is highly incomplete --- most of the left side of the camel model is simply missing as shown in Figure \ref{camel2}(c), which displays the meshed surface output by the RIMLS reconstruction \cite{oztireli2009feature} implemented in Meshlab. As we showed in Figure \ref{camel}, automatic meshing solutions may produce unpredictable and unintended meshing results, which in this case is a mesh model that conjoins the legs at the knees (Figure \ref{camel}). This is a form of topological noise. 

In Figure \ref{camel2}(e)-(f), the camel point cloud model has had model-scale Gaussian noise ($\mu=\epsilon/2$, $p=0.125$) added and the SPCL and HKS have then been computed, augmenting the evidence for spectral methods' robustness against noise. The HKS vector t-values recommended for database matching are based on the eigenvalues computed during HKS computation (see Section \ref{HKS}), but lower and higher HKS t-values are useful for emphasizing either global (e.g., ends of a shape, projections from a core) or local (e.g., smaller features, local curvatures) shape.

Running Algorithm \ref{algofig_clust} on the chosen HKS vector with an appropriate $\tau$ value, currently obtained by examining the persistence diagram of this cluster, outputs automatically a seven-segment (body, head, four legs, and tail is a reasonable guess for a segmentation of a camel) segmentation at $\tau = 29$ as shown in Figure \ref{camel2}(g-i). Note the \textit{distinct legs} for both front and back pairs of legs.

\begin{figure}[!ht]
\centering
   \begin{tabular}{ccc}
        \includegraphics[width=0.14\textwidth]{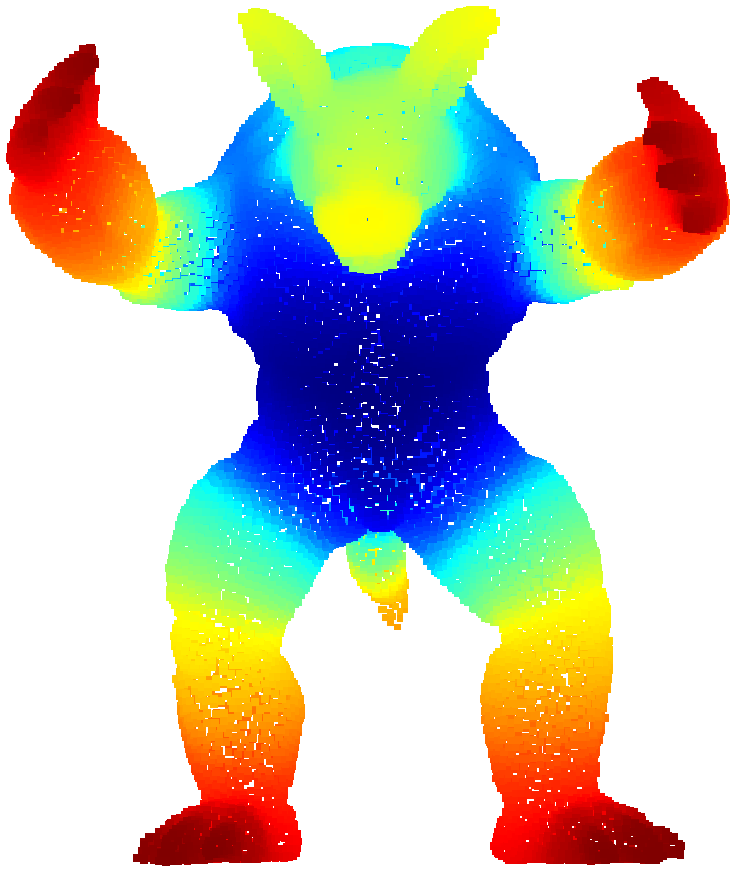} &
        \includegraphics[width=0.14\textwidth]{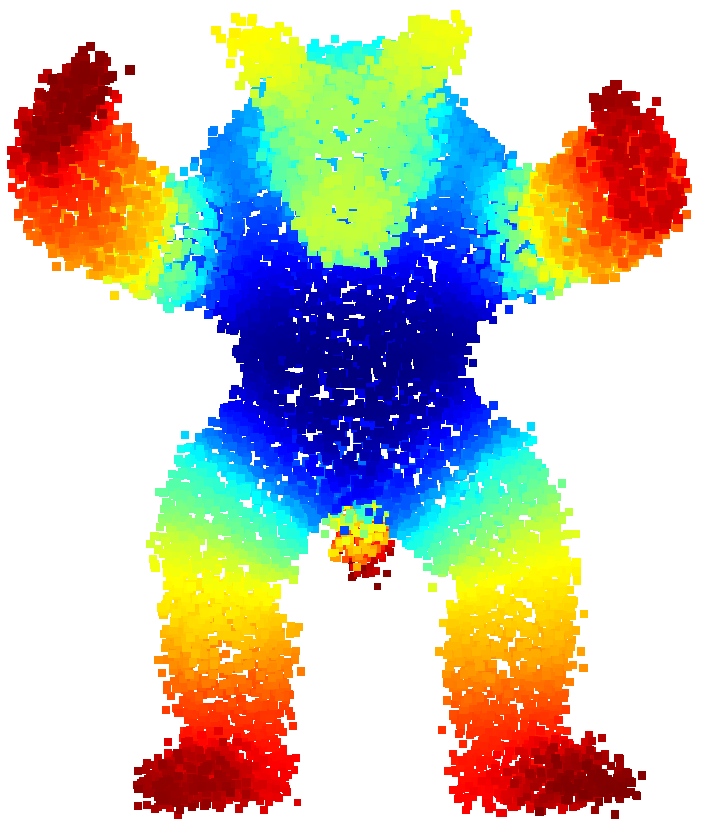} &  
        \includegraphics[width=0.14\textwidth]{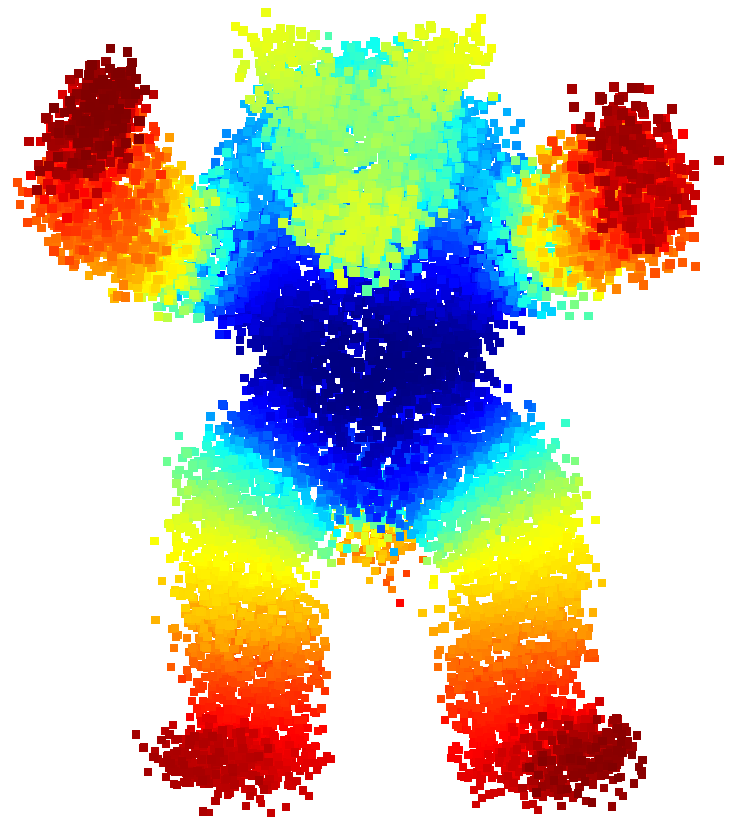} \\
        a) &(b) & c)  \vspace{-5pt}\\       
        \includegraphics[width=0.14\textwidth]{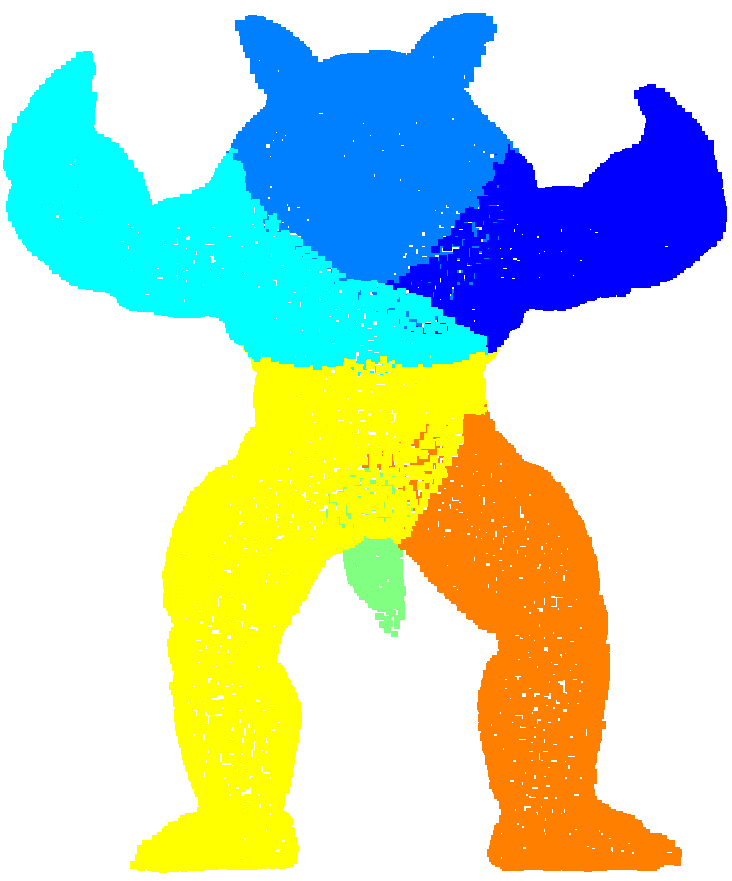} &
        \includegraphics[width=0.14\textwidth]{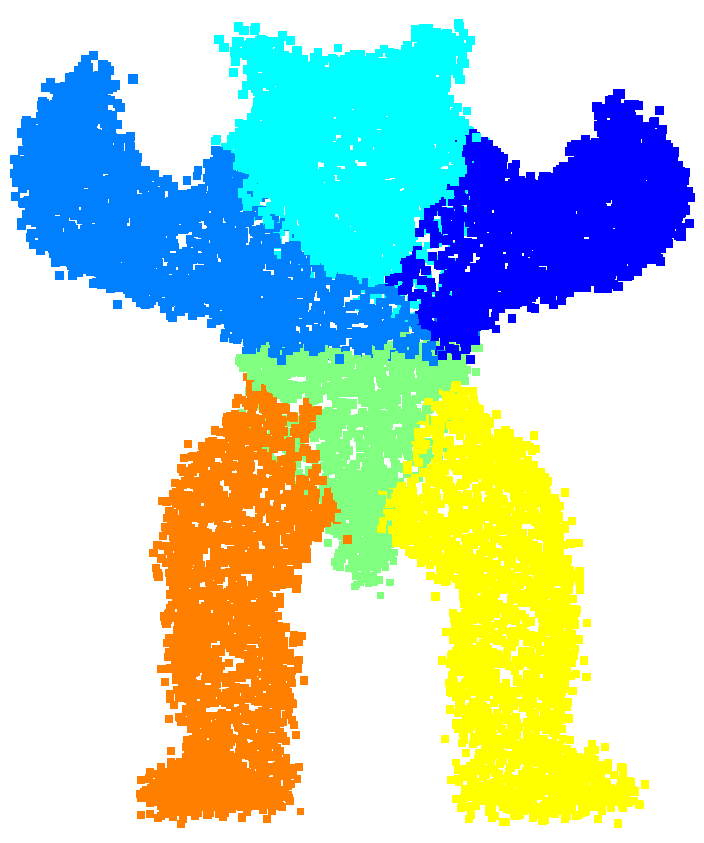} &
        \includegraphics[width=0.14\textwidth]{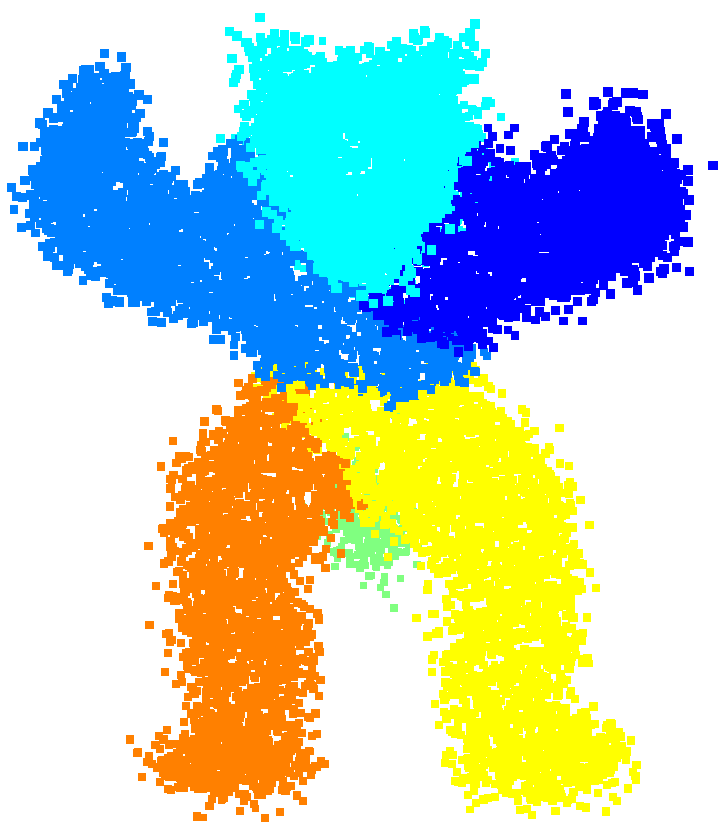} \\
        d) & e) & f)  \vspace{-5pt}\\
    \end{tabular}       
    \caption{(a) The $t = 0.1$ HKS vector for the approximately 15,000 point Armadillo model with no added noise; (b \& (c) The $t = 0.1$ HKS vector computed on the 50\% ($\mu=0$ and $\sigma=0.5\epsilon$) and 100\% ($\mu=0$ and $\sigma=\epsilon$) noised Armadillo PC model; (d-f) The PD-type segmentation of the model into 7 segments for the noiseless and noised models of (a-c).} \label{arma_noise}
\end{figure}

The example of the Armadillo model which follows the camel example demonstrates the impressive robustness against sensor/geometrical noise which the HKS displays. In rows two and three of Figure \ref{arma_noise}, the point cloud of the Armadillo was noised by adding random numbers to each coordinate of each point from a normal distribution with a mean of zero and a standard deviation of approximately 50\% and 100\% of the average inter-point distance respectively. As can be seen in the second and third columns, despite very high levels of random noise, the HKS vectors appear virtually unchanged and the PD-type segmentations naively computed from them are also highly consistent with the non-noised result.

\subsection{Comparing Mesh-based and Point Cloud-based Segmentations}
\label{ExR:pcmesh}

In order to demonstrate the quality of the proposed point cloud shape analysis framework against the known quality of mesh model-based shape analysis, we compare the database segmentation and shape matching of the Girl Dancing Samba dataset of \cite{vlasic2008articulated}.

\begin{figure}[!htb]
\begin{center}
    \begin{tabular}{cc}
      \includegraphics[dpi=96,trim=0px 0px 0px 0px,totalheight=180pt]{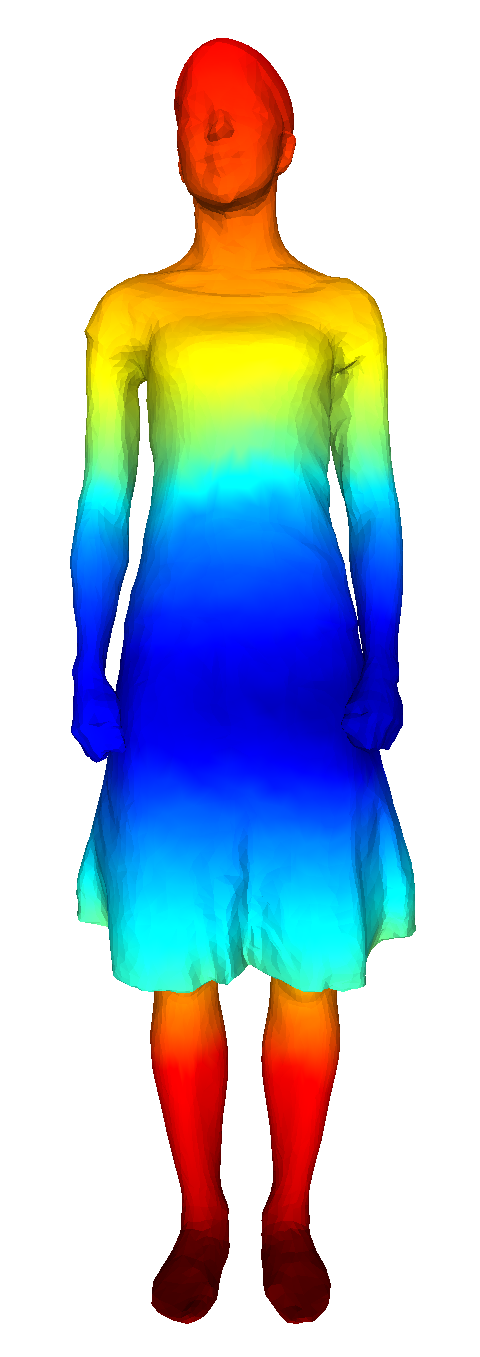} & 
            \includegraphics[dpi=96,trim=0px 0px 0px 0px,totalheight=180pt]{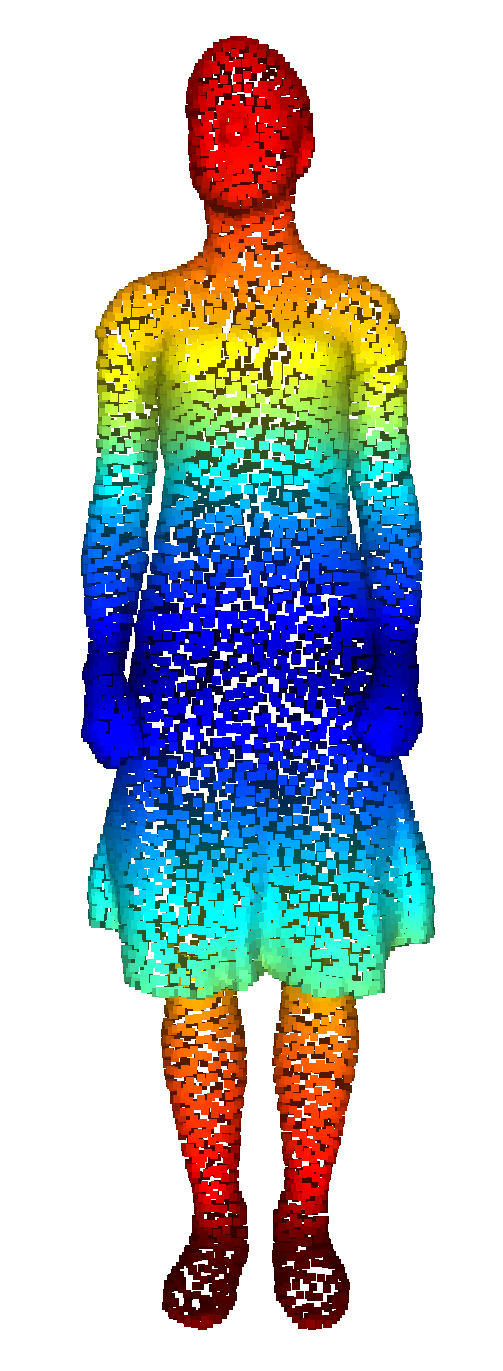}  
        \\   (a)  &  (b) \vspace{-10pt}
     \end{tabular}
\end{center}
\caption{The first model from the Girl Dancing Samba dataset colored by HKS vector $k_{t=0.1}$ as computed on (a) mesh model and(b) point cloud. Note the similarity between computed HKS vectors regardless of model type.} \label{GirlDancing1}
\end{figure}

Figure \ref{GirlDancing1} shows the first model in the dataset colored with HKS vectors. Each model in the dataset has 9971 points in the point cloud model or 9971 vertices in the mesh model. Since the 150 models in the database are all of the same physical body, a person, and the spectral signature we choose, the HKS, is to a large degree pose-invariant, we should expect to see only small differences between feature vectors and very similar shape segmentations between models. 

\begin{figure}[!htb]
\begin{center}
  \begin{tabular}{cc}
    \includegraphics[dpi=96,trim=0px 0px 0px 0px,totalheight=130pt]{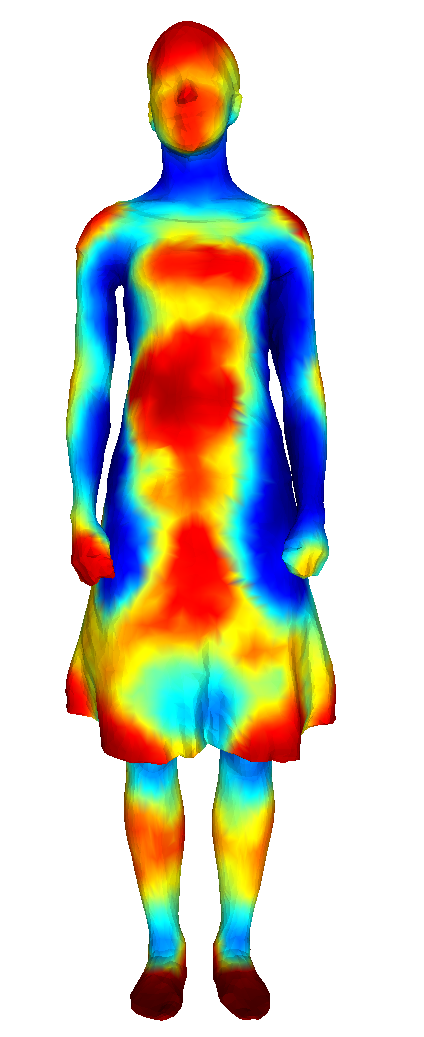} & 
        \includegraphics[dpi=96,trim=0px 0px 0px 0px,totalheight=130pt]{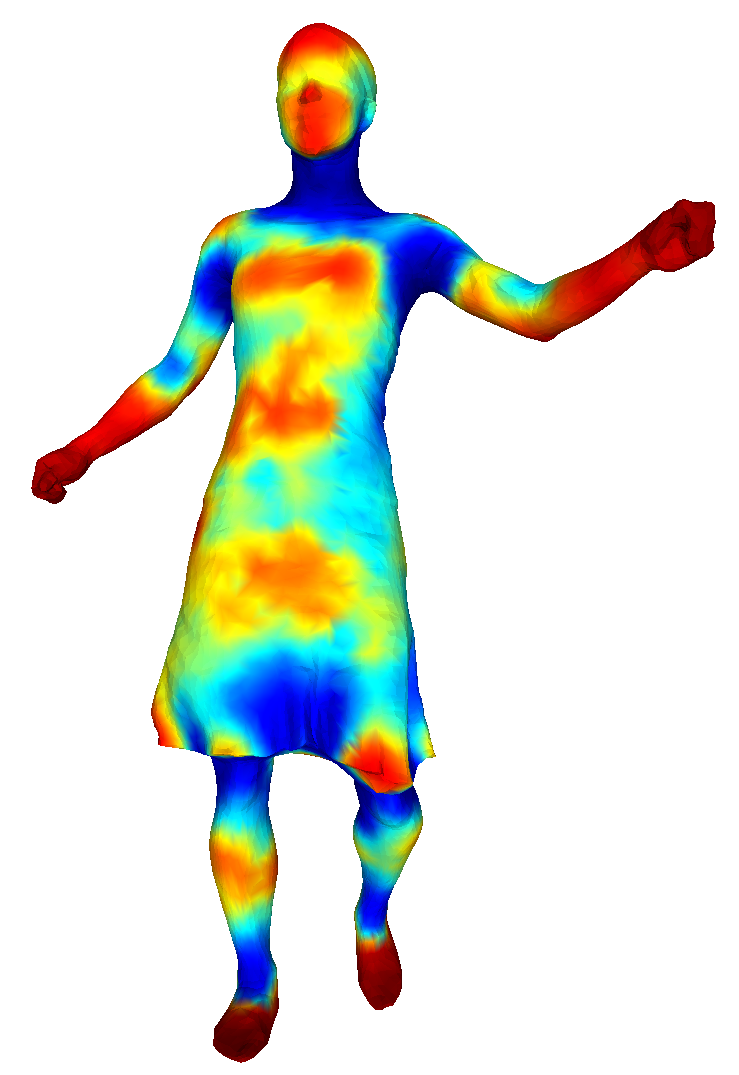}
    \\   (a)  &  (b) 
    \\
        \includegraphics[dpi=96,trim=0px 0px 0px 0px,totalheight=130pt]{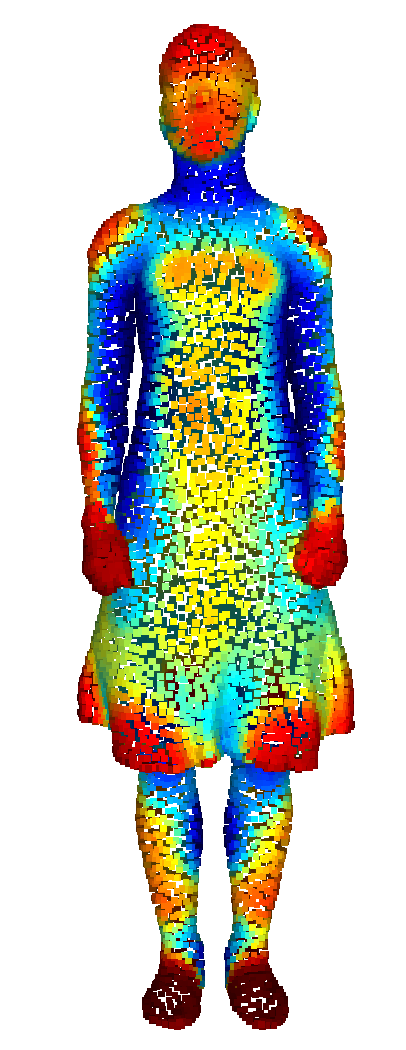} & 
        \includegraphics[dpi=96,trim=0px 0px 0px 0px,totalheight=130pt]{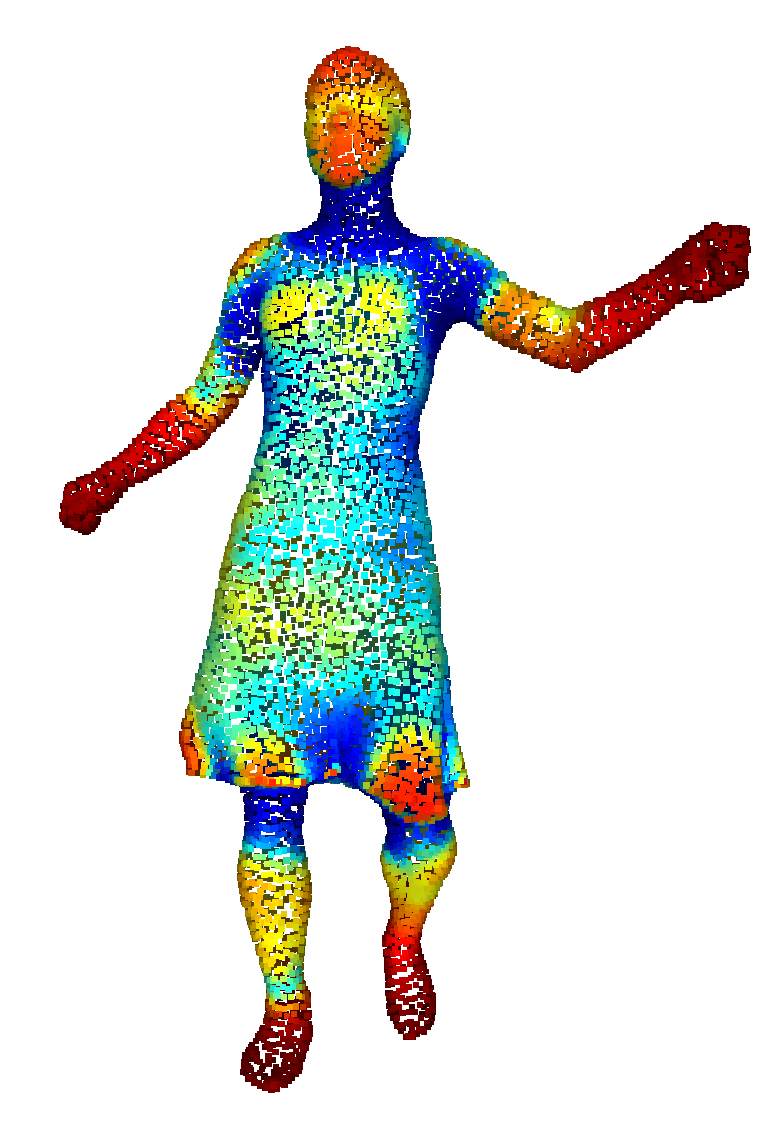} 
        \\ (c)  & (d) \vspace{-10pt}
  \end{tabular}
\end{center}
\caption{a-d) The 1st and 45th models from the Girl Dancing Samba dataset colored by the ``first'' HKS vector as computed on mesh model (a\& b) and point cloud model (c\& d). Note in (c) the (correct) distinction of the figure's left hand compared with the mesh model in (a).
} \label{girls}
\end{figure}

Figure \ref{girls} shows that, as we expect, different models of the same figure in different poses have highly similar HKS vectors in both mesh and point cloud space. The point cloud HKS for the first pose actually differentiates the left hand of the model to a greater degree than the mesh version does, again demonstrating the resistance of point cloud methods to topological disturbance.

\begin{figure}[!htb]
\begin{center}
    \begin{tabular}{cc}
      \includegraphics[dpi=96,trim=70px 70px 70px 70px,totalheight=130pt]{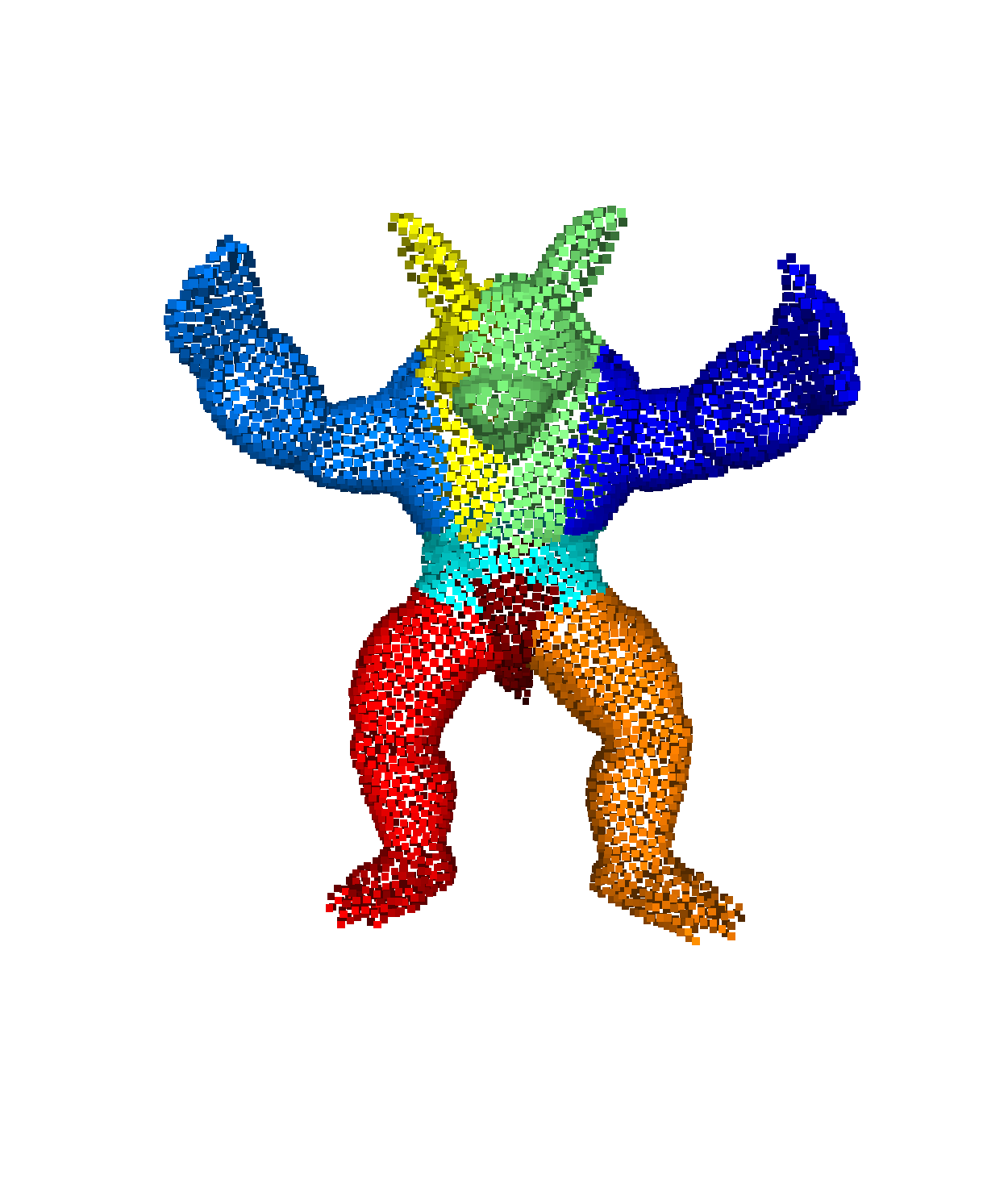}
        & 
        \includegraphics[dpi=96,trim=70px 80px 70px 70px,totalheight=130pt]{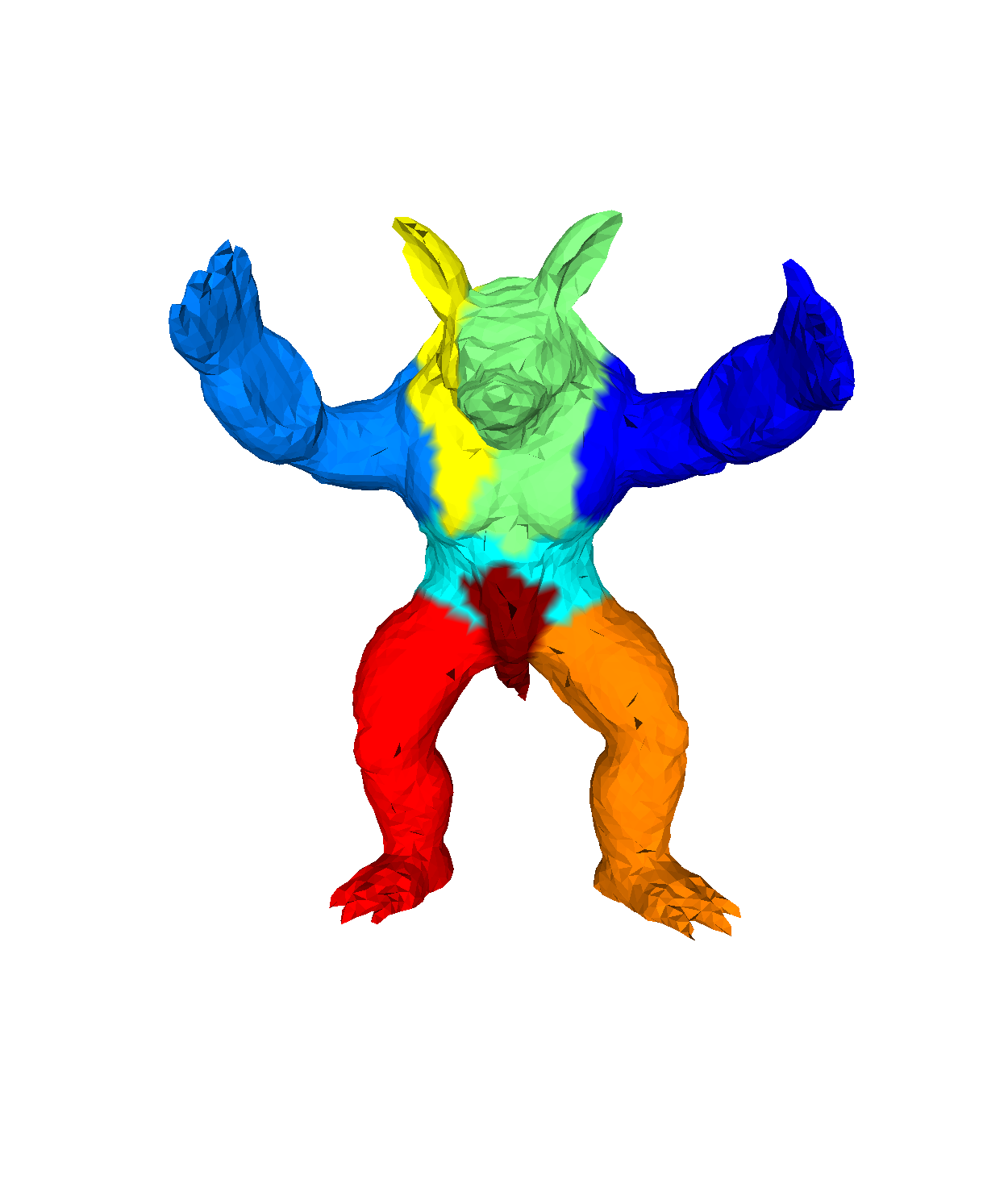}
        \\   (a)  &  (b) \vspace{-10pt}
     \end{tabular}
\end{center}
\caption{The Armadillo model with the Heat Walk segmentation on (a) the point cloud version of the model and (b) the equivalent mesh model. Note the near identical results between model types.} \label{heatpcmesh}
\end{figure}

Additionally, Figure \ref{heatpcmesh} shows the point cloud and mesh versions of the Heat Walks segmentation for the Armadillo model. Note the high degree of similarity. In each case, every limb is segmented away from the core of the model as is the tail and each ear-dominated head segment and the midsection is identified as a dissipator region. 

\subsubsection{Mesh Segmentation Benchmark dataset}
\begin{figure*}[!htb]
\begin{center}
    \begin{tabular}{cccc}
      \includegraphics[dpi=72,trim=10px 0px 100px 0px,totalheight=75pt]{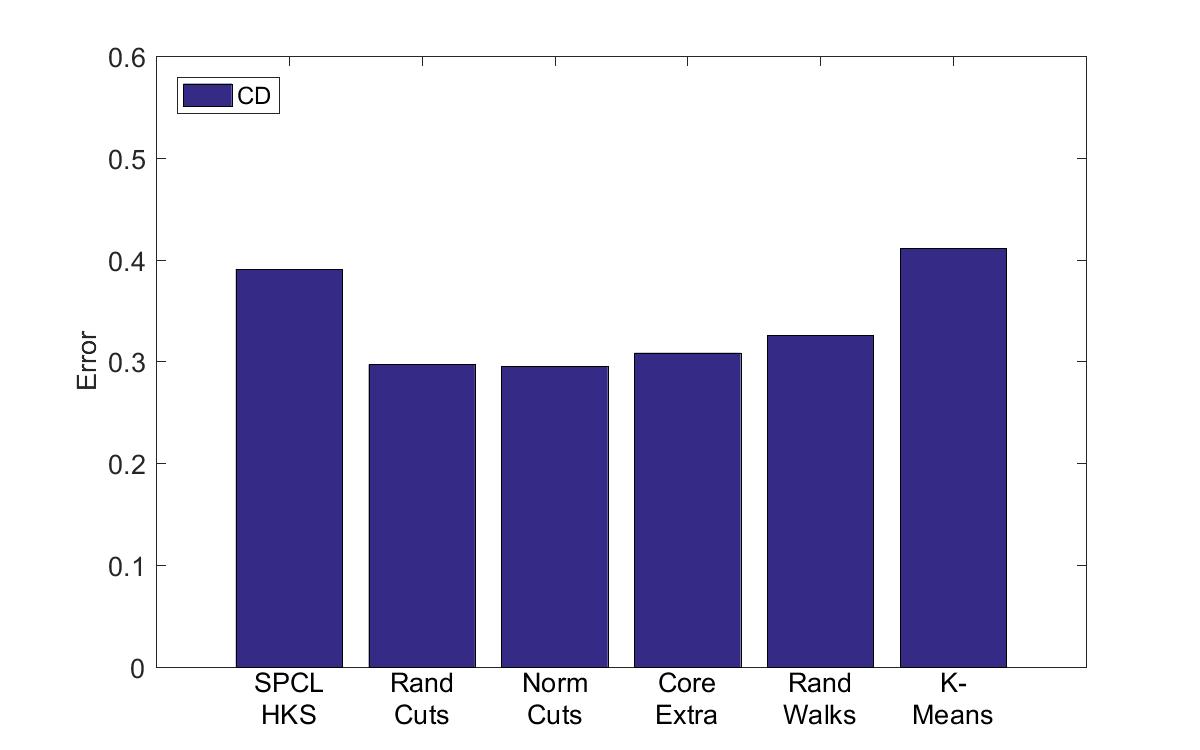}
        & 
      \includegraphics[dpi=72,trim=10px 0px 100px 0px,totalheight=75pt]{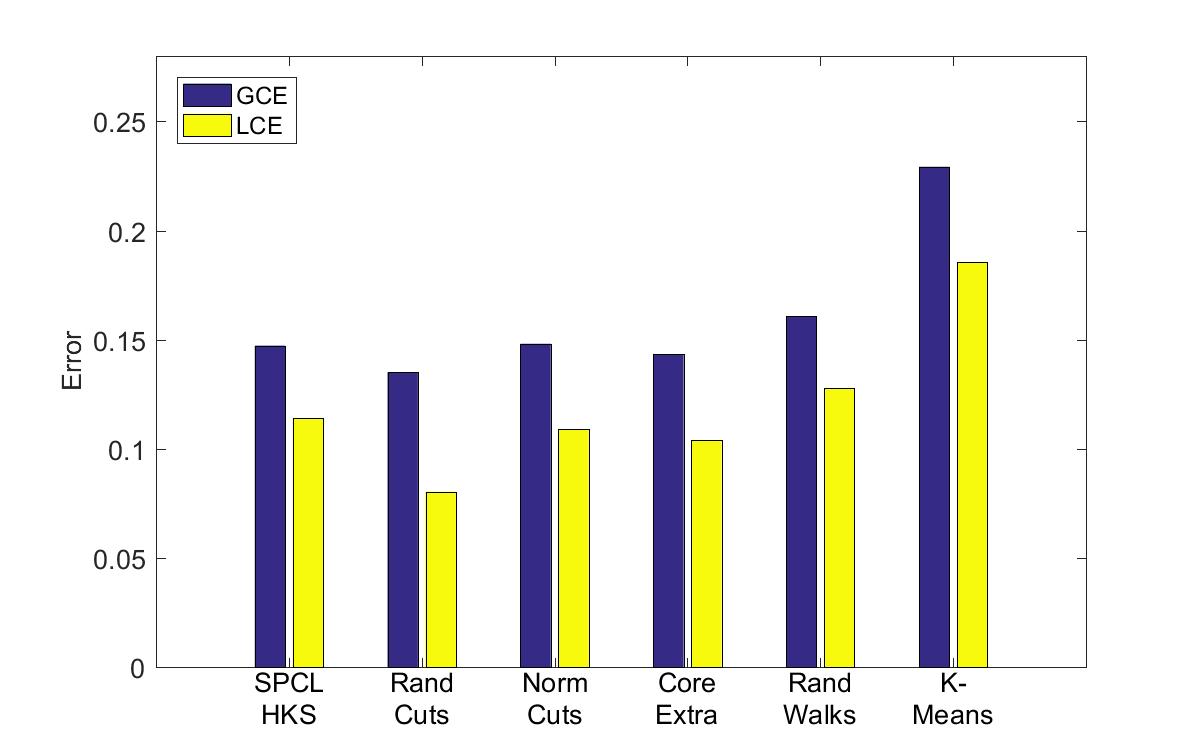}
		 &
      \includegraphics[dpi=72,trim=10px 0px 100px 0px,totalheight=75pt]{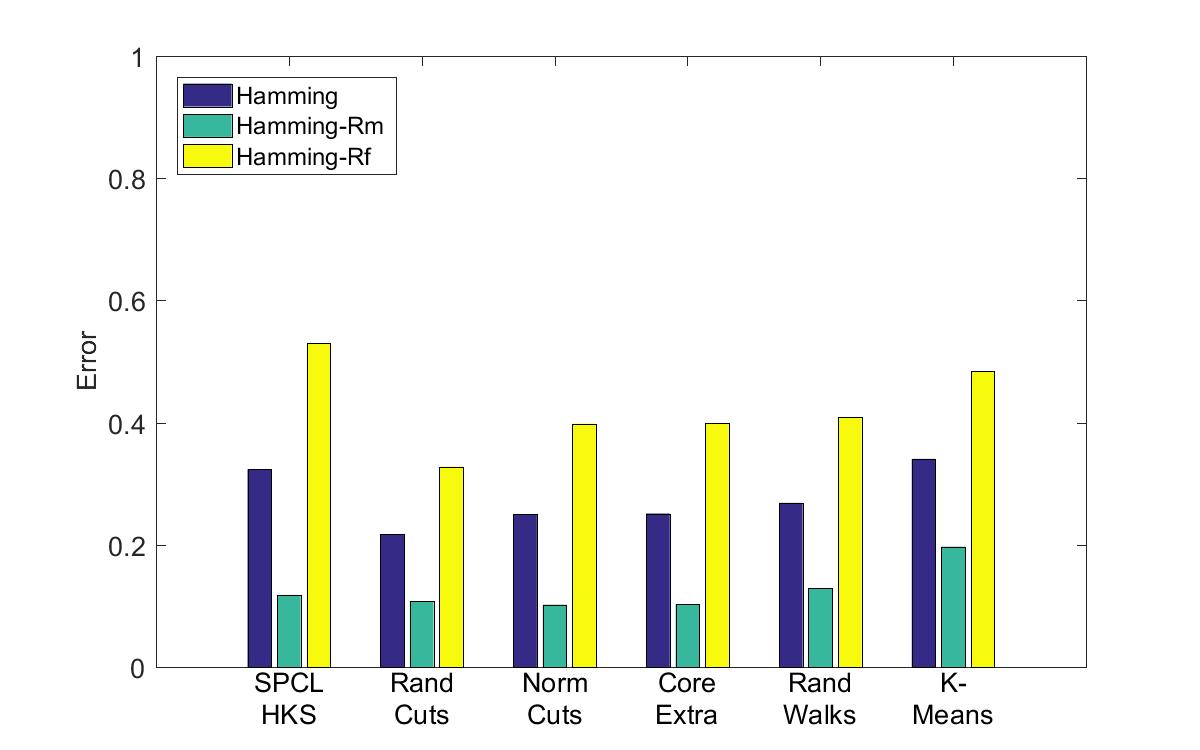}
        & 
      \includegraphics[dpi=72,trim=10px 0px 100px 0px,totalheight=75pt]{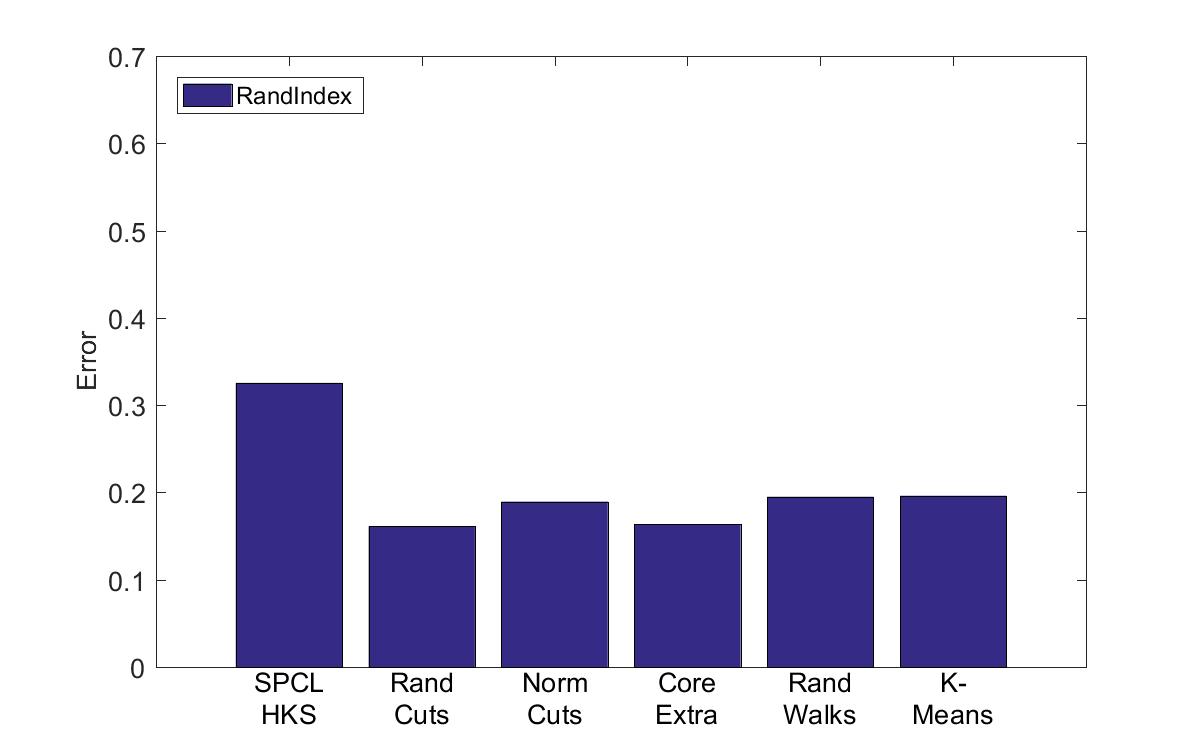}
      \\
      (a) & (b) & (c) & (d)\\
      \includegraphics[dpi=72,trim=10px 0px 100px 0px,totalheight=75pt]{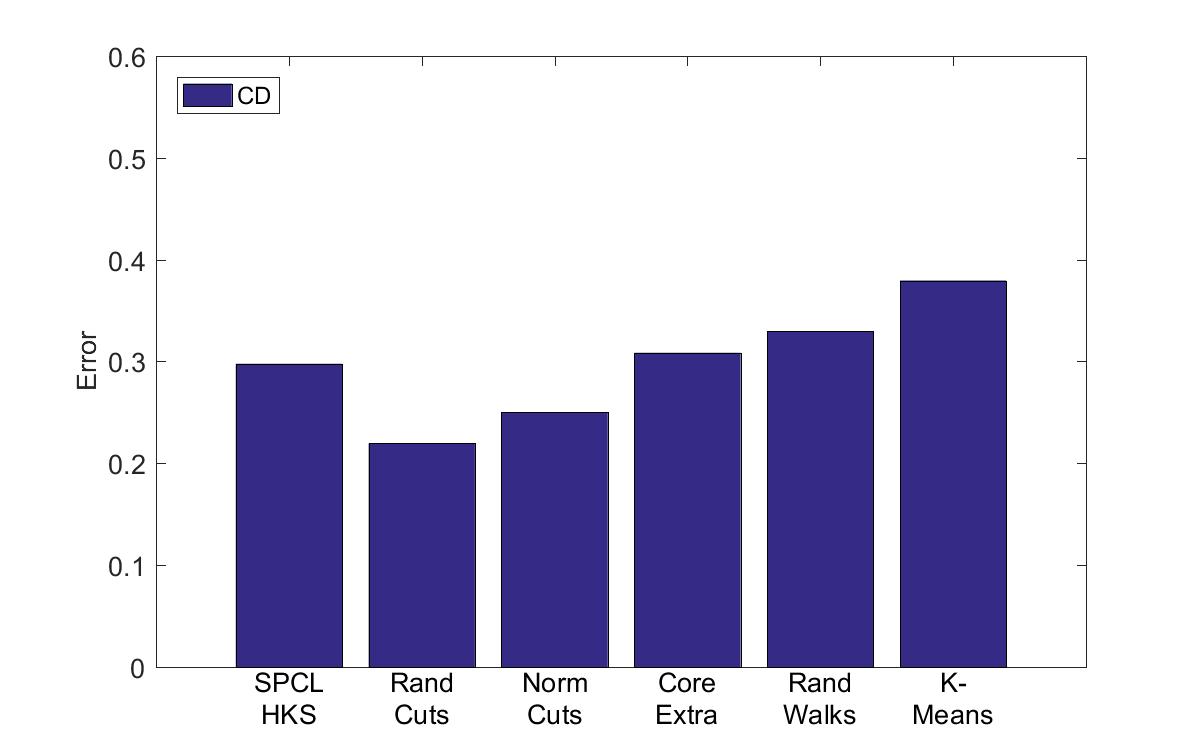}
        & 
      \includegraphics[dpi=72,trim=10px 0px 100px 0px,totalheight=75pt]{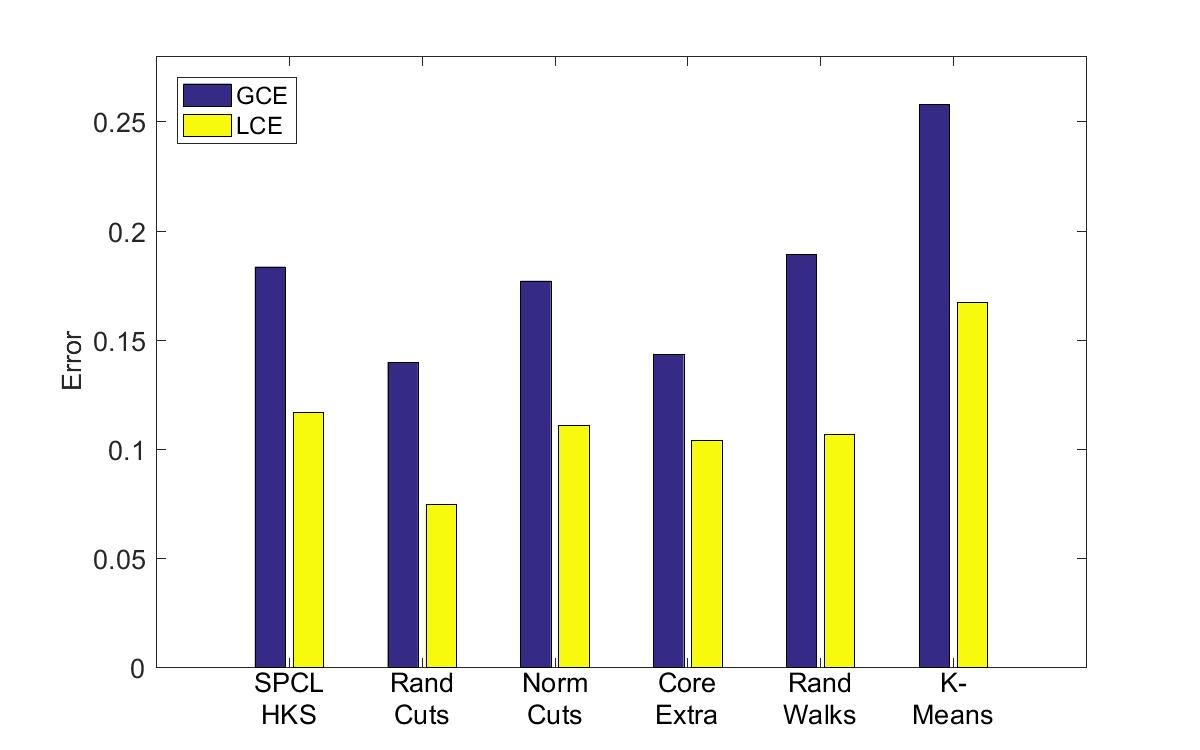}
		 &
      \includegraphics[dpi=72,trim=10px 0px 100px 0px,totalheight=75pt]{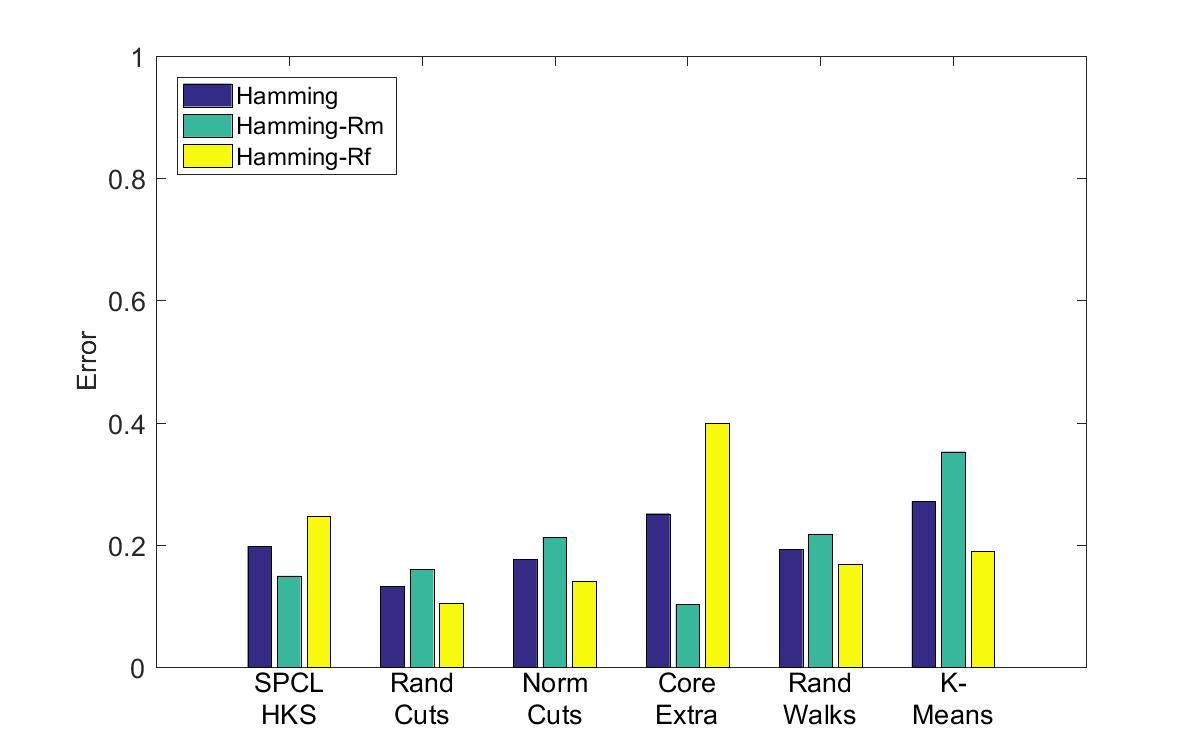}
        & 
      \includegraphics[dpi=72,trim=10px 0px 100px 0px,totalheight=75pt]{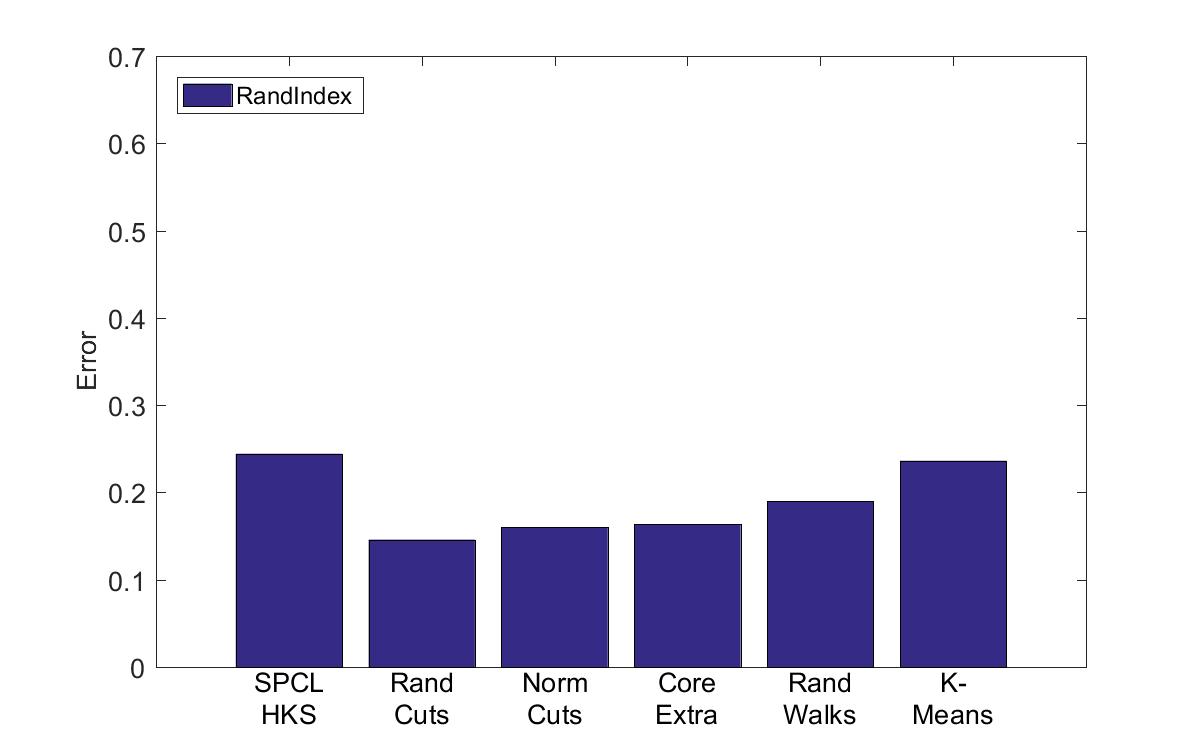}
      \\
      (e) & (f) & (g) & (h)\\
		\vspace{-10pt}
     \end{tabular}
\end{center}
\caption{For SPCL+HKS with naive persistent segmentations on point clouds vs other segmentation methods on mesh models from the Mesh Segmentation Benchmark of \cite{chen2009benchmark}: (a-d) Plots by Category of Cut Discrepency, Consistency Error, Hamming Distance, and Rand Index; (e-h) Plots by Segmentation of Cut Discrepency, Consistency Error, Hamming Distance, and Rand Index.} \label{MeshBenchmark}
\end{figure*}

To compare the output produced by using the SPCL and HKS with automatic persistence-based segmentations on point cloud models without a surface mesh, we sample the vertices from the models in \cite{chen2009benchmark}. We run our methods on this dataset, map our segments from points back to facets in the mesh models, for compatibility, by means of the \texttt{mode} function in Matlab over the vertices associated to each facet. The resultant plots in Figure \ref{MeshBenchmark} show that even the naive segmentations produced {\em automatically} by our example persistence-based segmentation display error metrics reasonably similar to those of the other segmentation methods tested in \cite{chen2009benchmark}, despite making no use of the surface mesh connectivity information exploited by the other algorithms. 

\begin{figure*}[!ht]
\begin{center}
    \begin{tabular}{cccccc}
    \includegraphics*[width=0.12\textwidth]{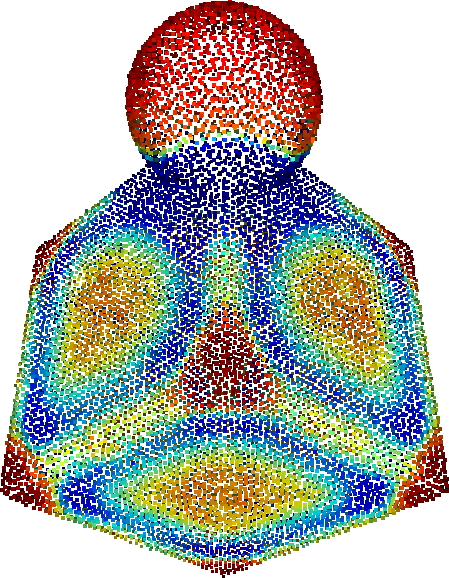} & 
    \includegraphics*[width=0.12\textwidth]{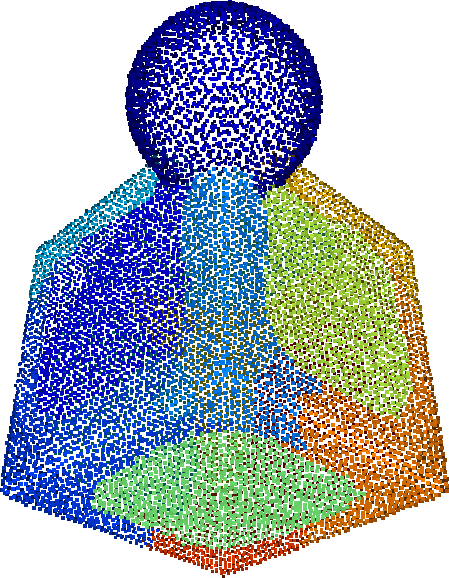} & 
    \includegraphics*[width=0.12\textwidth]{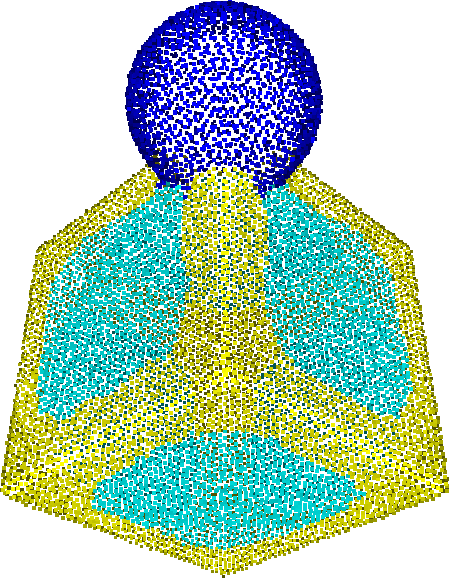} &
    \includegraphics*[width=0.12\textwidth]{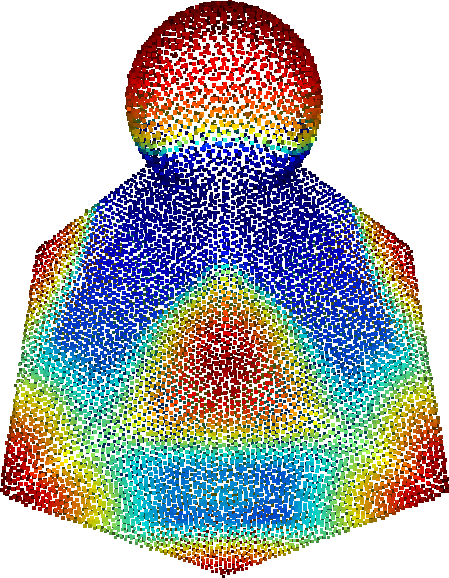} & 
    \includegraphics*[width=0.12\textwidth]{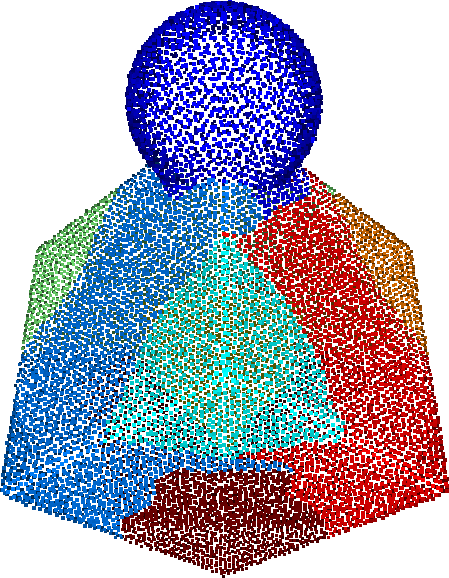} & 
    \includegraphics*[width=0.12\textwidth]{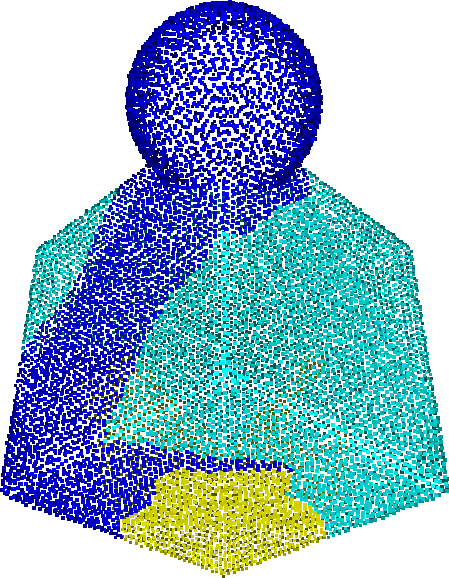}
    \\ \vspace{10 pt} (a) & (b) & (c) & (d) & (e) & (f) \\
    \includegraphics*[width=0.12\textwidth]{10a__00} & 
    \includegraphics*[width=0.12\textwidth]{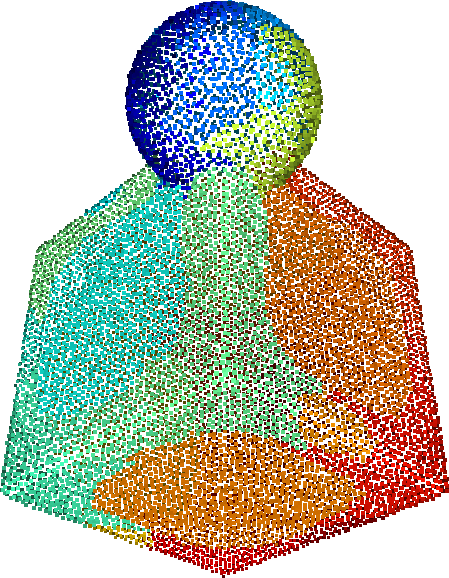} & 
    \includegraphics*[width=0.12\textwidth]{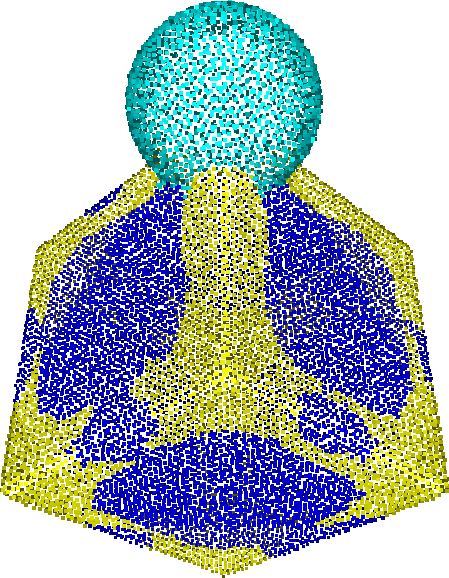} &
    \includegraphics*[width=0.12\textwidth]{10a__00} & 
    \includegraphics*[width=0.12\textwidth]{14b__09} & 
    \includegraphics*[width=0.12\textwidth]{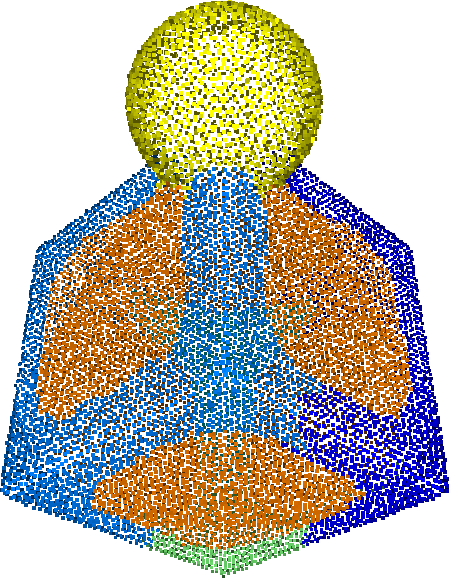}
    \\ \vspace{10 pt} (g) & (h) & (i) & (j) & (k) & (l)\vspace{-15pt} \\
    \end{tabular}
\end{center}
\caption{(a) HKS at $t=0.25$ with 300 SPCL eigs; (b) segmentation with $\tau=0.007$; (c) Re-clustered segmentation with $n=3$; (d) HKS at $t=1.0$ with 300 SPCL eigs, (e) segmentation with $\tau=0.007$, (f) Re-clustered segmentation with $n=3$; (g) HKS at $t=0.25$ with 300 SPCL eigs, (h) segmentation with $\tau=0.002$, (i) Re-clustered segmentation with $n=3$; (j) HKS at $t=0.25$ with 300 SPCL eigs, (k) segmentation with $\tau=0.007$; (l) Re-clustered segmentation with $n=6$.} \label{fig_ex0}
\end{figure*}


\begin{table*}[!htb]
\begin{center}
\scalebox{0.8}{
    \begin{tabu} to \textwidth {cX[r]X[r]X[r]X[r]X[r]}
    \textbf{model} & \multicolumn{1}{r}{\textbf{\# pts}} & \multicolumn{1}{r}{\textbf{\# eigs}} & \multicolumn{1}{r}{\textbf{SPCL time}} & \multicolumn{1}{r}{\textbf{eigs time}} & \multicolumn{1}{r}{\textbf{Total run time}} \\
\hline
Armadillo & 5528 & 300 & 1.4s & 9.9s & 0.25 min \\
Armadillo & 15228 & 300 & 4.0s & 22.7s & 0.59 min \\
camel     & 6815 & 300 & 1.6s & 10.0s & 0.22 min\\
camel     & 17036 & 300 & 4.15s & 21.3s & 0.45 min\\
turborotor & 24503 & 150 & 8.8s & 23.6s & 0.59 min\\
turborotor & 24503 & 300 & 8.7s & 50.7s & 1.05 min\\
turborotor & 24503 & 500 & 8.8s & 100.7s & 1.91 min\\
turborotor & 79723 & 300 & 21.2s & 159.8s & 3.21 min\\
turborotor & 150163 & 300 & 42.0s & 322.5s & 6.46 min\\
robot & 8446 & 300 & 2.4s & 16.6s & 0.34 min \\
robot & 39889 & 150 & 10.1s & 26.5s & 0.73 min \\
robot & 39889 & 300 & 10.1s & 62.3s & 1.35 min \\
c2u & 4494 & 150 & 1.1s & 3.12s & 0.11 min\\
c2u & 4494 & 300 & 1.1s & 5.6s & 0.16 min\\
c2u & 35836 & 150 & 9.5s & 35.6s & 1.07 min \\
c2u & 35836 & 300 & 9.5s & 76.5s & 1.78 min \\
    \end{tabu}}
\end{center}
\caption{Running times for various models by number of points and number of eigenvalues computed. Running times measured using Matlab's \texttt{tic} and \texttt{toc} functions; eigenpairs computed with Matlab's \texttt{eigs} function. All computations performed on a Dell XPS 13 9350 with an Intel Core i5-6200U @ 2.30 GHz and 8GB RAM running Windows 10 x64.} \label{runtimes}
\end{table*}


\subsection{A Brief Overview of Parameter Dependence}\label{param-dependence-appendix}

There are a number of parameters that control the computations of the SPCL, HKS as well as the clustering and segmentation methods described above. It is not within the scope of this paper to provide a detailed discussion of the influences of each parameter on the outputs or to make an attempt to optimize the various parameters. However, the results obtained from our methods are dependent upon careful selection of these parameters, so we will briefly demonstrate the effect upon the result of each parameter manipulation, in order to make applying these methods as straightforward as possible. Figure \ref{fig_ex0} shows the synthetic 14,124 point ``c4u'' ``ball \& cube'' model with model-scale Gaussian noise ($\mu=\epsilon/2$, $p=0.125$) added and the baseline parameter values against which we will be comparing parameter changes throughout the Section. 

\subsubsection*{HKS $t$-scale}
The $t$-scale(s) parameter determines the extent to which the HKS represents the surface either locally or globally. Figure \ref{fig_ex0}(d)-(f) shows the effect on clustering of choosing a higher $t$-value (and therefore equivalently characterizing the model by a larger neighborhood around each point) for HKS computation. Note that we do not adjust the clustering parameter to find a better segmentation of the surface for this higher $t$-value HKS in order to show the interaction between the parameters in each step.

\subsubsection*{Clustering parameter $\tau$}
The third parameter, $\tau$, controls region merging in the agglomerative clustering procedure. Higher $\tau$ values cause more regions to merge, resulting in fewer clusters whereas lower $\tau$ values prevent regions from merging, resulting in a larger number of clusters. Figure \ref{fig_ex0}(g)-(i) exemplify the result of an excessively low $\tau$ on the clustering output. Notice, however, that despite a larger-than-desired number of segments (57 rather than the 14 shown in Figure \ref{fig_ex0}(b), the reclustered model appears very similar to that in the baseline example in Figure \ref{fig_ex0}(c).

\subsubsection*{Unique cluster-type estimate $n$}
Finally, the proposed re-clustering procedure requires a user-defined estimate of the number of unique sub-shapes output by the re-clustering algorithm. More advanced clustering procedures which rely less on \textit{a priori} information, such as spectral clustering \cite{specclust} or clustering by reference to the gap statistic \cite{gap}, or assisted by machine learning algorithms may be used to reduce this dependence of final sub-shape grouping on user input. These and other methods for improving segmentation of point cloud models are the subject of future investigations. Figure \ref{fig_ex0}(l) shows the poorer sub-shape grouping, compared to that of Figure \ref{fig_ex0}(c) produced by sub-shape family estimation when a higher-than-optimal estimate of the number of present sub-shape families is used.

Additionally, we note that two factors primarily contribute to longer running times for a given model: the number of points in the model (equivalently, the sampling of the model) and the number of SPCL eigenvalues computed for the HKS calculation. As shown in Table \ref{runtimes}, for a given model, increasing sampling density and increasing the number of computed eigenvalues both appear to increase running time near linearly. This is as we should expect: increasing the number of points in a model increases the number of rows of SPCL linearly (and each row's computation is constant in number of model points) and the increased number of SPCL rows increases the eigenvalue computation runtime, which in general is $O(n^3)$. In practice, however, for sparse matrices like the SPCL, larger numbers of points seem to effect runtime of Matlab's \texttt{eigs} function linearly. Similarly, increasing the number of eigenvalues to be computed linearly increases the number of eigenvalue solution steps taken, the runtime for each of which depends only on number of model points.

\section{Conclusions}

Depth camera technology is becoming ubiquitous, allowing easy capture of dense, noisy point clouds models of real objects. The obvious questions of how we can automatically understand shapes from general 3D point cloud inputs have been traditionally answered by first performing a surface reconstruction on the input point cloud, followed by mesh processing. Unfortunately, meshing point clouds is itself a non-trivial, often application-dependent task with limited guarantees on the validity of the resulting mesh. 

On the other hand, useful analysis methods for describing and segmenting point clouds {\em{without surface reconstruction}}, especially those having engineering relevance, are quite limited in the literature. ``Mesh model'' analysis methods are well-developed, yet creating a quality surface mesh from general noisy point cloud inputs automatically remains challenging.

We presented an integrated analysis framework for shape description, similarity, and segmentation on point cloud models of real objects. We have shown that our symmetric Laplace-Beltrami operator estimate, the SPCL, does not worsen the error terms of the PCDL, which is known to converge in the limit to the manifold Laplacian. This, in turn, allows us to apply existing or develop new physics-based spectral shape signature methods directly on the point cloud.  We showed that our construction provides the neighborhood graph implied by the SPCL, which in turn can be used to find mesh $n$-ring--equivalent local neighborhoods and to apply  algorithms from the mesh literature (such as those for finding feature vectors) to point cloud models, affording a compact similarity-based tool for comparing the shapes represented by point clouds. At the same time, the proposed framework supports in principle any other convergent estimate of the LBO. Thanks to its reliance on models of physical phenomena, this comparison tool is highly robust against noise; various mesh based signatures, which now can be applied to point clouds within our framework, have even been shown to resist mis-categorizing incomplete or damaged models.  Consequently, we illustrated the effectiveness of the proposed framework by analyzing a database of point cloud models output by Kinect, which contain defects and noise that makes them resistant to meshing. 

The framework proposed here, which enables the capability to perform similarity and segmentation directly on point clouds, can be adapted to most application domains that require point cloud processing, including robotics, design and manufacturing, and opens the doors to a number of engineering applications. For example, this framework can easily be applied to compare point cloud of physical artifacts to CAD model databases, irrespective of the native CAD format, obviating the need to perform solid model reconstruction or to deal with the difficult CAD interoperability issues. (Models for which a Hermitian LBO estimate is available can have spectral signatures computed on them in the same way we here describe for point cloud models; for those that do not or for which implementing code for such procedure would require excessive effort or time for some group, a Monte Carlo sampling of the surface allows the point cloud methods we present to be used to make comparisons). The concept of \textit{clustering the clusters} introduced here, which gives the ability to determine the number of similar features in a given point cloud model, is critical in manufacturing planning as well as geometric reasoning. Our method of re-clustering by signature values lets us extract further salience information at low computational cost, and is a feature unique to segmentations that retain similarity information from the signature(s) on which they are based. Finally, to facilitate real-world application of this new analysis procedure, we have included appendices in which we provide a guide to constructing the SPCL and examine the manner in which application results depend on the parameters in our methods. 

Finally, our method can be viewed as a general purpose point cloud analyzer, although there are several challenges that need to be addressed next, including: understanding the relationship between the number of eigenpairs and the quality of the spectral shape similarity measures; choosing automatically an appropriate number of segments for a given model; a better understanding of how to automatically set the available parameters which influence the similarity and segmentation performance; and finally GPU acceleration of the eigensystem and clustering computations. 


\section*{Acknowledgments}
This work was supported in part by the National Science Foundation grants CMMI-1462759, IIS-1526249, and CMMI-1635103. Reed Williams was also partially supported through the General Electric Fellowship for Innovation.

\bibliography{bib_CAD_0_jab.mar25}
\bibliographystyle{elsarticle-num}

%
%
%

\end{document}